\documentclass[letterpaper,journal]{IEEEtran}

\PassOptionsToPackage{table}{xcolor}

\usepackage{tikz}
\usepackage{amsmath}
\usepackage{pifont}
\usepackage{subcaption}
\usepackage{multirow}
\usepackage{filecontents}
\usepackage{siunitx}
\usepackage{graphicx}
\usepackage{xcolor} % No need to specify options here anymore
\definecolor{IEEEblue}{RGB}{188, 32, 110}

\usepackage{enumitem}
\usepackage{amsmath}
\usepackage{makecell}
\usepackage{tabularx,booktabs}
\usepackage{cite}               % order multiple entries in \cite{...}
\usepackage{breakurl}           % break too-long urls in refs
\usepackage{url}                % allow \url in bibtex for clickable links
\usepackage[]{hyperref}         % ...clickable refs within pdf...
\hypersetup{                    % ...like so
  colorlinks,
  linkcolor={IEEEblue!80!black},
  citecolor={IEEEblue!70!black},
  urlcolor={IEEEblue!70!black}
}

\usepackage{tablefootnote}
\usepackage{threeparttable}
\usepackage{comment}
\usepackage{adjustbox}
\usepackage{cleveref}
\crefname{figure}{Fig.}{Figs.}
\crefname{table}{Tab.}{Tabs.} 
\crefname{algorithm}{Alg.}{Algs.}
\crefformat{table}{\textcolor{IEEEblue}{#2Tab.~#1#3}}
\Crefformat{table}{\textcolor{IEEEblue}{#2Tab.~#1#3}}
\crefmultiformat{table}
    {\textcolor{IEEEblue}{Tabs.~#2#1#3}}
    {\textcolor{IEEEblue}{, #2#1#3}}
    {\textcolor{IEEEblue}{, #2#1#3}}
    {\textcolor{IEEEblue}{ and #2#1#3}}
\crefrangeformat{table}{\textcolor{IEEEblue}{Tabs.~#3#1#4 to~#5#2#6}}
\crefformat{figure}{\textcolor{IEEEblue}{#2Fig.~#1#3}}
\Crefformat{figure}{\textcolor{IEEEblue}{#2Fig.~#1#3}}
\crefmultiformat{figure}
    {\textcolor{IEEEblue}{Figs.~#2#1#3}}
    {\textcolor{IEEEblue}{, #2#1#3}}
    {\textcolor{IEEEblue}{, #2#1#3}}
    {\textcolor{IEEEblue}{ and #2#1#3}}
\crefrangeformat{figure}{\textcolor{IEEEblue}{Figs.~#3#1#4 to~#5#2#6}}
\crefformat{algorithm}{\textcolor{IEEEblue}{#2Alg.~#1#3}}
\Crefformat{algorithm}{\textcolor{IEEEblue}{#2Alg.~#1#3}}
\crefmultiformat{algorithm}
    {\textcolor{IEEEblue}{Algs.~#2#1#3}}
    {\textcolor{IEEEblue}{, #2#1#3}}
    {\textcolor{IEEEblue}{, #2#1#3}}
    {\textcolor{IEEEblue}{ and #2#1#3}}
\crefrangeformat{algorithm}{\textcolor{IEEEblue}{Algs.~#3#1#4 to~#5#2#6}}

\usepackage{amsfonts}
\usepackage{algorithm}
\usepackage{algorithmic}
\usepackage{amsmath}
\usepackage{soul}
\usepackage{mathtools}
\usepackage{diagbox}
\newcommand{\ie}{\textit{i}.\textit{e}.}

\usepackage{ulem}

\hypersetup{linkcolor=IEEEblue}

\begin{document}
%-------------------------------------------------------------------------------

\title{Exploiting Vulnerabilities in Speech Translation Systems through Targeted Adversarial Attacks}

\author{
	\IEEEauthorblockN{
        Chang~Liu\textsuperscript{1},
        Haolin~Wu\textsuperscript{2},
        Xi~Yang\textsuperscript{3},
        Kui~Zhang\textsuperscript{4},
        Cong~Wu\textsuperscript{5},
        Weiming~Zhang\textsuperscript{1},\\
        Nenghai~Yu\textsuperscript{1},
        Tianwei~Zhang\textsuperscript{5},
        Qing~Guo\textsuperscript{6},
        and~Jie~Zhang\textsuperscript{6}}
        
	\IEEEauthorblockA{\textsuperscript{1}University of Science and Technology of China, Hefei, China}\\
	\IEEEauthorblockA{\textsuperscript{2}Wuhan University, Wuhan, China}\\
	\IEEEauthorblockA{\textsuperscript{3}The Hong Kong University of Science and Technology, Hong Kong, China}\\
	\IEEEauthorblockA{\textsuperscript{4}Huawei Noah's Ark Lab, Shanghai, China}\\
	\IEEEauthorblockA{\textsuperscript{5}Nanyang Technological University, Singapore}\\
	\IEEEauthorblockA{\textsuperscript{6}CFAR and IHPC, A*STAR, Singapore}\\
}

\maketitle

\begin{abstract}
As speech translation (ST) systems become increasingly prevalent, understanding their vulnerabilities is crucial for ensuring robust and reliable communication. However, limited work has explored this issue in depth. This paper explores methods of compromising these systems through imperceptible audio manipulations. Specifically, we present two innovative approaches: (1) the injection of perturbation into source audio, and (2) the generation of adversarial music designed to guide targeted translation, while also conducting more practical over-the-air attacks in the physical world.

Our experiments reveal that carefully crafted audio perturbations can mislead translation models to produce targeted, harmful outputs, while adversarial music achieve this goal more covertly, exploiting the natural imperceptibility of music. These attacks prove effective across multiple languages and translation models, highlighting a systemic vulnerability in current ST architectures.

The implications of this research extend beyond immediate security concerns, shedding light on the interpretability and robustness of neural speech processing systems. Our findings underscore the need for advanced defense mechanisms and more resilient architectures in the realm of audio systems. More details and samples can be found at \url{https://adv-st.github.io}.
\end{abstract}

\begin{IEEEkeywords}
Speech translation, Targeted adversarial attack, Adversarial music
\end{IEEEkeywords}

\section{Introduction}
The world's languages and indigenous tongues have diverse origins, with speech being the most widely recognized tool of the information exchange. On average, a person speaking over 11,000 words daily~\cite{dhawan2022speech}. However, communication becomes ineffective when the parties involved do not share a common language. As the Internet, smart devices, and the metaverse advance~\cite{wang2022survey}, cross-cultural interactions have become increasingly convenient and more frequent.
Yet, language remains a significant obstacle to effective information transmission as in this increasingly interconnected world.

Translation systems play a crucial role in bridging linguistic gaps by accurately conveying meaning and context across languages. Effective translation requires understanding semantic content to preserve intent and nuances, ensuring true comprehension and efficient information exchange~\cite{seamless-blog}. This is particularly important in the digital age, where the demand for translating multimedia content, including streaming videos, entertainment platforms, and educational resources, continues to grow. Advanced translation systems are key to maintaining semantic fidelity and enhancing global accessibility.

\begin{figure}[t]
\includegraphics[width=\columnwidth]{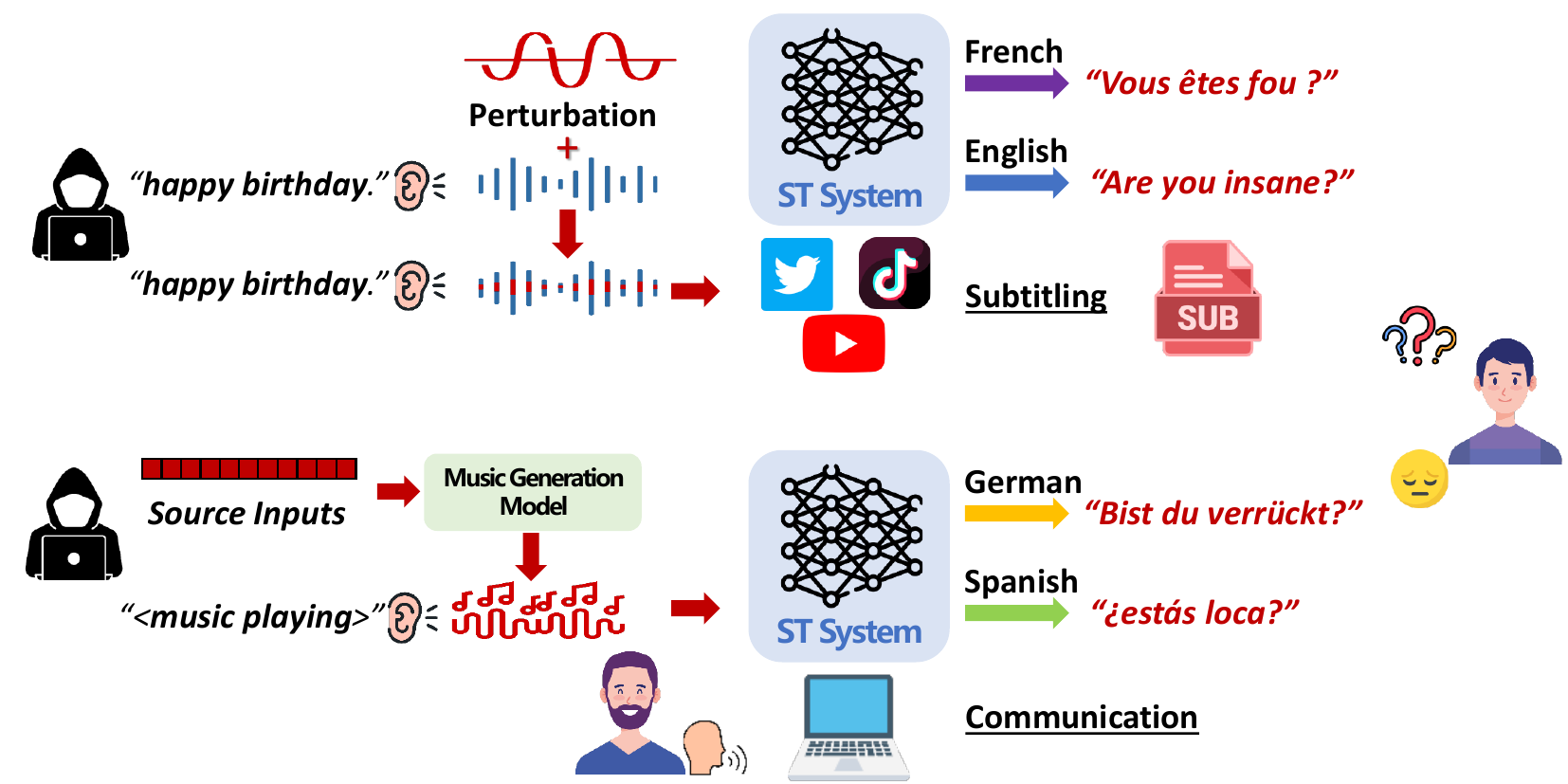}
\caption{Two attack methods on the speech translation (ST) system: 1) adding imperceptible perturbation to audio, and 2) generating adversarial music. Both methods cause malicious translations ``Are you insane?" across languages in this case.}
\label{fig:main}
\end{figure}

Fortunately, speech translation (ST)~\cite{lavie1997janus, wahlster2013verbmobil, nakamura2006atr, inaguma2020espnet, barrault2023seamless, wang2024end, canary} is emerging as a transformative technology.
At its core, ST technology converts spoken words from one language into texts and speech in another, effectively bridging communication gaps between speakers of different languages. 
Multilingual ST systems extend this capability by supporting translation between multiple language pairs, creating new opportunities for global interaction.
These systems preserves the linguistic information contained in the source speech and reproduces it as text and speech in the target language, maintaining the nuances and intent of the original message.

Early ST systems focused on speech-to-text tasks, relying on cascaded architectures, which combines Automatic Speech Recognition (ASR) and Machine Translation (MT) modules~\cite{lavie1997janus, wahlster2013verbmobil}. While modular designs allowed component-level optimization, they suffered from error propagation~\cite{ney1999speech, matusov2005integration}. Afterwards, end-to-end methods integrated ASR and MT into a single neural network to achieve direct speech-to-text translation~\cite{berard2016listen}. With the advances in encoder-decoder architectures~\cite{bansal2018low} and large-scale datasets~\cite{iranzo2020europarl},  these speech-to-text models can be integrated with text-to-speech modules for the whole speech-to-speech translation. These new generation ST systems, such as Seamless model family~\cite{barrault2023seamless, barrault2023seamlessm4t}, showcase the transformative impact of large language models (LLMs) on ST. These systems leverage joint pre-training and large-scale alignment to support many languages, including low-resource ones, achieving speech-to-text and speech-to-speech translation. Such advancements represent a critical step toward building intelligent, efficient, and accessible speech translation technologies.

However, as with any emerging technology, ST systems are not immune to vulnerabilities. As these systems become more prevalent in our daily lives, understanding and addressing their potential weaknesses becomes crucial for ensuring robust and reliable communication. 
In parallel, the field of adversarial attacks on speech systems has rapidly developed, addressing security issues in areas such as Voice Conversion (VC)~\cite{huang2021defending, yu2023antifake, liu2023protecting},  ASR~\cite{carlini2016hidden, yuan2018commandersong, abdullah2021hear, chen2020metamorph, schonherr2018adversarial, yu2023smack}, and Speaker Recognition (SR)~\cite{chen2021real, li2023enrollment, chen2023qfa2sr}. 
Despite the growing importance of ST models, security concerns have not been sufficiently explored, particularly for leading models like Seamless~\cite{seamless-blog}. 
Moreover, these models follow a paradigm similar to large language models, progressively predicting the next token via autoregressive methods \cite{barrault2023seamless, canary}, which makes existing adversarial techniques for end-to-end ASR models ineffective \cite{yakura2018robust, carlini2016hidden, yuan2018commandersong, carlini2018audio, chen2020metamorph}.

To address this gap, this paper investigates methods of compromising ST systems through imperceptible audio manipulations. As shown in \autoref{fig:main}, our research explores two innovative \textbf{\textit{targeted}} adversarial attack approaches that expose potential vulnerabilities in current ST models:
\noindent{\textbf{1) Injection of imperceptible perturbations into source audio}}: We design our core attack using teacher-forcing goal supervision \cite{vaswani2017attention} and enhance its impact on the model’s semantic understanding through a Multi-language Enhancement scheme, improving its generalizability. To increase the effectiveness of targeted semantic attacks, we employ Target Cycle Optimization. Additionally, we improve adversarial perturbation imperceptibility and generalization by constraining the noise within the mid-frequency range using filtering techniques.
\noindent{\textbf{2) Generation of adversarial music}}:
Interestingly, we observe that ST models translate pure music into specific sentences, which differs from human perception. Based on this observation, we present a technique for creating music designed to trigger targeted mistranslations. By optimizing the diffusion-based music generation process, we demonstrate the feasibility of guiding ST system towards predetermined malicious outputs. This novel attack expands the attack surface to include communication environments with background music, raising concerns about the vulnerability of ST systems in real-world scenarios. 

Our experiments reveal that these two attacks are effective across multiple languages and ST models, indicating a systemic vulnerability in current state-of-the-art ST architectures. The implications of this research extend beyond immediate security concerns, shedding light on the interpretability and robustness of neural speech processing systems. These findings underscore the urgent need for developing more resilient ST models and implementing robust defense mechanisms against such sophisticated attacks.

In summary, our key contributions are as follows:

\begin{itemize}[leftmargin=*]
  \item To the best of our knowledge, this is the first attempt to investigate adversarial attacks on speech translation (ST) models. Our work pioneers the exploration of vulnerabilities in deep speech models that utilize a novel paradigm combining large language model structures with discrete token encoding and autoregressive prediction.
  \item We develop a targeted attack scheme by thoroughly analyzing the structure and operational mode of the speech translation model. Specifically, we enhance the semantic impact of attacks through Multi-language Enhancement to improve generalization and further boost performance using Target Cycle Optimization.
  \item We introduce an innovative adversarial music attack based on a diffusion music generation model, enabling more covert and naturalistic attacks. This is the first application of music generation models in speech adversarial attack research, demonstrating their capability to reduce the perceptibility of adversarial examples effectively.
  \item Experimental results demonstrate that the proposed methods can effectively carry out targeted attacks and achieve cross-lingual semantic attack transfer. 
\end{itemize}

\section{Related Work}
\subsection{Speech Translation Systems}

The goal of speech translation (ST) is to convert speech from one language into text and speech in another language, enabling cross-linguistic information understanding. 

Early ST primarily relied on cascaded systems, which implemented cross-lingual conversion by sequentially combining ASR and MT modules~\cite{lavie1997janus, wahlster2013verbmobil, nakamura2006atr}. While such modular approaches allowed independent optimization of individual components, their inherent error propagation significantly constrained performance. 

\begin{figure}[t]
\includegraphics[width=\columnwidth]{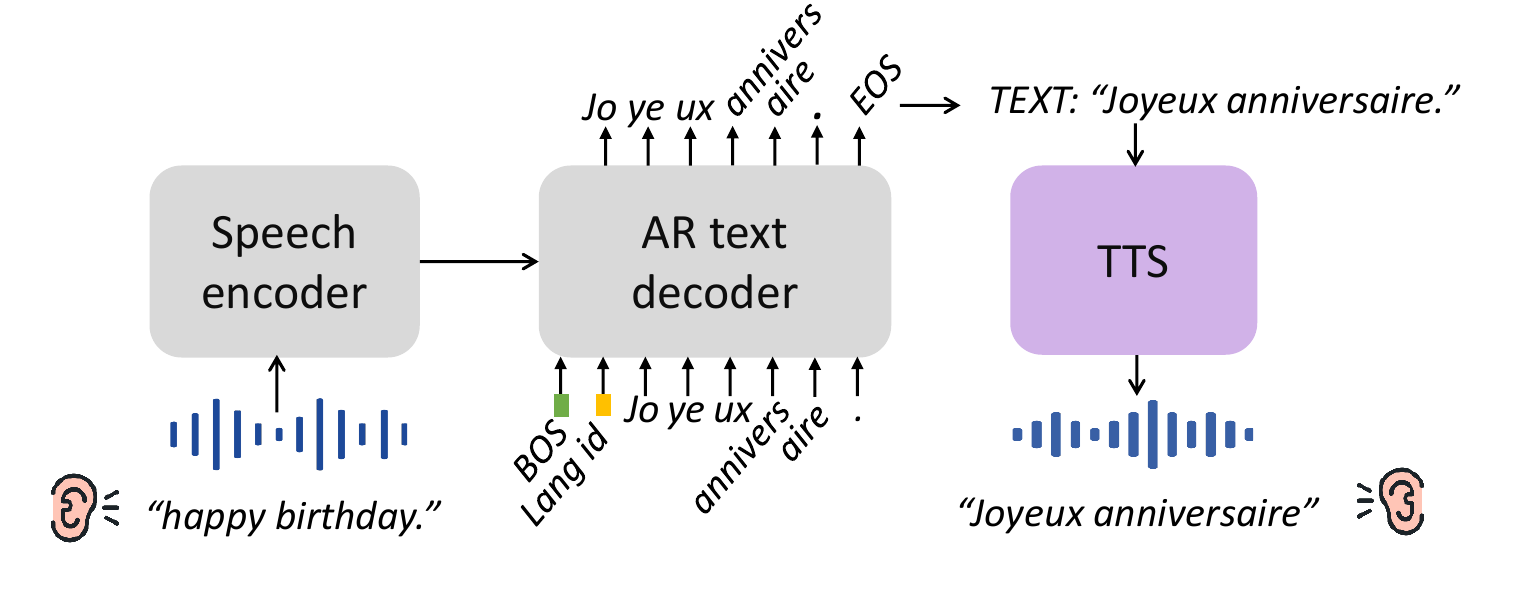}
\vspace{-2em}
\caption{A Standard E2E ST framework features a speech encoder and an autoregressive text decoder that generate translated text in the target language end-to-end (French, in this case). An additional TTS module can be used to convert the translated text into speech, providing full functionality for a ST system.}
\label{fig:s2st}
\end{figure}

To overcome the limitations of cascaded systems, end-to-end (E2E) speech translation methods emerged. These approaches integrate ASR and MT into a single neural network, directly converting source language speech into target language text~\cite{berard2016listen}. Advances in encoder-decoder architectures~\cite{bansal2018low} and the development of specialized end-to-end datasets~\cite{iranzo2020europarl} have significantly enhanced the performance of E2E models. The Canary system introduced an innovative tokenizer design and leveraged large-scale training, achieving groundbreaking results in multilingual translation tasks. Furthermore, these standard end-to-end speech-to-text translation models can incorporate an additional TTS module to achieve the goal of speech-to-speech translation, as illustrated in \autoref{fig:s2st}.

\begin{figure}[t]
\includegraphics[width=\columnwidth]{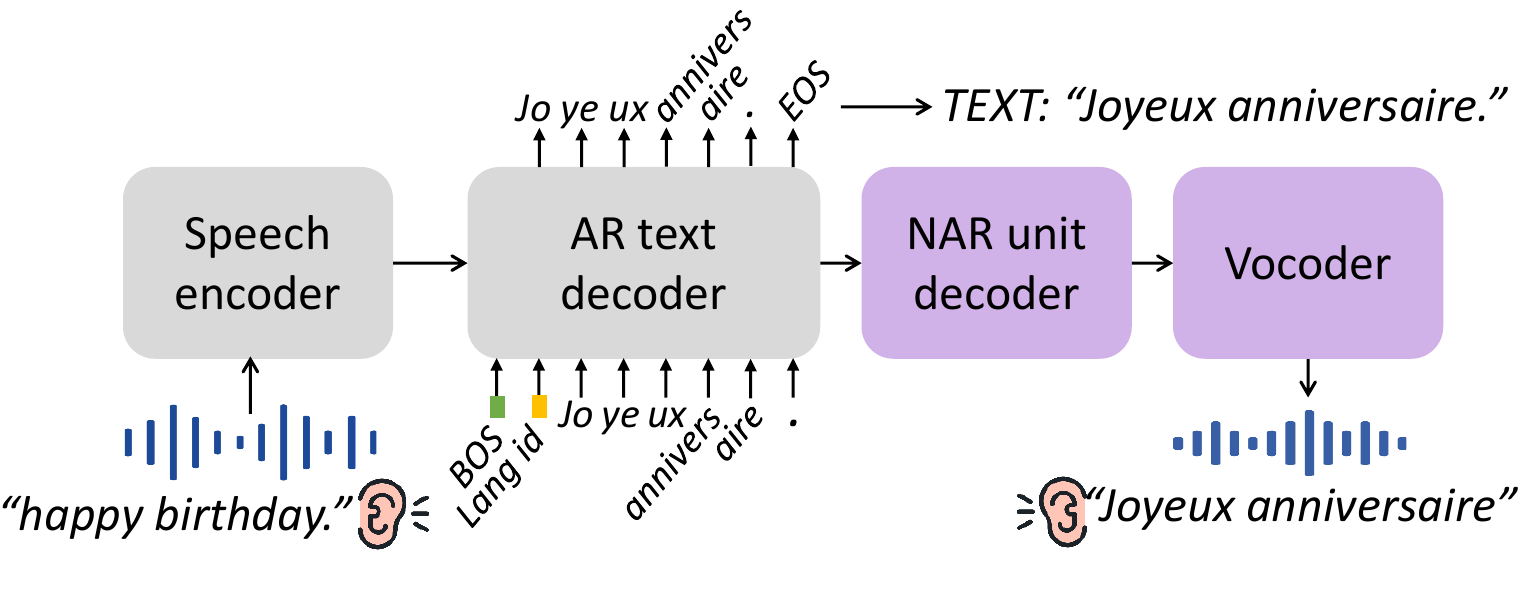}
\vspace{-2em}
\caption{Speech-to-any translation framework, where the features generated during the decoding of the target language text  (French, in this case) are subsequently leveraged to predict audio features.}
\vspace{-1em}
\label{fig:ds2st}
\end{figure}

Despite these significant advances, challenges persist in 
achieving robust multilingual semantic understanding. Research efforts continue to focus on developing more
generalizable translation systems to achieve truly seamless cross-lingual communication. 
With the advent of large language models (LLMs), speech translation has entered a transformative era. 
Speech-LLaMA~\cite{wu2023decoder} highlights the potential of transformer-based~\cite{vaswani2017attention} LLM architectures for speech understanding and translation. Language modeling-based joint pre-training of speech and text data~\cite{bapna2022mslam} has delivered substantial performance improvements across diverse tasks. 
Comprehensive frameworks like Seamless model family~\cite{barrault2023seamless, barrault2023seamlessm4t, seamless-blog}, built on the UnitY2 framework, leverage large-scale training and alignment to support a wide range of languages, including many low-resource ones. Notably, Seamless achieves true speech-to-any translation, marking a milestone in cross-lingual communication. As shown in \autoref{fig:ds2st}, modern systems seamlessly handle both speech-to-text and direct speech-to-speech translation, demonstrating exceptional versatility and robustness. These advancements mark a critical step toward more intelligent, efficient, and accessible speech translation technology.

\subsection{Adversarial Attacks on Speech Systems}
Currently, adversarial attacks on speech processing systems primarily target Automatic Speech Recognition (ASR), Automatic Speaker Verification (ASV), and Voice Conversion (VC) systems.

\noindent
\textbf{Adversarial Attacks on ASR.}
Adversarial attacks on ASR systems primarily craft waveforms that sound like original speech to human listeners but deceive the ASR model \cite{schonherr2018adversarial}. These attacks can lead to hidden voice commands being issued without detection, resulting in various real-world threats \cite{xiong2023fundamentals}.
Recent research has shifted towards black-box adversarial attacks, which require only the final transcription from ASR systems. However, these attacks often involve numerous queries to the ASR, leading to substantial costs and increased detection risk. To address these limitations, novel approaches like ALIF~\cite{cheng2023alif} have been developed, leveraging the reciprocal process of Text-to-Speech (TTS) and ASR models to generate perturbations in the linguistic embedding space.

\noindent
\textbf{Adversarial Attacks on ASV.}
Adversarial attacks on ASV systems have evolved from targeting binary systems to more complex x-vector systems, considering practical scenarios such as over-the-air attacks \cite{kreuk2018fooling, li2020practical, xie2020real}. To overcome the challenge of obtaining gradient information in real-world scenarios, researchers have developed query-based adversarial attacks like FakeBob~\cite{chen2021real} and SMACK~\cite{yu2023smack}.
More recent approaches include transfer-based adversarial attacks and speech synthesis spoofing attacks. A notable development is the Adversarial Text-to-Speech Synthesis (AdvTTS) method, which combines the strengths of transfer-based adversarial attacks and speech synthesis spoofing attacks \cite{zuo2024advtts}.

\noindent
\textbf{Adversarial Attacks on VC.}
Voice Conversion (VC) technology transforms the speaker characteristics of an utterance without altering its linguistic content, raising concerns about privacy and security. Recent works have introduced adversarial attacks on VC systems to prevent unauthorized voice conversion. For instance, adversarial noise can be introduced into a speaker's utterances, making it difficult for VC models to replicate the speaker’s voice \cite{huang2021defending}. 
To address the growing threat of deepfake speech, the AntiFake system \cite{yu2023antifake} was developed as a defense mechanism against unauthorized speech synthesis. This system applies adversarial perturbations to a speaker's audio to protect against deepfake generation, achieving high protection rates against state-of-the-art synthesizers. Additionally, efforts have been made to safeguard public audio from being exploited by attackers, with methods designed to degrade the performance of speech synthesis systems while maintaining the utility of the original speaker’s voice \cite{liu2023protecting}.

\textit{This paper is the first to investigate adversarial attacks on speech translation (ST) models. }
\section{Attack Overview}
\subsection{Threat Model}
\label{sec:tm}
In this paper, we examine an attacker's attempt to create audio Adversarial Examples (AEs) designed to deceive a speech translation model. The goal is to manipulate the model into recognizing the AE as a sentence with targeted semantics. 
Since the target model has a large number of parameters and has acquired a relatively strong understanding of semantics through large-scale pretraining~\cite{barrault2023seamless}, attacking such a model is challenging. We assume that the attacker has access to the model's parameters and can obtain gradients in our white-box investigation.

\begin{figure}[t]
\includegraphics[width=\columnwidth]{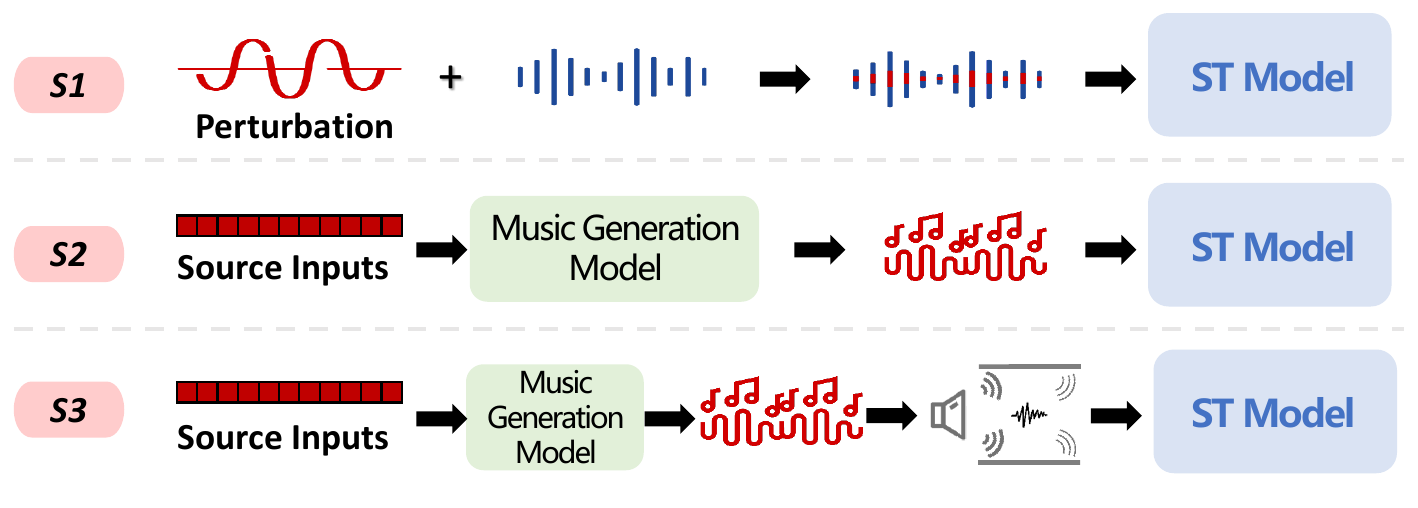}
\vspace{-1em}
\caption{Three threat model scenarios discussed: \textit{S1}: Cover-Related Attack, \textit{S2}: Cover-Independent Attack, \textit{S3}: Over-the-Air Attack.}
\label{fig:tm}
\vspace{-1em}
\end{figure}

In this threat model, we explore scenarios where an attacker attempts to manipulate a speech translation model $\mathbf{M}$ to produce targeted translations. The attack focuses on exploiting automatic translation systems used by international video platforms (e.g., YouTube) and in real-time multilingual settings, such as international conferences. As illustrated in \autoref{fig:tm}, we outline three distinct attack scenarios, each with a unique approach to achieving the desired malicious output:

\noindent
{\textit{\ul{S1: Cover-Related (Cover-Based Perturbation).}}}
In this scenario, the attacker targets a specific piece of audio, such as a segment of a video or a spoken sentence in a recording, and applies adversarial perturbations. These small, carefully crafted changes to the audio are undetectable to human listeners but force the translation model $\mathbf{M}$ to recognize it as a predefined, malicious semantic meaning.

For instance, an attacker could replace a segment of audio in a YouTube video with modified adversarial audio. When the platform’s automatic translation subtitling feature processes this audio, it may translate it into the target language based on the attacker’s intended meaning, potentially misleading viewers or injecting inappropriate content into the subtitles, as shown in \autoref{fig:main}.

\noindent
{\textit{\ul{S2: Cover-Independent (Synthetic Audio).}}}
Here, the attacker does not start with a specific audio recording but instead synthesizes a piece of music engineered to carry an adversarial signal. The adversarially crafted sound is designed so that, when processed by the translation model $\mathbf{M}$, it will be recognized as a specific phrase or meaning, even though it sounds like harmless background audio to human listeners.

This type of audio can be embedded in various media, such as videos or podcasts, and disseminated into international video platforms like YouTube. When the platform’s translation model processes the embedded sound, it produces the attacker’s intended semantic meaning in the target language.

This method allows the attacker to covertly manipulate content without relying on pre-existing speech recordings. Furthermore, a pre-generated piece of music can be reused multiple times, in contrast to T1, where optimization is required for each individual sample. This significantly reduces the cost of launching the attack.

\noindent
{\textit{\ul{S3: Over-the-Air Attack.}}}
In the third scenario, the attacker further enhances the adversarial robustness of the crafted audio, creating an audio signal that can survive over-the-air distortions, such as playback over speakers and capture by microphones in a conference environment. The adversarial audio is designed so that when it is played out and captured by any microphone, the model $\mathbf{M}$ will interpret it as a specific inappropriate or misleading phrase.

This technique allows an attacker to influence real-time translation systems used in multilingual conferences and conversations. For instance, the attacker could play this adversarial audio during a session, causing the translation system to deliver inappropriate or misleading messages to attendees in various languages. This poses a severe risk to the integrity of international communication and could lead to misunderstandings or conflicts in high-stakes settings.

These scenarios demonstrate various attack pathways on speech translation systems, from targeted content manipulations to generalized audio signals causing malicious translations. They highlight vulnerabilities and emphasize the need for robust defenses against adversarial attacks in real-world settings.

\subsection{Attack Strategy}
\label{sec:attackScenarioTestbed}
As mentioned above, we explore two types of attacks to investigate the vulnerabilities of the ST model and propose an enhancement strategy that improves the adversarial robustness of the crafted audio, making it resilient to real-world over-the-air distortions.

\vspace{1mm}
\noindent
\textbf{Perturbation-based Attack.} In this method, carefully crafted perturbations serve as the adversarial information. This approach requires an original speech sample to act as a carrier for the perturbations. 
As shown in \autoref{fig:main}, the attacker adds adversarial perturbations to the original speech so that the adversarial example conveys the target semantics to the model, rather than the original semantics.

\vspace{1mm}
\noindent
\textbf{Adversarial Music-based Attack.} 
Here, the music itself carries the adversarial information, disguised as semantic camouflage. This method does not require an original speech sample and can stand alone as the attack vector.
As shown in \autoref{fig:main}, the attacker optimizes the input embedding of the music generation model so that the synthesized music conveys the target semantics to the model.

\vspace{1mm}
\noindent
\textbf{Enhance Stratetgy.} 
By simulating over-the-air distortions during the adversarial music generation process, we guide the music to resist specific distortions, thus enhancing its robustness in real-world environments.

\subsection{Target Victim Model}
In this paper, we consider two kinds of target models: Standard End-to-end ST Model and Speech-to-any ST Model.

\noindent\textbf{Standard End-to-end Speech Translation Model.}
As shown in \autoref{fig:s2st}, a basic end-to-end speech translation system maps a speech signal in the source language \(L_o\), consisting of \(N\) frames, \(\mathbf{x} = x_{1:N}\), to the target text \(\mathbf{z} = z_{1:M}\), representing the linguistic information in the target language. A Text-to-Speech (TTS) model can then be adopted to further generate the target speech \(\mathbf{y} = y_{1:T}\), consisting of \(T\) frames in the target language \(L_t\), which contains the semantic information in the target text, thereby enabling a broader range of application scenarios.

For example, the Canary model~\cite{canary} employs a Speech Encoder $\mathbf{SE}$ and a Text Decoder $\mathbf{TD}$, which auto-regressively predicts the next token \(z_m\) by computing the probability distribution over the token vocabulary \(\mathcal{V}\).
The speech encoder processes the input speech \(\mathbf{x} = x_{1:N}\), extracting features necessary for the text decoder to generate the corresponding text:
\begin{equation}
\mathbf{h} = \mathbf{SE}(\mathbf{x}),
\end{equation}
while the text decoder uses the previously decoded tokens \(\mathbf{z}^*_{<m}\) and speech features as input:
\begin{equation}
P(z_m = z \mid \mathbf{h}, \mathbf{z}^*_{<m}) = \mathbf{TD}(\mathbf{h}, \mathbf{z}^*_{<m})_{z}, \quad z \in \mathcal{V},
\label{eq:text_decoder}
\end{equation}
where a greedy decoding process selects the most likely token:
\begin{equation}
\mathbf{z}^*_m = \arg\max_z P(z_m = z \mid \mathbf{h}, \mathbf{z}^*_{<m}).
\label{eq:greedy_text_decoder}
\end{equation}
The initial token sequence must include the Begin of Sentence (BOS) token and the language token represented by the language ID $tgt\_lang$. Through iterative processes, we can obtain the complete token prediction sequence.

\vspace{1mm}
\noindent\textbf{Speech-to-Any Translation Model.}
In the Standard E2E ST model, the conversion from source speech to target text is performed in an end-to-end manner. Once the translated text sequence is obtained, an additional independent speech synthesis stage can be employed to further generate target speech, enhancing convenience.
\begin{equation}
\mathbf{y} = \text{TTS}(\mathbf{z}),
\end{equation}
where \(\mathbf{z}\) is the translated text sequence, and \(\mathbf{y}\) is the synthesized target speech.
Differently, Speech-to-Any Translation model uses text as an intermediate output. The features generated during the decoding of the target language text are subsequently utilized to predict audio features. In Seamless~\cite{barrault2023seamless}, these intermediate features are employed to predict discrete audio units, which are then converted into audio waveforms using a vocoder (see \autoref{fig:ds2st}).

This indicates that, to carry out adversarial attacks on ST and introduce targeted semantic meaning, we can use the intermediate output text as the optimization objective.

\section{Method}

\subsection{Attack ST with Perturbation}
\label{sec:attack-with-perturbation}

\begin{figure}[]
\includegraphics[width=\columnwidth]{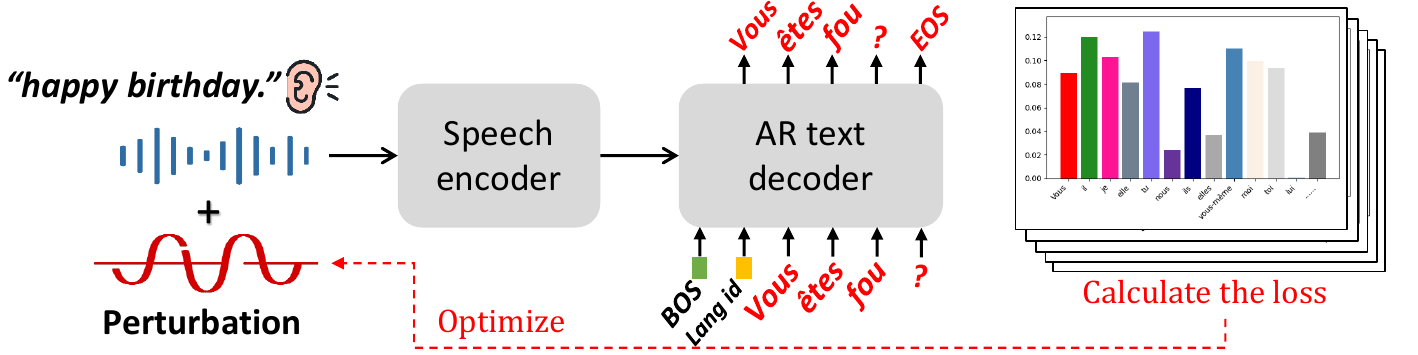}
% \vspace{-em}
\caption{Overview of our perturbation-based attack on ST.}
\label{fig:fr}
\vspace{-1em}
\end{figure}

\autoref{fig:fr} illustrates the main framework of perturbation-based attack strategy, which uses a teacher forcing mechanism~\cite{vaswani2017attention} to align the text translation results from the autoregressive prediction model with the target sentences. This alignment guides the adversarial perturbations to shift towards approximating the semantics of the target sentence.

\noindent
\textbf{Formalizing the adversarial objective.}
The goal of the attack is to generate an adversarial perturbation \(\delta\) that, when added to the original speech signal \(x_{orig}\), causes the translation output to match a specified target text \(tgt\_text\) across multiple attack languages \(\mathbf{L}_{attack}\).
For a speech translation (ST) system with a Speech Encoder $\mathbf{SE}$ and a Text Decoder $\mathbf{TD}$, the loss function can be formulated as:
\begin{align}
    \mathcal{L}(\delta) &= \sum_{l \in \mathbf{L}_{attack}} \sum_{m=1}^{M_l} \text{CrossEntropy}(
    \notag \\
    &\quad\textbf{TD}(\textbf{SE}(x_{orig} + \delta), \mathbf{z}^*_{<m}), tgt\_text_{l}[m]),
    \label{eq:ce_loss}
\end{align}
where \(x_{orig}\) is the original speech input and \(\delta\) is the adversarial perturbation. \(tgt\_text_{l}\) denotes the target text in language \(l \in \mathbf{L}_{attack}\) and \(M_l\) is the length of the target text in language \(l\). \(\mathbf{z}^*_{<m}\) represents the sequence of predicted tokens before position \(m\), and \(\text{CrossEntropy}\) is the cross-entropy loss between the target token and the predicted token.

\begin{algorithm}[t]
\caption{Attack ST with Perturbation}
\label{alg:perturbation_attack}
\begin{algorithmic}[1]
\REQUIRE Original speech $x_{orig}$, target model (SpeechEncoder $\mathbf{SE}$ and TextDecoder $\mathbf{TD}$), target text $tgt\_text$, set of attack languages $\mathbf{L}_{attack}$, perturbation strength $\epsilon$, maximum iterations $max\_iteration$
\STATE Initialize adversarial perturbation $\delta$ randomly
\STATE $iteration \leftarrow 0$
\STATE $target\_list \leftarrow \emptyset$
\FOR{$tgt\_lang \in \mathbf{L}_{attack}$}
    \STATE $tgt\_text_{l} \leftarrow \text{Translate}(tgt\_text, tgt\_lang)$
    \STATE $target\_list.\text{append}(tgt\_text_{l})$
\ENDFOR
\STATE $translated\_list \leftarrow \emptyset$
\WHILE{$translated\_list \neq target\_list$ \AND $iteration \leq max\_iteration$}
    \STATE $\delta \leftarrow \epsilon \cdot \tanh(\delta)$ \textit{// Limit the perturbation strength}
    \STATE $\delta \leftarrow \text{bandpass}_{[1k,4k]}(\delta)$ \textit{// Avoid excessively high or low frequencies }
    \STATE $loss \leftarrow 0$
    \STATE $x_{adv} \leftarrow x_{orig} + \delta$ 
    \STATE $\mathbf{h} \leftarrow \mathbf{SE}(x_{adv})$ 
    \STATE $translated\_list \leftarrow \emptyset$
    \FOR{$id = 0$ \text{ to } $len(\mathbf{L}_{attack})-1$}
        \STATE $tgt\_lang \leftarrow \mathbf{L}_{attack}[id]$
        \STATE $tgt\_text_{l} \leftarrow target\_list[id]$
        \STATE $count = 0$, $\mathbf{z}^*_0 \leftarrow \text{BOS}$, $\mathbf{z}^*_1 \leftarrow \text{token}(tgt\_lang)$
        \WHILE{$\mathbf{z}^*_m \neq \text{end\_of\_sentence}$}
            \STATE $P(z_m \mid \mathbf{h}, \mathbf{z}^*_{<m}) = \mathbf{TD}(\mathbf{h}, \mathbf{z}^*_{<m})$
            \STATE $\mathbf{z}^*_m = \arg\max_z P(z_m = z \mid \mathbf{h}, \mathbf{z}^*_{<m})$
            $\autoref{eq:greedy_text_decoder}$
            \STATE $loss \mathrel{+}= \text{CrossEntropy}(\textbf{TD}(\mathbf{h}, \mathbf{z}^*_{<m}), \mathclap{\begin{array}{l}\\[1em]
                \hspace{-4.5em} tgt\_text_{l}[count]) \autoref{eq:ce_loss} \end{array}}$
            \STATE $count \leftarrow count + 1$
        \ENDWHILE
        \STATE $translated\_list.\text{append}(\mathbf{z}^*)$
    \ENDFOR
    \STATE Optimize $\delta$ with $loss$
    \STATE $iteration \leftarrow iteration + 1$
\ENDWHILE
\end{algorithmic}
\end{algorithm}

The goal of the attack is to minimize this loss function \(\mathcal{L}(\delta)\) with respect to the perturbation \(\delta\), such that the adversarial input \(x_{adv} = x_{orig} + \delta\) forces the model to produce the desired translation across all targeted languages.
It is worth noting that the perturbation nextly undergoes two processing steps: (1) Scaling, where \(\delta \leftarrow \epsilon \cdot \tanh(\delta)\) is used to limit the perturbation strength~\cite{carlini2017towards}. (2) Filtering, where \(\delta \leftarrow \text{bandpass}_{[1k,4k]}(\delta)\) applies a bandpass filter to the scaled perturbation, focusing on the 1k-4kHz range to avoid excessively high or low frequencies.
Further details of the algorithm are provided in \cref{alg:perturbation_attack}.

\begin{figure}[t]
    \centering
    \includegraphics[width=\columnwidth]{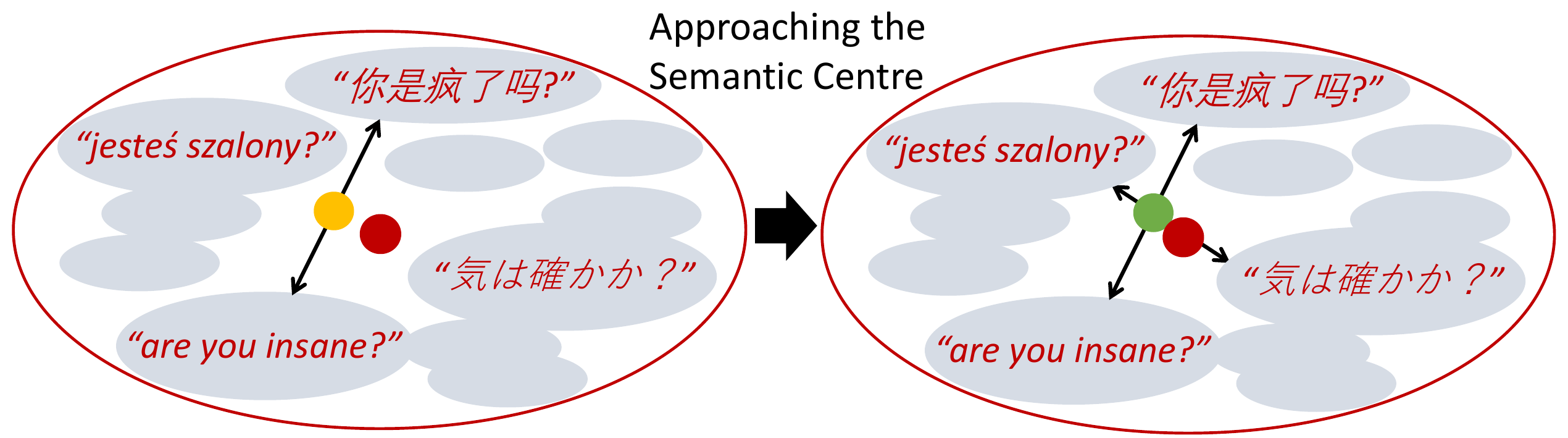}
    % \vspace{-1em}
    \caption{Increasing the number of Seen languages brings the semantics in adversarial perturbations closer to the global semantic center.}
    \label{fig:more-lang}
\end{figure}

\noindent\textbf{Multi-language Enhancement.}
\label{sec:more-lang-enhance}
As described in \cref{alg:perturbation_attack}, adversarial perturbation optimization can be enhanced by incorporating multiple target languages (\(tgt\_lang\)). This approach strengthens the semantic alignment between the perturbation and the target sentence while improving the generalizability of the attack to Unseen languages. As illustrated in \autoref{fig:more-lang}, optimizing perturbations using more languages helps align the target semantics closer to the actual semantic center.

\begin{figure}[htbp]
    \centering
    \includegraphics[width=.8\columnwidth]{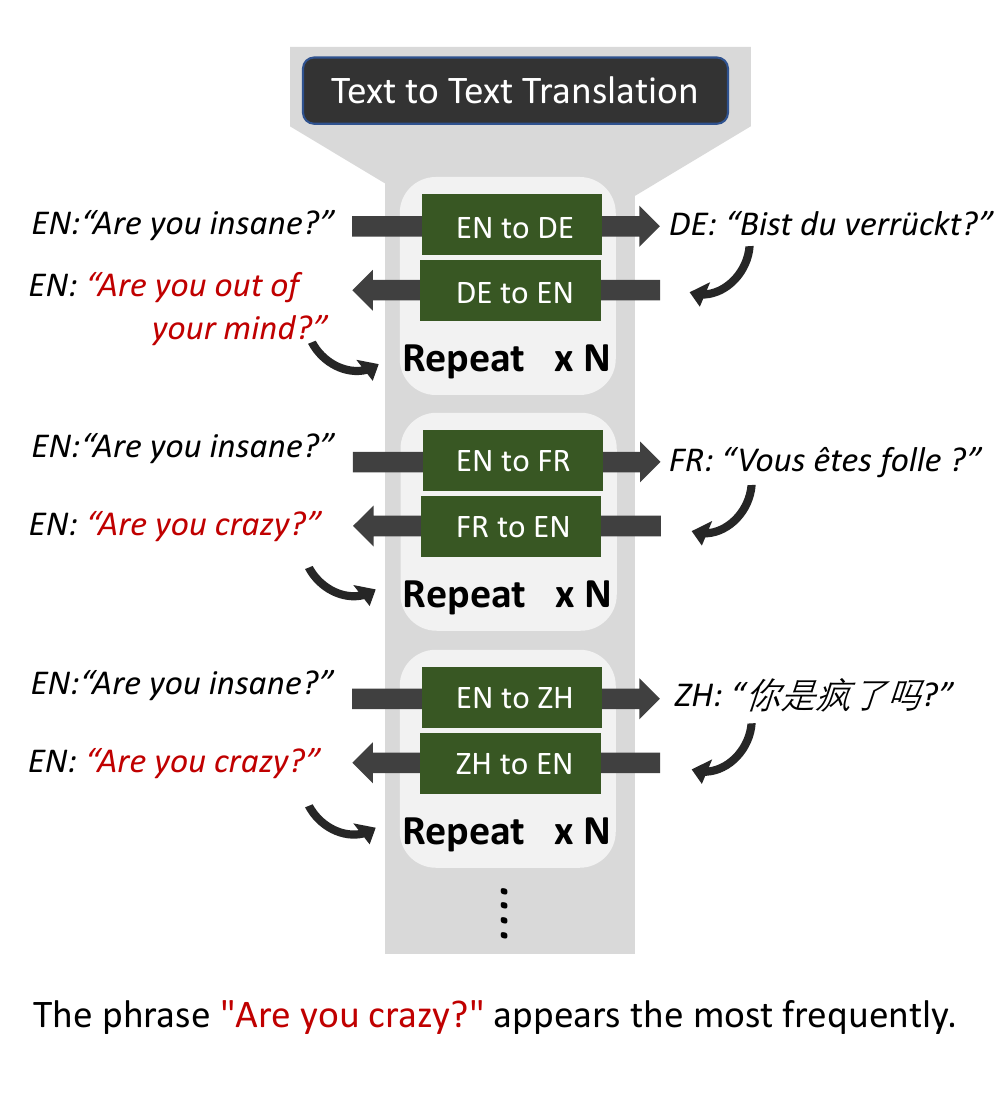}
    % \vspace{-2em}
    \caption{Illustration of Target Cycle Optimization. In this case, in the cycle translation targeting multiple languages, ``Are you crazy?" appears most frequently as the retranslated result and is selected as the updated sentence.}
    \label{fig:method-cycle-enhance}
\end{figure}

\noindent\textbf{Target Cycle Optimization.}
\label{sec:cycle}
For speech translation models that rely on semantic understanding, we can further explore the adaptability of the target text to the model before adversarial optimization. This involves identifying whether an alternative text exists in the model's semantic space that conveys the intended meaning more effectively than the original target text.
Different models may exhibit semantic preferences due to imbalances in their training dataset~\cite{zhang2018mitigating,bender2018data,beinborn2020semantic}. Therefore, we can first optimize the target text to select a more suitable alternative for the model. Seamless~\cite{seamless-blog}, being a model based on semantic understanding, allows text inputs to be processed through a Text Encoder that maps them into the semantic space. To find an alternative target text, we employ a cycle translation method, as illustrated in \autoref{fig:method-cycle-enhance}.
By repeatedly performing Text-to-Text Translation (T2TT) with the target model and recording the intermediate translations across multiple languages, we identify the text that appears most frequently, which is then selected as the new target text.

\subsection{Attack ST with Adversarial Music}
\label{sec:attack-with-music}

Using adversarial music presents a more covert and effective method for attacks, as it eliminates the need to constrain the amplitude of the music, unlike perturbation-based attacks where the perturbation magnitude $\epsilon$ must be controlled. In this section, we introduce the proposed adversarial music generation approach for attacking ST systems.

\begin{figure}[t]
    \centering
    \includegraphics[width=.9\columnwidth]{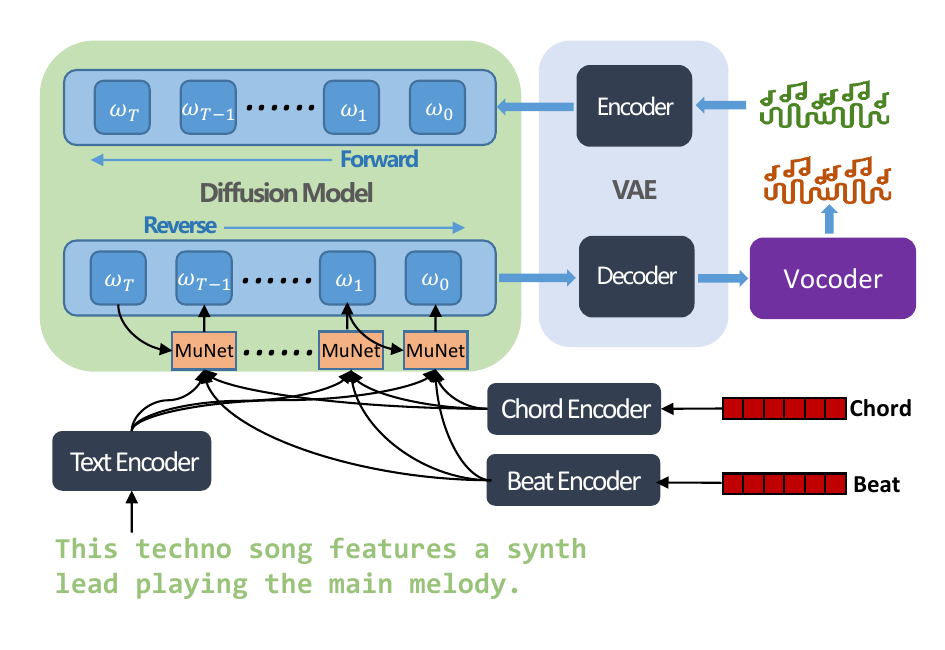}
    % \vspace{-1em}
    \caption{Diffusion-based music generation pipeline, where text, chord, and beat encoders guide MuNet in the reverse diffusion process to create music from random sampled initial noise.}
    \label{fig:dmg}
\end{figure}

\noindent\textbf{Diffusion-based Music Generation.}
\label{sec:dmg}
Recent diffusion-based music generation~(DMG) techniques are inspired by diffusion-based general audio generation~(DGAG), such as \texttt{Tango}~\cite{ghosal2023tango} and \texttt{AudioLDM}\allowbreak\cite{liu2023audioldm}, \cite{melechovsky2023mustango}, which leverages the latent diffusion model (LDM)~\cite{rombach2022high} to reduce computational complexity while maintaining the expressiveness of the diffusion model. 
As shown in \autoref{fig:dmg}, the music generation process (reverse diffusion process) requires three types of information: (1) Text information, which consists of a textual description of the music and is encoded by the text encoder to extract features; (2) Chord and beat information, which are processed by the chord encoder and be\underline{}at encoder, respectively, to produce corresponding embeddings; and (3) Initial noise \(\omega_T\), which serves as the starting point for the reverse diffusion process.

\vspace{1mm}
\noindent
{\textit{\ul{Forward Diffusion.}}}
In DMG~\cite{melechovsky2023mustango}, the latent audio prior $\omega_0$ is extracted using a variational autoencoder (VAE) on condition $\mathcal{C}$, which refers to a joint music and text condition. The VAE is borrowed from the pre-trained model in \texttt{AudioLDM}~\cite{liu2023audioldm} to obtain the latent code of the audio.
During the forward diffusion process (Markovian Hierarchical VAE), the latent audio prior $\omega_0$ is gradually transformed into standard Gaussian noise $\omega_T\sim \mathcal{N}(\mathbf{0}, \mathbf{I})$, as shown in \cref{eq:forward_diffusion}.
At each step of the forward process, pre-scheduled Gaussian noise($0<\beta_1<\beta_2<\dots<\beta_T<1$) is is progressively added:
\begin{flalign}
q(\omega_t|\omega_{t-1}) &= \mathcal{N}(\sqrt{1-\beta_t} \omega_{t-1}, \beta_t \mathbf{I}).  \label{eq:forward_diffusion}
\end{flalign}

\begin{figure}[htbp]
\includegraphics[width=\columnwidth]{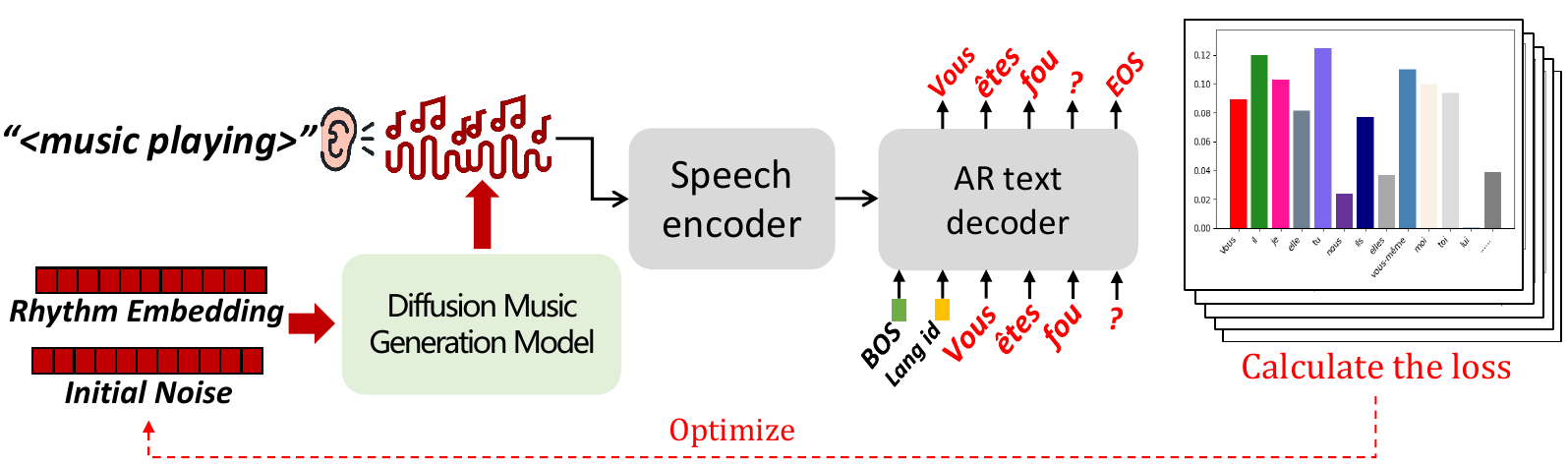}
\vspace{-1em}
\caption{Overview of our music-based attack on ST.}
\label{fig:fr-music}
\end{figure}

\noindent
{\textit{\ul{Reverse Diffusion.}}}
In the reverse diffusion process, which reconstructs $\omega_0$ from Gaussian noise $\omega_N\sim \mathcal{N}(\mathbf{0}, \mathbf{I})$, MuNet~\cite{melechovsky2023mustango} is used to steer the generated music towards the given condition $\mathcal{C}$, which consists of musical attributes (beat $b$ and chord $c$) and text $\tau$ ($\mathcal{C}:=\{\tau, b, c\}$). This is realized through the Music-Domain-Knowledge-Informed UNet (MuNet) denoiser. After the Chord Encoder \(\mathbf{CE}\) and Beat Encoder \(\mathbf{BE}\) encode the chord and beat information, respectively, MuNet takes the chord embedding, beat embedding, the encoded text from the Text Encoder \(\mathbf{TE}\), and the output from the previous step \(\omega_{t+1}\) to generate the current step's output \(\omega_t\):
\begin{equation}
    \omega_t = \text{MuNet}(\mathbf{CE}(\text{chord}), \mathbf{BE}(\text{beat}), \mathbf{TE}(\text{text}), \omega_{t+1}).
    \label{eq:reverse-diffusion}
\end{equation}
\autoref{fig:fr-music} presents the framework of our music-based attack scheme. Similar to \autoref{sec:attack-with-perturbation}, our goal is to align the translation results obtained by the autoregressive prediction model with the target sentences. However, unlike the previous approach, here we focus on optimizing the control inputs for music generation, specifically the inputs to \autoref{eq:reverse-diffusion}.

\begin{algorithm}[t]
\caption{Attack ST with Adversarial Music}
\label{alg:music_attack}
\begin{algorithmic}[1]
\REQUIRE Target model (SpeechEncoder $\mathbf{SE}$, TextDecoder $\mathbf{TD}$), target text $tgt\_text$, attack languages $\mathbf{L}_{attack}$, max iterations $max\_iteration$, DMG (Chord Encoder $\mathbf{CE}$ with $\theta_c$, Beat Encoder $\mathbf{BE}$ with $\theta_b$, Text Encoder $\mathbf{TE}$ with embedding $\theta_t$, Chord $c$, Beat $b$, Prompt Text $\tau$), initial noise $\omega_T$, diffusion steps $ds$
\STATE Initialize $\omega_T$ randomly, $iteration \leftarrow 0$
\STATE $target\_list \leftarrow \emptyset$
\FOR{$tgt\_lang \in \mathbf{L}_{attack}$}
    \STATE $tgt\_text_{l} \leftarrow \text{Translate}(tgt\_text, tgt\_lang)$
    \STATE $target\_list.\text{append}(tgt\_text_{l})$
\ENDFOR
\STATE $translated\_list \leftarrow \emptyset$
\WHILE{$translated\_list \neq target\_list$ \AND $iteration \leq max\_iteration$}
    \STATE $\omega_{ds} \leftarrow \omega_T$
    \FOR{$t = ds-1$ \text{ to } $0$}
        \STATE $\omega_t \leftarrow \text{Reverse}(\mathbf{CE}(c), \mathbf{BE}(b), \mathbf{TE}(\tau), \omega_{t+1})$
        \autoref{eq:reverse-diffusion} 
    \ENDFOR
    \STATE $loss \leftarrow 0$
    \STATE $x_{adv} \leftarrow \omega_0$, $\mathbf{h} \leftarrow \mathbf{SE}(x_{adv})$
    \STATE $translated\_list \leftarrow \emptyset$
    \FOR{$id = 0$ \text{ to } $len(\mathbf{L}_{attack})-1$}
        \STATE $tgt\_lang \leftarrow \mathbf{L}_{attack}[id]$
        \STATE $tgt\_text_{l} \leftarrow target\_list[id]$
        \STATE $count \leftarrow 0$, $\mathbf{z}^*_0 \leftarrow \text{BOS}$, $\mathbf{z}^*_1 \leftarrow \text{token}(tgt\_lang)$
        \WHILE{$\mathbf{z}^*_m \neq \text{end\_of\_sentence}$}
            \STATE $P(z_m \mid \mathbf{h}, \mathbf{z}^*_{<m}) = \mathbf{TD}(\mathbf{h}, \mathbf{z}^*_{<m})$
            \STATE $\mathbf{z}^*_m = \arg\max_z P(z_m = z \mid \mathbf{h}, \mathbf{z}^*_{<m})$
            $\autoref{eq:greedy_text_decoder}$
            \STATE $loss \mathrel{+}= \text{CrossEntropy}(\textbf{TD}(\mathbf{h}, \mathbf{z}^*_{<m}), \mathclap{\begin{array}{l}\\[1em]
                \hspace{-4.5em} tgt\_text_{l}[count]) \autoref{eq:ce_loss} \end{array}}$
            \STATE $count \leftarrow count + 1$
        \ENDWHILE
        \STATE $translated\_list.\text{append}(\mathbf{z}^*)$
    \ENDFOR
    \STATE $loss \mathrel{+}= \text{KL\_Divergence}(z_T', z_T)$
    \STATE Optimize $\theta_c$, $\theta_b$, and $\omega_T$ with $loss$
    \STATE $iteration \leftarrow iteration + 1$
\ENDWHILE
\end{algorithmic}
\end{algorithm}

\noindent\textbf{Attack with Diffusion-based Music Generation (DMG).}
During the audio generation phase, \ie, the denoising process of the LDM, we set the initial noise $z_T$ as the optimization target. This noise is optimized through gradient backpropagation to ensure that the final denoised music contains adversarial elements. To enhance the effectiveness of the adversarial attack, we also include rhythm (beat and chord) in the optimization target set. We optimize the parameters $\theta_c$ of the Chord Encoder $\mathbf{CE}$ and $\theta_b$ of the Beat Encoder $\mathbf{BE}$ to refine the fundamental music properties. The goal is to make the audio generated by the LDM adversarial, ensuring it translates into specific semantic content. The complete algorithm is presented in \autoref{alg:music_attack}.

Since music is not part of the typical data distribution for speech translation models, the model tends to interpret adversarial music as target semantics with lower confidence during the optimization process. This can compromise the stability of the optimization and the cross-lingual generalization of adversarial samples. To adress this challenge, we employ \textbf{SharpnessLoss} (see \autoref{alg:sharpness_loss}) as a replacement for Cross-Entropy Loss in this context. Specifically, we enhance the standard cross-entropy loss by optimizing the logits of the translation results. The objective is to ensure that each step of the translation process has a high probability of generating the target text, thereby sharpening the predicted distribution at each step of the autoregressive prediction process.

\begin{algorithm}[t]
\caption{SharpnessLoss}
\label{alg:sharpness_loss}

\textbf{Input:} Logits $logits$, Targets $targets$, Sharpness coefficient $\alpha$ \\
\textbf{Output:} Loss value $loss$

\begin{algorithmic}[1]
\STATE $ce\_loss = \text{CrossEntropy}(logits, targets)$
\STATE $sharpness\_penalty = logits[targets]$
\STATE $loss = ce\_loss - \alpha \cdot \text{mean}(sharpness\_penalty)$
\RETURN $loss$
\end{algorithmic}
\end{algorithm}

\noindent\textbf{Enhance with Simulated Over-the-air Process.}
As outlined in \autoref{sec:tm}, a more potent attack strategy involves transmitting adversarial music through an over-the-air channel, as depicted in \autoref{fig:tm}. 
To ensure over-the-air robustness, we simulate air transmission distortions and environmental noise by overlaying speech from the Librispeech dataset\cite{panayotov2015librispeech} and applying random impulse responses from \cite{jeub2009binaural} for reverberation. Small random noise is also added. Details are provided in \autoref{sec:ota-simulate}.
\section{Evaluations}
\label{sec:evaluation}

\subsection{Experimental Setup}
\vspace{1mm}
\noindent
\textbf{Target Models.} 
To thoroughly investigate the vulnerability of ST models to adversarial attacks, we selected two types of models: Standard End-to-end ST Model and Speech-to-any ST Model. Canary~\cite{canary} and Seamless~\cite{barrault2023seamless} are chosen as the state-of-the-art representatives for each category. Additionally, as Seamless is currently one of the most advanced speech translation models, we conducted extensive experiments on its various versions (Large, Expressive, M4tv2, and Medium), which differ in model architecture and parameter sizes.

\vspace{1mm}
\noindent
\textbf{Languages.} 
We selected different target language sets as the optimization targets during the attack and conducted tests across multiple target languages. These target languages included both \textbf{Seen} languages (encountered during adversarial optimization) and \textbf{Unseen} languages (not encountered), allowing for a comprehensive evaluation of semantic attack effectiveness and the weaknesses of multilingual speech translation models in resisting semantic adversarial attacks.

For Seamless~\cite{barrault2023seamless}, as shown in \autoref{tb:languages}, we selected \textit{English(EN)}, \textit{Mandarin(ZH)}, \textit{German(DE)}, \textit{French(FR)}, \textit{Italian(IT)}, and \textit{Spanish(ES)} as the target test languages, since this set represents the largest intersection of languages supported by all Seamless models.
This selection ensures consistency across the different model versions. For the attacks, we tested four language sets: \textit{EN}, \textit{EN+ZH}, \textit{EN+ZH+DE}, and \textit{EN+ZH+DE+FR}, to evaluate how the number of \textbf{Seen} languages influence the attack's performance.

For Canary~\cite{canary}, which only supports \textit{English}, \textit{French}, \textit{German}, and \textit{Spanish}, we set up four attack combinations: \textit{EN}, \textit{EN+FR}, \textit{EN+FR\allowbreak+DE}, and \textit{EN+FR+DE+ES}. These four languages were also used as the target test languages.

\begin{table}[t]
    \centering
    \caption{Settings of Seen and Unseen languages.}
    \vspace{-.5em}
    \resizebox{\columnwidth}{!}{
        
        \begin{tabular}{c|c|c|c}
        \hline
        Model & Attack Lang. & Test Lang. (Seen) & Test Lang. (Unseen) \\ \hline
        \multirow{4}{*}{Seamless} & EN & EN & ZH, DE, FR, IT, ES \\ 
         & EN + ZH & EN, ZH & DE, FR, IT, ES \\ 
         & EN + ZH + DE & EN, ZH, DE & FR, IT, ES \\ 
         & EN + ZH + DE + FR & EN, ZH, DE, FR & IT, ES \\ \hline
        \multirow{4}{*}{Canary} & EN & EN & FR, DE, ES \\ 
         & EN + FR & EN, FR & DE, ES \\ 
         & EN + FR + DE & EN, FR, DE & ES \\ 
         & EN + FR + DE + ES & EN, FR, DE, ES & None \\ \hline
        \end{tabular}
    }
    \footnotesize Note: EN=English, ZH=Mandarin, DE=German, FR=French, IT=Italian, ES=Spanish
    \label{tb:languages}
\end{table}

\begin{table}[t]
    \centering
    \caption{Examples of ESIM scores between semantically related(successful) and unrelated(unsuccessful) text pairs.}
    \vspace{-.5em}
\resizebox{\columnwidth}{!}{%

\begin{tabular}{c|c|c|c}
\hline
Target   semantic                  & Adversarial output						& ESIM            		& NSCORE          		\\ \hline
\multirow{2}{*}{Shame on you.}     & Have you no shame?						& \textbf{0.5644} 		& \textbf{0.6581} 		\\ 
                                   & The bus is running late today.         & \underline{0.0492}	& \underline{0.0077}    \\ \hline
\multirow{2}{*}{You make me sick.} & You revolt me.							& \textbf{0.7134} 		& \textbf{0.4215} 		\\ 
                                   & We need to buy more coffee. 			& \underline{0.0942}    & \underline{0.0072}    \\ \hline
\end{tabular}
}
    \label{tb:threshold}
\vspace{-.5em}
\end{table}

\vspace{1mm}
\noindent
\textbf{Target semantics.}
We conduct experiments with 5 target semantics: ``You make me sick.",``Shame on you.", ``Are you insane?", ``Who do you think you're talking to?", and ``Don't waste my time anymore.", representing the malicious semantics that attackers may inject in speech translation scenarios.

\vspace{1mm}
\noindent
\textbf{Carrier Speech Set.}
For the perturbation carriers in perturbation based attack, we select one speech from two speakers in each of the following six languages: English (from VCTK \cite{veaux2017cstr}), Mandarin (from AISHELL \cite{shi2020aishell}), German, French, Italian (from CommonVoice \cite{ardila2019common}), and Spanish (from VoxPopuli \cite{wang-etal-2021-voxpopuli}). This results in 60 test cases for each attack configuration (attack method, target language, target semantics).

\vspace{1mm}
\noindent
\textbf{Attack Method.}
We explore the vulnerability of speech translation models using two different strategies: Perturbation-based Attack and Music-based Attack. For the perturbation-based attack, as outlined in \autoref{sec:attack-with-perturbation}
, we applied adversarial perturbations to the carrier speech through gradient optimization. For the music-based attack, as described in \autoref{sec:attack-with-music}, we introduced a novel adversarial music optimization scheme based on diffusion-based music generation. This approach is more covert, because it imitates background music and environmental noise that are not easily noticeable.
By employing both strategies, we conducted a more comprehensive evaluation of the vulnerabilities in S2ST models.

% \vspace{1mm}
\noindent
\textbf{Evaluation Metrics.} 
We used a variety of metrics to comprehensively evaluate the two perspectives of the attack: adversarial audio quality and attack effectiveness.

To evaluate the quality of adversarial speech, we utilize three objective metrics:
Perceptual Evaluation of Speech Quality (${\textbf{PESQ}}$)~\cite{rix2001perceptual}, Speaker Vector Cosine Similarity (${\textbf{VSIM}}$)~\cite{casanova2021sc}, and Speaker Vector Cosine Similarity specific to Seamless (${\textbf{VSIM-E}}$)~\cite{barrault2023seamless}, along with a subjective metric, Mean Opinion Score\allowbreak(\textbf{MOS}).
In detail, PESQ assesses speech quality (\ie, imperceptibility) by taking into account the nuances of the human auditory system. VSIM measures speaker similarity to evaluate fidelity, with higher values indicating greater similarity. Following prior works \cite{casanova2021sc, timbrewatermarking-ndss2024}, we compute VSIM using the speaker encoder from the Resemblyzer package~\cite{jemine2019real}. For VSIM-E, we use the speaker encoder of Seamless~\cite{barrault2023seamless}. 

To evaluate the model's vulnerability, \ie, the effectiveness of adversarial attack schemes, we employed two metrics. Since we need to explore the semantic similarity between the model output and the target at a deeper semantic level, methods that can measure the semantic similarity distance between text pairs are necessary. Traditional metrics like Word Error Rate (WER) are not suitable here, as the target translation model is designed to map the source and target languages within a semantic space. Metrics like WER are overly rigid for such models; for example, while ``shame on you" and ``you should be ashamed of yourself" would yield a high WER, their semantic meanings are nearly identical. 

The first metric we use is the semantic similarity between the translation output and the target, measured by the embedding similarity (\textbf{ESIM}) from a pre-trained BERT model, as outlined by \cite{reimers2019sentence}. This metric is widely used in machine translation evaluations \cite{rei2020comet, barrault2023seamlessm4t}. The second metric is \textbf{NSCORE}, which assesses the semantic entailment relationship between the translation result and the target text, following the approach in Natural Language Inference (NLI) tasks \cite{liu2019roberta, wang2018glue}.

To establish appropriate semantic similarity thresholds for measuring Attack Success Rate (\textbf{ASR}), we leveraged sentence embedding similarity scores, which typically yield very low values between semantically unrelated sentences (examples shown in Table \ref{tb:threshold}). For each target semantic, we used ChatGPT-4 to generate six different expressions with the same sematic. This process produced semantically consistent but structurally varied text pairs, such as ``shame on you" and ``you should be ashamed of yourself." We then calculated \textbf{ESIM} and \textbf{NSCORE} values between the original text and these variations to obtain the lowest similarity scores, denoted as \(\Gamma_e\) and \(\Gamma_n\), which serve as thresholds to determine semantic consistency between target semantic and adversarial output. The prompts and examples used for generating similar texts are shown in \autoref{fig:prompt} in Appendix.

\subsection{Perturbation-based Attack} 
In this section, we first conduct a detailed analysis of the attack effectiveness of the generated perturbations, followed by an evaluation of their perceptual quality.

\subsubsection{Attack Effectiveness}
\label{attack-effectiveness}
% \vspace{-.5em}
We begin by assessing the attack's fundamental effectiveness on the default model, Seamless Large, before extending our analysis to other models.

We begin with preliminary experiments in a single-language attack scenario, specifically investigating the effectiveness of targeted adversarial attacks in a translation task from language A to language B. The experimental results, shown in \autoref{tb:one2one}, evaluate the impact of attacks under varying perturbation levels. The results demonstrate that adversarial attacks can be effectively applied to any target language. Notably, we observed that the attack's effectiveness is closely related to the target language but shows minimal dependence on the source language. This phenomenon arises because LLM models like Seamless employ a paradigm that maps input languages into a language-agnostic semantic space, eliminating the need to specify the source language token during translation. Therefore, in subsequent experiments, we focus on analyzing scenarios involving different target languages.

\begin{table*}[htbp]
\centering
\caption{Effectiveness of adversarial attacks at different perturbation levels in single-language translation scenarios, measured by average ESIM, NSCORE and ASR. ``Src" denotes the source language, while ``Tgt" refers to the target language.}
\vspace{-.5em}
\resizebox{\textwidth}{!}{
\begin{tabular}{c|c|ccc|ccc|ccc|ccc|ccc|ccc}
\hline
\multirow{2}{*}{$\epsilon$} & \multicolumn{1}{l|}{\multirow{2}{*}{\begin{tabular}[c]{@{}l@{}}\diagbox{Src}{Tgt}\end{tabular}}} & \multicolumn{3}{c|}{EN}                                                          & \multicolumn{3}{c|}{ZH}                                                          & \multicolumn{3}{c|}{DE}                                                          & \multicolumn{3}{c|}{FR}                                                          & \multicolumn{3}{c|}{IT}                                                          & \multicolumn{3}{c}{ES}                                                          \\ \cline{3-20} 
                            & \multicolumn{1}{l|}{}                                                             & \multicolumn{1}{l}{ESIM} & \multicolumn{1}{l}{NSCORE} & \multicolumn{1}{l|}{ASR} & \multicolumn{1}{l}{ESIM} & \multicolumn{1}{l}{NSCORE} & \multicolumn{1}{l|}{ASR} & \multicolumn{1}{l}{ESIM} & \multicolumn{1}{l}{NSCORE} & \multicolumn{1}{l|}{ASR} & \multicolumn{1}{l}{ESIM} & \multicolumn{1}{l}{NSCORE} & \multicolumn{1}{l|}{ASR} & \multicolumn{1}{l}{ESIM} & \multicolumn{1}{l}{NSCORE} & \multicolumn{1}{l|}{ASR} & \multicolumn{1}{l}{ESIM} & \multicolumn{1}{l}{NSCORE} & \multicolumn{1}{l}{ASR} \\ \hline
\multirow{6}{*}{0.5}        & EN                                                                                & 0.9449                   & 0.8869                     & 10/10                    & 0.9791                   & 0.8100                     & 10/10                    & 0.8705                   & 0.7105                     & 9/10                     & 0.9987                   & 0.9850                     & 10/10                    & 0.8078                   & 0.5928                     & 9/10                     & 0.9359                   & 0.8827                     & 9/10                    \\
                            & ZH                                                                                & 0.9951                   & 0.9844                     & 10/10                    & 0.9358                   & 0.7873                     & 10/10                    & 0.8749                   & 0.8098                     & 9/10                     & 1.0000                   & 0.9851                     & 10/10                    & 0.9155                   & 0.8850                     & 9/10                     & 0.9416                   & 0.8855                     & 9/10                    \\
                            & DE                                                                                & 1.0000                   & 0.9846                     & 10/10                    & 0.9648                   & 0.8817                     & 10/10                    & 0.8643                   & 0.7881                     & 8/10                     & 0.8952                   & 0.8008                     & 10/10                    & 0.7476                   & 0.5949                     & 7/10                     & 1.0000                   & 0.9848                     & 10/10                   \\
                            & FR                                                                                & 0.8455                   & 0.6939                     & 9/10                     & 1.0000                   & 0.9859                     & 10/10                    & 0.9346                   & 0.7927                     & 9/10                     & 0.8651                   & 0.8794                     & 9/10                     & 0.9621                   & 0.9200                     & 9/10                     & 0.9808                   & 0.8880                     & 10/10                   \\
                            & IT                                                                                & 0.9827                   & 0.9842                     & 10/10                    & 0.9536                   & 0.9616                     & 10/10                    & 0.7374                   & 0.6343                     & 9/10                     & 0.9586                   & 0.9474                     & 10/10                    & 0.9571                   & 0.8038                     & 10/10                    & 0.8970                   & 0.8603                     & 10/10                   \\
                            & ES                                                                                & 0.9987                   & 0.9846                     & 10/10                    & 0.9400                   & 0.7908                     & 9/10                     & 0.7621                   & 0.6094                     & 8/10                     & 0.7602                   & 0.7320                     & 8/10                     & 0.9629                   & 0.8843                     & 10/10                    & 0.9023                   & 0.8080                     & 10/10                   \\ \hline
\multirow{6}{*}{0.1}        & EN                                                                                & 0.8574                   & 0.7137                     & 8/10                     & 0.9552                   & 0.8915                     & 9/10                     & 0.7748                   & 0.7261                     & 8/10                     & 0.9321                   & 0.8861                     & 9/10                     & 0.9047                   & 0.8192                     & 9/10                     & 0.9006                   & 0.9768                     & 10/10                   \\
                            & ZH                                                                                & 1.0000                   & 0.9846                     & 10/10                    & 0.8809                   & 0.7884                     & 9/10                     & 0.8042                   & 0.6101                     & 9/10                     & 0.8441                   & 0.8627                     & 9/10                     & 0.9360                   & 0.8192                     & 10/10                    & 0.8341                   & 0.7908                     & 8/10                    \\
                            & DE                                                                                & 0.8229                   & 0.5927                     & 9/10                     & 0.8387                   & 0.8253                     & 9/10                     & 0.5300                   & 0.3321                     & 5/10                     & 0.9679                   & 0.9802                     & 10/10                    & 0.7321                   & 0.5960                     & 8/10                     & 0.8626                   & 0.7182                     & 8/10                    \\
                            & FR                                                                                & 0.8579                   & 0.6916                     & 8/10                     & 0.9591                   & 0.8860                     & 10/10                    & 0.7057                   & 0.5094                     & 6/10                     & 0.8588                   & 0.7350                     & 9/10                     & 0.8954                   & 0.8228                     & 10/10                    & 0.6614                   & 0.4015                     & 5/10                    \\
                            & IT                                                                                & 0.9373                   & 0.8967                     & 9/10                     & 0.9490                   & 0.8879                     & 10/10                    & 0.7960                   & 0.6590                     & 8/10                     & 0.9302                   & 0.9653                     & 10/10                    & 0.9118                   & 0.8858                     & 9/10                     & 0.9461                   & 0.9249                     & 10/10                   \\
                            & ES                                                                                & 0.9084                   & 0.8334                     & 8/10                     & 0.8570                   & 0.7996                     & 7/10                     & 0.6418                   & 0.4208                     & 5/10                     & 0.7899                   & 0.6153                     & 10/10                    & 0.8959                   & 0.8514                     & 9/10                     & 0.6179                   & 0.4348                     & 5/10                    \\ \hline
\multirow{6}{*}{0.01}       & EN                                                                                & 0.7672                   & 0.5545                     & 7/10                     & 0.7832                   & 0.6041                     & 7/10                     & 0.5118                   & 0.1111                     & 5/10                     & 0.6584                   & 0.6014                     & 6/10                     & 0.6312                   & 0.5751                     & 7/10                     & 0.6114                   & 0.3465                     & 5/10                    \\
                            & ZH                                                                                & 0.6899                   & 0.6057                     & 6/10                     & 0.5134                   & 0.3060                     & 5/10                     & 0.5136                   & 0.2168                     & 6/10                     & 0.9188                   & 0.7878                     & 10/10                    & 0.6919                   & 0.5069                     & 6/10                     & 0.9270                   & 0.8465                     & 10/10                   \\
                            & DE                                                                                & 0.4083                   & 0.2915                     & 5/10                     & 0.7132                   & 0.5945                     & 7/10                     & 0.3274                   & 0.2452                     & 5/10                     & 0.6078                   & 0.5147                     & 7/10                     & 0.5650                   & 0.4992                     & 6/10                     & 0.3720                   & 0.1101                     & 4/10                    \\
                            & FR                                                                                & 0.3923                   & 0.2326                     & 2/10                     & 0.6216                   & 0.3099                     & 7/10                     & 0.5256                   & 0.4209                     & 5/10                     & 0.7566                   & 0.6969                     & 8/10                     & 0.3142                   & 0.2258                     & 3/10                     & 0.4552                   & 0.4745                     & 5/10                    \\
                            & IT                                                                                & 0.5082                   & 0.4083                     & 6/10                     & 0.3361                   & 0.2649                     & 3/10                     & 0.2327                   & 0.1717                     & 5/10                     & 0.5172                   & 0.3997                     & 6/10                     & 0.3502                   & 0.2024                     & 4/10                     & 0.5275                   & 0.4367                     & 6/10                    \\
                            & ES                                                                                & 0.5362                   & 0.4320                     & 6/10                     & 0.2591                   & 0.1814                     & 2/10                     & 0.2413                   & 0.1721                     & 4/10                     & 0.3383                   & 0.2292                     & 5/10                     & 0.4166                   & 0.2470                     & 4/10                     & 0.4250                   & 0.2633                     & 4/10                    \\ \hline
\end{tabular}
}
\small Note: EN=English, ZH=Mandarin, DE=German, FR=French, IT=Italian, ES=Spanish
\label{tb:one2one}
\end{table*}

\begin{figure}[]
\centering
\includegraphics[width=\columnwidth]{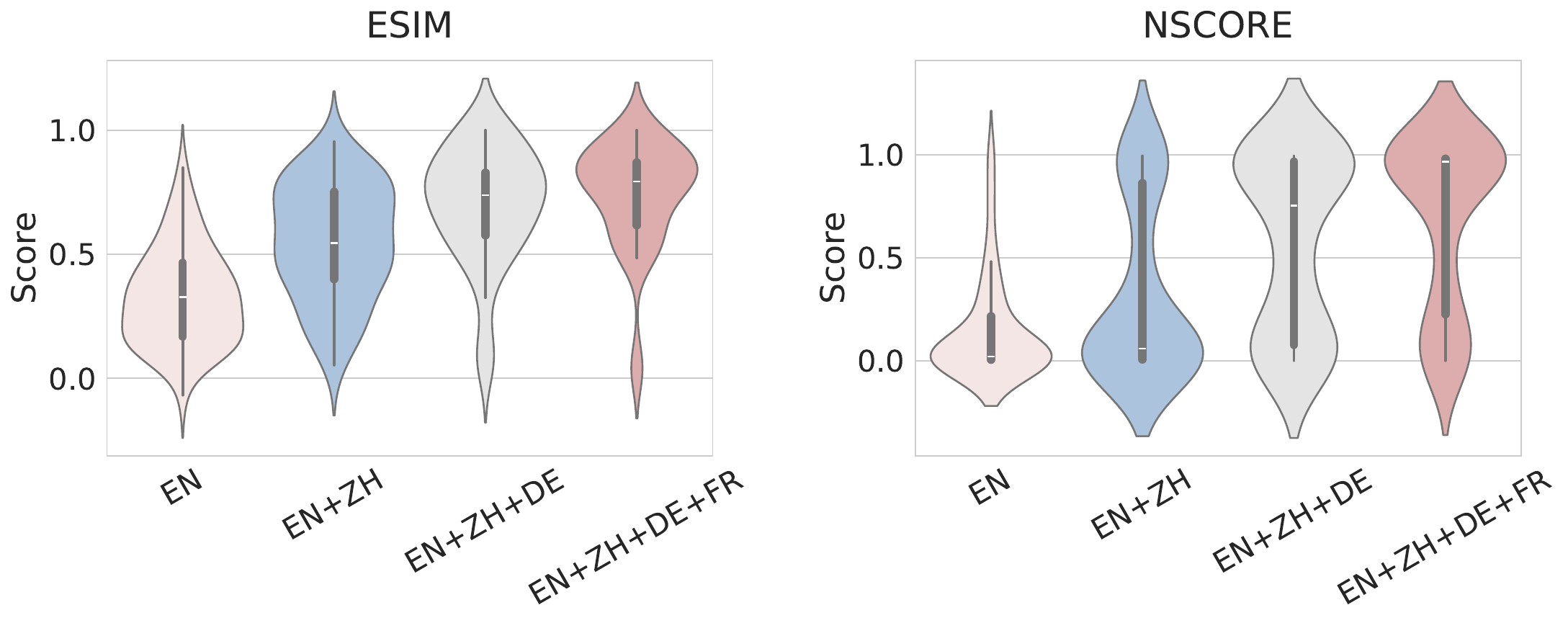}
% \vspace{-1em}
\footnotesize Note: EN=English, ZH=Mandarin, DE=German, FR=French, IT=Italian, ES=Spanish
\caption{Standard violin plot showing the distribution of ESIM and NSCORE scores for Spanish (Unseen) translation under varying numbers of attack languages ($\epsilon = 0.1$).}
\label{fig:distribution-spa-more-enhance}
\end{figure}

\subsubsection{Enhancement based on More Seen Languages}
As briefly mentioned in \autoref{sec:attack-with-perturbation}, introducing more Seen languages during the generation of adversarial perturbations enhances cross-language generalization. 

As shown in \autoref{tb:perturbation-seamless-enhance-eps0.1}, increasing the number of Seen languages enhances the attack transferability to Unseen languages.
This indicates that multilingual translation models exhibit semantic alignment across different languages, and optimizing perturbations with more Seen languages results in perturbations that more closely align with the target semantics, as shown in \autoref{fig:more-lang}. More results under different attack intensities refer to \cref{tb:perturbation-seamless-enhance-eps0.5,tb:perturbation-seamless-enhance-eps0.01} in Appendix.

In \autoref{fig:distribution-spa-more-enhance}, we illustrate the data distributions of ESIM and NSCORE, using Spanish as the target Unseen language, under the scenario of increasing the number of Seen languages. Combining the insights from \autoref{tb:perturbation-seamless-enhance-eps0.1} and \autoref{fig:distribution-spa-more-enhance}, we observe that incorporating more attack languages improves the generalizability of adversarial perturbations to Unseen languages.

\begin{table}[t]
\centering
\caption{Attack effectiveness of adversarial perturbations ($\epsilon = 0.1$), measured by average ESIM, NSCORE, and ASR. Blue-highlighted areas indicate tests conducted on \textbf{Seen} languages.}
\vspace{-.5em}
\fontsize{8}{10}\selectfont
\setlength{\tabcolsep}{9.8pt} 
\begin{tabular}{c|c|cc|c}
\hline
                                                                                                              &                                  & \multicolumn{2}{c|}{Similarity with Target}                                          &                               \\ \cline{3-4}
\multirow{-2}{*}{Attack   with}                                                                               & \multirow{-2}{*}{Target}         & \multicolumn{1}{c}{ESIM}                           & NSCORE                         & \multirow{-2}{*}{ASR}         \\ \hline
                                                                                                              & \cellcolor[HTML]{DDEBF7}English  & \multicolumn{1}{c}{\cellcolor[HTML]{DDEBF7}0.8973} & \multicolumn{1}{c|}{\cellcolor[HTML]{DDEBF7}0.7855} & \cellcolor[HTML]{DDEBF7}52/60 \\ 
                                                                                                              & Mandarin                         & \multicolumn{1}{c}{0.3900}                         & \multicolumn{1}{c|}{0.1717}                         & 19/60                         \\ 
                                                                                                              & German                           & \multicolumn{1}{c}{0.2999}                         & \multicolumn{1}{c|}{0.1214}                         & 27/60                         \\  
                                                                                                              & French                           & \multicolumn{1}{c}{0.3645}                         & \multicolumn{1}{c|}{0.1794}                         & 30/60                         \\ 
                                                                                                              & Italian                          & \multicolumn{1}{c}{0.3148}                         & \multicolumn{1}{c|}{0.1515}                         & 14/60                         \\ 
\multirow{-6}{*}{English}                                                                                     & Spanish                          & \multicolumn{1}{c}{0.3275}                         & \multicolumn{1}{c|}{0.1467}                         & 22/60                         \\ \hline
                                                                                                              & \cellcolor[HTML]{DDEBF7}English  & \multicolumn{1}{c}{\cellcolor[HTML]{DDEBF7}0.9234} & \multicolumn{1}{c|}{\cellcolor[HTML]{DDEBF7}0.9221} & \cellcolor[HTML]{DDEBF7}58/60 \\ 
                                                                                                              & \cellcolor[HTML]{DDEBF7}Mandarin & \multicolumn{1}{c}{\cellcolor[HTML]{DDEBF7}0.9844} & \multicolumn{1}{c|}{\cellcolor[HTML]{DDEBF7}0.9677} & \cellcolor[HTML]{DDEBF7}59/60 \\ 
                                                                                                              & German                           & \multicolumn{1}{c}{0.5823}                         & \multicolumn{1}{c|}{0.4216}                         & 37/60                         \\ 
                                                                                                              & French                           & \multicolumn{1}{c}{0.6036}                         & \multicolumn{1}{c|}{0.3901}                         & 39/60                         \\  
                                                                                                              & Italian                          & \multicolumn{1}{c}{0.4854}                         & \multicolumn{1}{c|}{0.3459}                         & 34/60                         \\  
\multirow{-6}{*}{\makecell{English\\Mandarin}}                             & Spanish                          & \multicolumn{1}{c}{0.5492}                         & \multicolumn{1}{c|}{0.3254}                         & 35/60                         \\ \hline
                                                                                                              & \cellcolor[HTML]{DDEBF7}English  & \multicolumn{1}{c}{\cellcolor[HTML]{DDEBF7}0.9290} & \multicolumn{1}{c|}{\cellcolor[HTML]{DDEBF7}0.9010} & \cellcolor[HTML]{DDEBF7}58/60 \\ 
                                                                                                              & \cellcolor[HTML]{DDEBF7}Mandarin & \multicolumn{1}{c}{\cellcolor[HTML]{DDEBF7}0.9479} & \multicolumn{1}{c|}{\cellcolor[HTML]{DDEBF7}0.8877} & \cellcolor[HTML]{DDEBF7}56/60 \\  
                                                                                                              & \cellcolor[HTML]{DDEBF7}German   & \multicolumn{1}{c}{\cellcolor[HTML]{DDEBF7}0.8771} & \multicolumn{1}{c|}{\cellcolor[HTML]{DDEBF7}0.8124} & \cellcolor[HTML]{DDEBF7}58/60 \\ 
                                                                                                              & French                           & \multicolumn{1}{c}{0.7281}                         & \multicolumn{1}{c|}{0.6866}                         & 53/60                         \\ 
                                                                                                              & Italian                          & \multicolumn{1}{c}{0.6415}                         & \multicolumn{1}{c|}{0.6593}                         & 49/60                         \\ 
\multirow{-6}{*}{\makecell{English\\      Mandarin \\      German}}              & Spanish                          & \multicolumn{1}{c}{0.6964}                         & \multicolumn{1}{c|}{0.5705}                         & 51/60                         \\ \hline
                                                                                                              & \cellcolor[HTML]{DDEBF7}English  & \multicolumn{1}{c}{\cellcolor[HTML]{DDEBF7}0.9238} & \multicolumn{1}{c|}{\cellcolor[HTML]{DDEBF7}0.9015} & \cellcolor[HTML]{DDEBF7}58/60 \\ 
                                                                                                              & \cellcolor[HTML]{DDEBF7}Mandarin & \multicolumn{1}{c}{\cellcolor[HTML]{DDEBF7}0.9222} & \multicolumn{1}{c|}{\cellcolor[HTML]{DDEBF7}0.8511} & \cellcolor[HTML]{DDEBF7}57/60 \\ 
                                                                                                              & \cellcolor[HTML]{DDEBF7}German   & \multicolumn{1}{c}{\cellcolor[HTML]{DDEBF7}0.8912} & \multicolumn{1}{c|}{\cellcolor[HTML]{DDEBF7}0.8335} & \cellcolor[HTML]{DDEBF7}56/60 \\ 
                                                                                                              & \cellcolor[HTML]{DDEBF7}French   & \multicolumn{1}{c}{\cellcolor[HTML]{DDEBF7}0.9356} & \multicolumn{1}{c|}{\cellcolor[HTML]{DDEBF7}0.9118} & \cellcolor[HTML]{DDEBF7}57/60 \\ 
                                                                                                              & Italian                          & \multicolumn{1}{c}{0.7026}                         & \multicolumn{1}{c|}{0.7450}                         & 53/60                         \\  
\multirow{-6}{*}{\makecell{English\\      Mandarin\\      German\\      French}} & Spanish                          & \multicolumn{1}{c}{0.7414}                         & \multicolumn{1}{c|}{0.6804}                         & 55/60                         \\ \hline
\end{tabular}
\label{tb:perturbation-seamless-enhance-eps0.1}
% \vspace{-1em}
\end{table}

\subsubsection{Enhancement based on Target Cycle Optimization}
\label{sec:cycle_enhancement_perturbation}
\begin{figure}[t]
    \centering
    \includegraphics[width=\columnwidth]{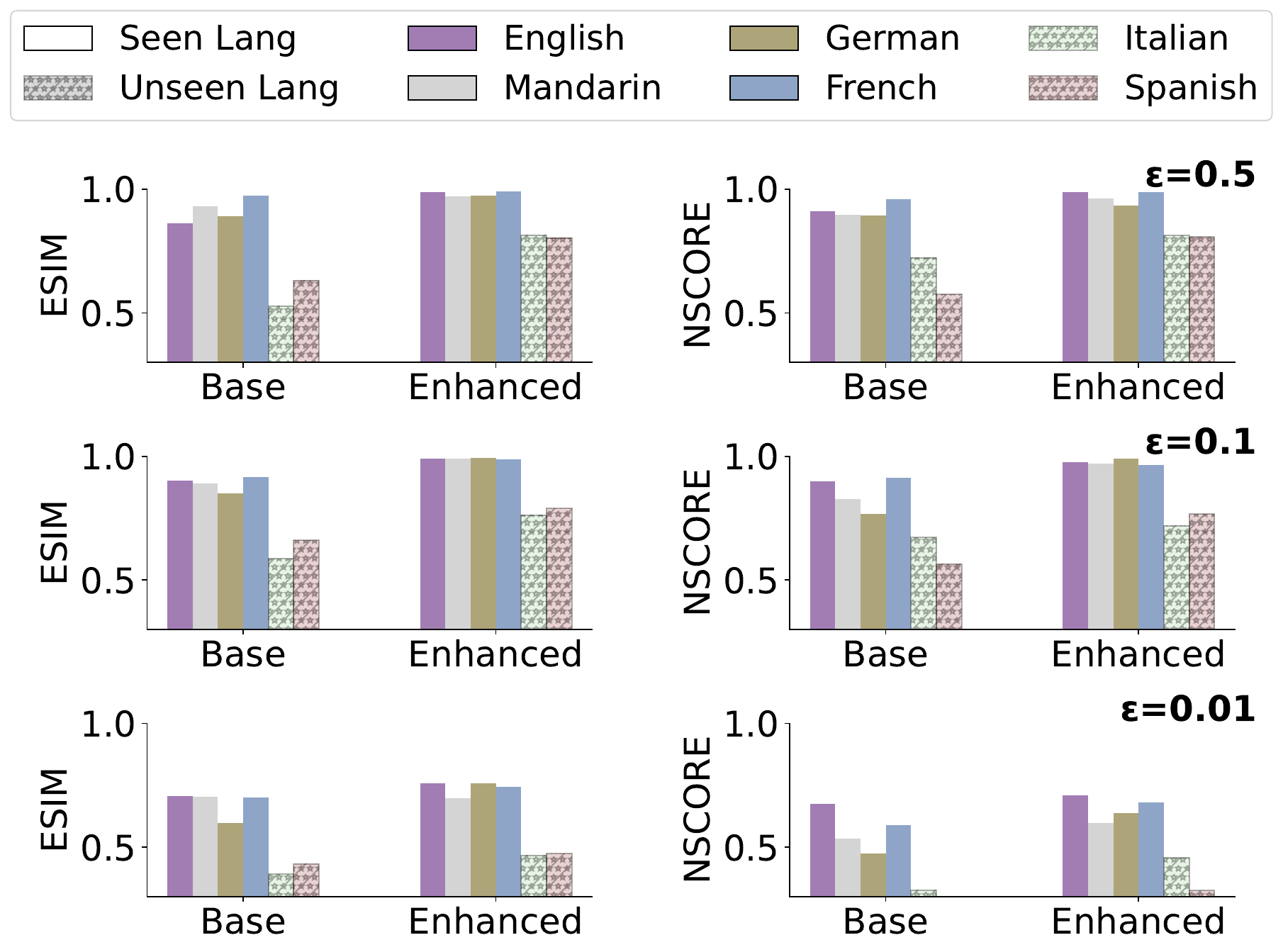}
    \caption{The attack effectiveness of adversarial perturbations before and after TCO enhancement. TCO significantly enhances attack effectiveness, particularly for Unseen languages, i.e., Italian and Spanish.}
    \label{fig:perturbation-seamless-cycle2}
\end{figure}

As described in \autoref{sec:cycle} and \autoref{fig:method-cycle-enhance}, we can perform a Target Cycle Optimization(TCO) on the attack targets to generate semantically similar targets that are easier to attack. We tested this approach on the default target, Seamless Large, and the results are shown in \autoref{fig:perturbation-seamless-cycle2}. Under different perturbation intensities ($\epsilon=0.5, 0.1, 0.01$), the effectiveness of adversarial attacks improves after applying TCO, as measured by the semantic similarity between the translation results and the target (ESIM and NSCORE). This improvement is particularly notable in the transferability to Unseen languages, which significantly outperforms the model before enhancement. This is because the target text generated through TCO is more compatible with different languages and is closer to the central semantics for the target model. The updated targets are presented in \autoref{tb:target-update} in Appendix, the semantic whose corresponding sentence changes during the updating are used for enhancement testing.

\subsubsection{Perceptual Evaluation}
% \vspace{-.5em}
To more comprehensively explore the effects of perturbation attacks, we applied different range constraints to the perturbations, as described in \autoref{sec:attack-with-perturbation}. Larger perturbation ranges are easier to perceive but tend to yield better attack results. In this study, we investigated perturbation ranges set to \( \epsilon = 0.5 \), \( 0.1 \), and \( 0.01 \). \autoref{tb:perception} presents the perceptual metrics for adversarially perturbed audio when the target model is Seamless Large. We optimized the perturbations using different numbers of attack languages as targets and compared the results with random perturbations of the same magnitude applied to the original speech. 

The results show that adversarial perturbations exhibit better perceptual quality than random perturbations of the same magnitude, particularly in maintaining speaker timbre and acoustic environment (VSIM-E). This is because our perturbations are specifically designed to avoid high or low frequency bands, as explained in \autoref{sec:attack-with-perturbation}. This approach significantly minimizes the impact on the core content of the speech (PESQ) and preserves speech style (VSIM, VSIM-E).
A more detailed analysis and discussion of the perceptual quality impact of adversarial perturbations are provided in \autoref{sec:perception-perturbation}.

\begin{table}[t]
\centering
\caption{Perception influence of adversarial perturbation. Indicators marked with “*” correspond to random perturbations.}
\vspace{-.5em}
\resizebox{\columnwidth}{!}{%
\begin{tabular}{c|c|ccc|ccc}
\hline
\multirow{2}{*}{$\epsilon$} & \multirow{2}{*}{Attack with} & \multicolumn{3}{c|}{Adversarial perturbation}                 & \multicolumn{3}{c}{Random perturbation}                          \\ \cline{3-8} 
                            &                              & PESQ($\uparrow$) & VSIM($\uparrow$) & VSIM-E($\uparrow$) & PESQ*($\uparrow$) & VSIM*($\uparrow$) & VSIM-E*($\uparrow$) \\ \hline
\multirow{4}{*}{0.5}        & EN                           & 1.1395           & 0.4661           & 0.2617             & 1.0658            & 0.4886            & -0.0942             \\
                            & EN+ZH                        & 1.0692           & 0.4756           & 0.2492             & 1.0052            & 0.4881            & -0.1103             \\
                            & EN+ZH+DE                     & 1.2289           & 0.4705           & 0.2413             & 1.0443            & 0.4831            & -0.1124             \\
                            & EN+ZH+DE+FR                  & 1.1191           & 0.4724           & 0.2369             & 1.0248            & 0.4768            & -0.1091             \\ \hline
\multirow{4}{*}{0.1}        & EN                           & 1.4102           & 0.6146           & 0.4172             & 1.4541            & 0.5911            & 0.1452              \\
                            & EN+ZH                        & 1.4077           & 0.6003           & 0.4037             & 1.4050            & 0.5746            & 0.1252              \\
                            & EN+ZH+DE                     & 1.3930           & 0.5915           & 0.3942             & 1.3687            & 0.5581            & 0.1096              \\
                            & EN+ZH+DE+FR                  & 1.3811           & 0.5881           & 0.3917             & 1.3579            & 0.5592            & 0.1049              \\ \hline
\multirow{4}{*}{0.01}       & EN                           & 2.3671           & 0.8346           & 0.6710             & 2.6614            & 0.8366            & 0.5107              \\
                            & EN+ZH                        & 2.3297           & 0.8286           & 0.6654             & 2.6300            & 0.8309            & 0.5009              \\
                            & EN+ZH+DE                     & 2.3089           & 0.8218           & 0.6600             & 2.6063            & 0.8259            & 0.4935              \\
                            & EN+ZH+DE+FR                  & 2.3045           & 0.8228           & 0.6577             & 2.6093            & 0.8257            & 0.4959              \\ \hline
\end{tabular}
}
\footnotesize Note: EN=English, ZH=Mandarin, DE=German, FR=French
\label{tb:perception}
\end{table}

\subsubsection{Generalizability}
% \vspace{-.5em}
We evaluated the effectiveness of the proposed method across different models. We conducted extensive tests on multiple models to evaluate the generalizability of the adversarial approach. As shown in \autoref{tb:generalizability-seamless}, we tested all examples and target semantics with English as both the attack language and the target language. As shown in \autoref{tb:generalizability-seamless}, the translated semantics of the audio (ESIM, NSCORE) after the attack closely align with the target semantics while significantly deviating from the original semantics.

\begin{table}[t]
\centering
\caption{Generalizability evaluation of adversarial perturbations across multiple models, with English used as the attack and target language.}
\vspace{-.5em}
\resizebox{\columnwidth}{!}{%
\begin{tabular}{c|c|cc|cc|c}
\hline
\multirow{2}{*}{$\epsilon$} & \multirow{2}{*}{Target model} & \multicolumn{2}{c|}{Similarity with Original} & \multicolumn{2}{c|}{Similarity with Target} & \multirow{2}{*}{ASR} \\ \cline{3-6}
                            &                               & ESIM                  & NSCORE                & ESIM                 & NSCORE               &                      \\ \hline
\multirow{4}{*}{0.5}        & Seamless Large                & 0.0285                & 0.0288                & 0.9612               & 0.9198               & 59/60                \\
                            & Seamless Medium               & 0.0332                & 0.0291                & 0.9927               & 0.9635               & 60/60                \\
                            & Seamless M4tv2                & 0.0383                & 0.0254                & 0.9201               & 0.8980               & 57/60                \\
                            & Seamless Expressive           & 0.0452                & 0.0374                & 0.8925               & 0.8224               & 55/60                \\ \hline
\multirow{4}{*}{0.1}        & Seamless Large                & 0.0317                & 0.0303                & 0.8973               & 0.7855               & 52/60                \\
                            & Seamless Medium               & 0.0422                & 0.0343                & 0.9240               & 0.8780               & 57/60                \\
                            & Seamless M4tv2                & 0.0412                & 0.0272                & 0.9022               & 0.8375               & 54/60                \\
                            & Seamless Expressive           & 0.0813                & 0.0386                & 0.7815               & 0.6840               & 49/60                \\ \hline
\multirow{4}{*}{0.01}       & Seamless Large                & 0.2154                & 0.0921                & 0.5503               & 0.4207               & 32/60                \\
                            & Seamless Medium               & 0.1223                & 0.0748                & 0.7514               & 0.6205               & 44/60                \\
                            & Seamless M4tv2                & 0.2449                & 0.1318                & 0.4980               & 0.3361               & 25/60                \\
                            & Seamless Expressive           & 0.2099                & 0.0996                & 0.6003               & 0.5031               & 36/60                \\ \hline
\end{tabular}
}
\label{tb:generalizability-seamless}
\end{table}

Additionally, to further investigate the generalization capability of the proposed method, we introduced an additional model, Canary~\cite{canary}, which is not part of the Seamless model family, for testing.
We conducted attacks on Canary using different numbers of languages, as shown in \cref{tb:perturbation-canary-enhance-eps0.1}. The proposed method demonstrates strong attack performance on the model of different categories outside the Seamless model family. Furthermore, the enhancement effect with more Seen languages remains consistent with the findings of previous experiments.
Combined with the results in \autoref{tb:generalizability-seamless}, these findings demonstrate that the proposed perturbation-base attack method effectively performs attacks across different models.
\subsection{Music-based Attack}
As described in \autoref{sec:attack-with-music}, we also explored the method of attacking using adversarial music. The Seamless model family do not require the specification of a source language during translation due to its inherent design. They process and analyze input audio by mapping it directly to a multilingual semantic space through speech understanding. 

For Canary model, preliminary study also show that source language has a limited influence on translation result. Therefore, we set English as the default source language for translation models during evaluation.

\begin{table}[t]
\centering
\caption{Attack ability of adversarial music in single-language translation scenarios.}
\vspace{-.5em}
\fontsize{8}{10}\selectfont
\setlength{\tabcolsep}{13pt} 
\renewcommand{\arraystretch}{1.1}
\begin{tabular}{cccc}
\hline
Target   Language & ESIM   & NSCORE & ASR   \\ \hline
English           & 0.7879 & 0.7507 & 9/10  \\ 
Mandarin          & 0.9669 & 0.9314 & 10/10 \\ 
German            & 0.7281 & 0.6139 & 8/10  \\ 
French            & 0.7783 & 0.7646 & 8/10  \\
Italian           & 0.6216 & 0.4871 & 6/10  \\
Spanish           & 0.8491 & 0.8823 & 9/10  \\ \hline
\end{tabular}
\label{tb:music-seamless-one}
\end{table}

\subsubsection{Attack Effectiveness}
To further investigate the impact of adversarial music, we expanded the target semantics based on the original set\footnote{The newly added target semantics are: ``This is unbelievable.", ``I can't stand you.", ``This is ridiculous.", ``Stop bothering me.", and ``What's wrong with you?"}.
\autoref{tb:music-seamless-one} presents the attack performance when targeting six different languages. The results demonstrate effectiveness comparable to the perturbation outcomes reported in \autoref{tb:one2one}.

The adversarial music generation process optimizes only the initial latent code and rhythm encoding during the diffusion process. To ensure experimental control, a fixed prompt was used as the text-to-music input. An exploration of different music generation prompts is detailed in \autoref{sec:prompts}.

\subsubsection{Enhancement based on More Seen Languages}
\cref{tb:music-seamless-enhance} show the results of attacks using different Seen languages. We observe that: (1) the generated adversarial music demonstrates strong attack capabilities on seen languages; (2) as the number of Seen languages increases, the adversarial music exhibits better generalization across multilingual scenarios; (3) overall, the adversarial music effectively attacks the target model. 
\begin{table}[t]
\centering
\caption{Attack ability of adversarial music with Seamless large as target model. Blue-highlighted areas indicate tests conducted on \textbf{Seen} languages.}
\vspace{-.5em}
\fontsize{8}{10}\selectfont
\setlength{\tabcolsep}{9.5pt} 
\renewcommand{\arraystretch}{1.1}
\begin{tabular}{c|c|cc|c}
\hline
                                                                                                              &                                  & \multicolumn{2}{c|}{Similarity With Target}                                          &                               \\ \cline{3-4}
\multirow{-2}{*}{Attack   with}                                                                               & \multirow{-2}{*}{Target}         & \multicolumn{1}{c}{ESIM}                           & NSCORE                         & \multirow{-2}{*}{ASR}         \\ \hline
                                                                                                              & \cellcolor[HTML]{DDEBF7}English  & \multicolumn{1}{c}{\cellcolor[HTML]{DDEBF7}0.7879} & \cellcolor[HTML]{DDEBF7}0.7507 & \cellcolor[HTML]{DDEBF7}9/10  \\ 
                                                                                                              & Mandarin                         & \multicolumn{1}{c}{0.5152}                         & 0.4257                         & 6/10                          \\ 
                                                                                                              & German                           & \multicolumn{1}{c}{0.5706}                         & 0.4236                         & 6/10                          \\ 
                                                                                                              & French                           & \multicolumn{1}{c}{0.4643}                         & 0.5759                         & 7/10                          \\ 
                                                                                                              & Italian                          & \multicolumn{1}{c}{0.4877}                         & 0.6616                         & 7/10                          \\ 
\multirow{-6}{*}{English}                                                                                     & Spanish                          & \multicolumn{1}{c}{0.4408}                         & 0.4661                         & 4/10                          \\ \hline
                                                                                                              & \cellcolor[HTML]{DDEBF7}English  & \multicolumn{1}{c}{\cellcolor[HTML]{DDEBF7}0.8434} & \cellcolor[HTML]{DDEBF7}0.7893 & \cellcolor[HTML]{DDEBF7}9/10  \\ 
                                                                                                              & \cellcolor[HTML]{DDEBF7}Mandarin & \multicolumn{1}{c}{\cellcolor[HTML]{DDEBF7}0.8362} & \cellcolor[HTML]{DDEBF7}0.6615 & \cellcolor[HTML]{DDEBF7}10/10 \\ 
                                                                                                              & German                           & \multicolumn{1}{c}{0.7633}                         & 0.6370                         & 9/10                          \\ 
                                                                                                              & French                           & \multicolumn{1}{c}{0.6396}                         & 0.5117                         & 8/10                          \\ 
                                                                                                              & Italian                          & \multicolumn{1}{c}{0.6199}                         & 0.6236                         & 7/10                          \\ 
\multirow{-6}{*}{\begin{tabular}[c]{@{}c@{}}English\\      Mandarin\end{tabular}}                             & Spanish                          & \multicolumn{1}{c}{0.6691}                         & 0.5366                         & 7/10                          \\ \hline
                                                                                                              & \cellcolor[HTML]{DDEBF7}English  & \multicolumn{1}{c}{\cellcolor[HTML]{DDEBF7}0.8493} & \cellcolor[HTML]{DDEBF7}0.8823 & \cellcolor[HTML]{DDEBF7}9/10  \\ 
                                                                                                              & \cellcolor[HTML]{DDEBF7}Mandarin & \multicolumn{1}{c}{\cellcolor[HTML]{DDEBF7}0.8516} & \cellcolor[HTML]{DDEBF7}0.8559 & \cellcolor[HTML]{DDEBF7}9/10  \\ 
                                                                                                              & \cellcolor[HTML]{DDEBF7}German   & \multicolumn{1}{c}{\cellcolor[HTML]{DDEBF7}0.9466} & \cellcolor[HTML]{DDEBF7}0.7901 & \cellcolor[HTML]{DDEBF7}10/10 \\ 
                                                                                                              & French                           & \multicolumn{1}{c}{0.6953}                         & 0.6626                         & 8/10                          \\ 
                                                                                                              & Italian                          & \multicolumn{1}{c}{0.7277}                         & 0.7611                         & 8/10                          \\ 
\multirow{-6}{*}{\begin{tabular}[c]{@{}c@{}}English\\      Mandarin \\      German\end{tabular}}              & Spanish                          & \multicolumn{1}{c}{0.7412}                         & 0.6852                         & 8/10                          \\ \hline
                                                                                                              & \cellcolor[HTML]{DDEBF7}English  & \multicolumn{1}{c}{\cellcolor[HTML]{DDEBF7}0.9267} & \cellcolor[HTML]{DDEBF7}0.9821 & \cellcolor[HTML]{DDEBF7}10/10 \\ 
                                                                                                              & \cellcolor[HTML]{DDEBF7}Mandarin & \multicolumn{1}{c}{\cellcolor[HTML]{DDEBF7}0.8899} & \cellcolor[HTML]{DDEBF7}0.8915 & \cellcolor[HTML]{DDEBF7}9/10  \\ 
                                                                                                              & \cellcolor[HTML]{DDEBF7}German   & \multicolumn{1}{c}{\cellcolor[HTML]{DDEBF7}0.8519} & \cellcolor[HTML]{DDEBF7}0.8866 & \cellcolor[HTML]{DDEBF7}9/10  \\ 
                                                                                                              & \cellcolor[HTML]{DDEBF7}French   & \multicolumn{1}{c}{\cellcolor[HTML]{DDEBF7}0.8804} & \cellcolor[HTML]{DDEBF7}0.8851 & \cellcolor[HTML]{DDEBF7}9/10  \\ 
                                                                                                              & Italian                          & \multicolumn{1}{c}{0.7434}                         & 0.8818                         & 9/10                          \\ 
\multirow{-6}{*}{\begin{tabular}[c]{@{}c@{}}English\\      Mandarin\\      German\\      French\end{tabular}} & Spanish                          & \multicolumn{1}{c}{0.8021}                         & 0.8704                         & 10/10                         \\ \hline
\end{tabular}
\label{tb:music-seamless-enhance}
\end{table}

\subsubsection{Enhancement based on Target Cycle Optimization}
\begin{figure}[!htb]
    \centering
    \includegraphics[width=\columnwidth]{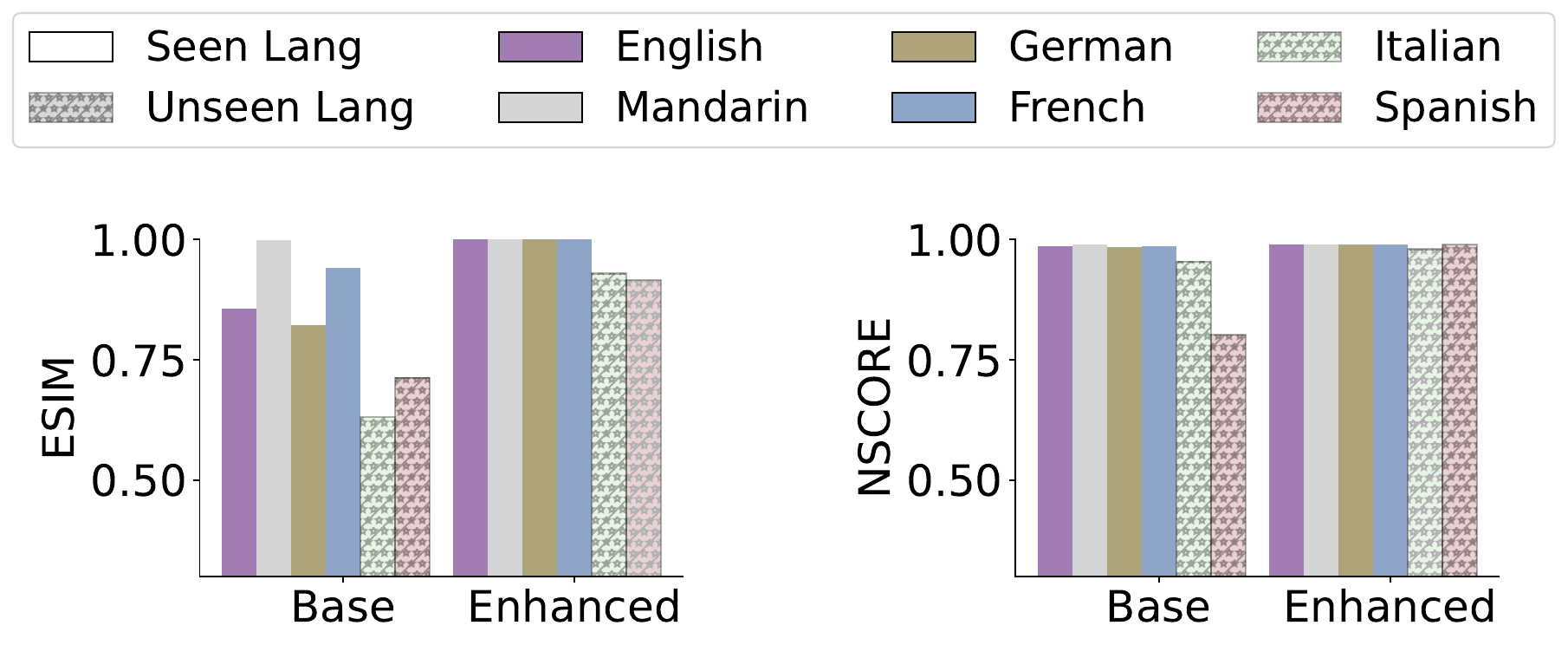}
    \caption{The attack effectiveness of adversarial music is enhanced through Target Cycle Optimization.}
    \label{fig:music-seamless-cycle2}
\end{figure}

As outlined in \autoref{sec:cycle} and \autoref{fig:method-cycle-enhance}, Target Cycle Optimization(TCO) can be applied to the attack targets, generating semantically similar targets that are more susceptible to attack. Similar to the experiments discussed in \autoref{sec:cycle_enhancement_perturbation}, we tested this approach on the default target, Seamless Large, and the results are presented in \autoref{fig:music-seamless-cycle2}. The application of TCO significantly improves the effectiveness of adversarial attacks, as indicated by higher semantic similarity between the translated results and the target (measured using ESIM and NSCORE). The improvement is especially evident in the transferability to Unseen languages, where attack performance improve significantly after applying TCO. The enhanced targets generated through TCO are better aligned with various languages and are closer to the central semantics of the target model. The updated targets are summarized in \autoref{tb:target-update}, the semantic whose corresponding sentence changes during the updating are used for enhancement testing.

\begin{table*}[t]
\centering
\caption{Evaluation of the impact of different audio signal processing techniques on the effectiveness of adversarial perturbations and adversarial music.}
\vspace{-.5em}
\resizebox{\textwidth}{!}{%
\begin{tabular}{cc|cccccc|cccccc}
\hline
\multicolumn{2}{c|}{Type}                            & \multicolumn{6}{c|}{Perturbation}                              & \multicolumn{6}{c}{Music}                                      \\ \hline
\multicolumn{2}{c|}{Processing}                      & None            & LPF    & MP3    & Quant  & Noise  & Resample & None            & LPF    & MP3    & Quant  & Noise  & Resample \\ \hline
\multirow{2}{*}{Similarity   With Original} & ESIM   & \textbf{0.0317} & 0.0932 & 0.0879 & 0.1367 & 0.0487 & 0.1037   & -               & -      & -      & -      & -      & -        \\
                                            & NSCORE & \textbf{0.0303} & 0.0345 & 0.0478 & 0.1166 & 0.0355 & 0.0556   & -               & -      & -      & -      & -      & -        \\ \hline
\multirow{2}{*}{Similarity   With Target}   & ESIM   & \textbf{0.8973} & 0.5803 & 0.5361 & 0.2675 & 0.7490 & 0.4175   & \textbf{0.7879} & 0.7196 & 0.6378 & 0.4175 & 0.5845 & 0.4638   \\
                                            & NSCORE & \textbf{0.7855} & 0.3872 & 0.3349 & 0.1345 & 0.5767 & 0.1723   & \textbf{0.7507} & 0.6585 & 0.6125 & 0.3704 & 0.5577 & 0.3034   \\ \hline
\multicolumn{2}{c|}{ASR}                             & \textbf{52/60}  & 35/60  & 27/60  & 12/60  & 45/60  & 21/60    & \textbf{9/10}   & 8/10   & 8/10   & 4/10   & 7/10   & 4/10     \\ \hline
\end{tabular}
}
\label{tb:defense-processing}
\end{table*}

\subsubsection{Generalizability}
\begin{table}[htbp]
\centering
\caption{Generalizability evaluation of adversarial Music across multiple models, with English used as the attack and target language.}
\vspace{-.5em}
\fontsize{8}{10}\selectfont
\setlength{\tabcolsep}{13pt} 
\renewcommand{\arraystretch}{1.1}
\begin{tabular}{cccc}
\hline
\multirow{2}{*}{Target   model} & \multicolumn{2}{c}{Similarity With Target} & \multirow{2}{*}{ASR} \\ \cline{2-3}
                                & \multicolumn{1}{c}{ESIM}       & NSCORE    &                      \\ \hline
Seamless   Large                & \multicolumn{1}{c}{0.7879}     & 0.7507    & 9/10                 \\ 
Seamless   Medium               & \multicolumn{1}{c}{0.9211}     & 0.8100    & 9/10                 \\ 
Seamless   M4tv2                & \multicolumn{1}{c}{0.8017}     & 0.8164    & 10/10                \\ 
Seamless   Expressive           & \multicolumn{1}{c}{0.9142}     & 0.9595    & 10/10                \\ \hline
\end{tabular}
\label{tb:music-seamless-general}
\end{table}

In addtion, we conducted additional experiments on Canary~\cite{canary}, and the results are shown in \autoref{tb:music-canary-enhance}. 

Consistent with the results in \autoref{tb:music-seamless-general}, this further demonstrates the generalization capability of adversarial music.
Furthermore, the enhancement effect with more Seen languages remains consistent with the findings of previous experiments.

\begin{table}[t]
\centering
\caption{Attack ability of adversarial music with Canary as target model. Blue-highlighted areas indicate tests conducted on \textbf{Seen} languages.}
\vspace{-.5em}
\fontsize{8}{10}\selectfont
\setlength{\tabcolsep}{9.5pt} 
\renewcommand{\arraystretch}{1.1}
\begin{tabular}{c|c|cc|c}
\hline
                                                                                                             &                                 & \multicolumn{2}{c|}{Similarity With Target}                                          &                               \\ \cline{3-4}
\multirow{-2}{*}{Attack   with}                                                                              & \multirow{-2}{*}{Target}        & \multicolumn{1}{c}{ESIM}                           & NSCORE                         & \multirow{-2}{*}{ASR}         \\ \hline
                                                                                                             & \cellcolor[HTML]{DDEBF7}English & \multicolumn{1}{c}{\cellcolor[HTML]{DDEBF7}0.7899} & \cellcolor[HTML]{DDEBF7}0.7588 & \cellcolor[HTML]{DDEBF7}7/10  \\ 
                                                                                                             & French                          & \multicolumn{1}{c}{0.5881}                         & 0.3989                         & 6/10                          \\ 
                                                                                                             & German                          & \multicolumn{1}{c}{0.5863}                         & 0.3620                         & 9/10                          \\ 
\multirow{-4}{*}{English}                                                                                    & Spanish                         & \multicolumn{1}{c}{0.5661}                         & 0.4407                         & 7/10                          \\ \hline
                                                                                                             & \cellcolor[HTML]{DDEBF7}English & \multicolumn{1}{c}{\cellcolor[HTML]{DDEBF7}0.9817} & \cellcolor[HTML]{DDEBF7}0.9543 & \cellcolor[HTML]{DDEBF7}10/10 \\ 
                                                                                                             & \cellcolor[HTML]{DDEBF7}French  & \multicolumn{1}{c}{\cellcolor[HTML]{DDEBF7}0.9397} & \cellcolor[HTML]{DDEBF7}0.9117 & \cellcolor[HTML]{DDEBF7}9/10  \\ 
                                                                                                             & German                          & \multicolumn{1}{c}{0.7567}                         & 0.7013                         & 9/10                          \\ 
\multirow{-4}{*}{\begin{tabular}[c]{@{}c@{}}English\\      French\end{tabular}}                              & Spanish                         & \multicolumn{1}{c}{0.6729}                         & 0.6010                         & 7/10                          \\ \hline
                                                                                                             & \cellcolor[HTML]{DDEBF7}English & \multicolumn{1}{c}{\cellcolor[HTML]{DDEBF7}0.9616} & \cellcolor[HTML]{DDEBF7}0.9730 & \cellcolor[HTML]{DDEBF7}10/10 \\ 
                                                                                                             & \cellcolor[HTML]{DDEBF7}French  & \multicolumn{1}{c}{\cellcolor[HTML]{DDEBF7}1.0000} & \cellcolor[HTML]{DDEBF7}0.9862 & \cellcolor[HTML]{DDEBF7}10/10 \\ 
                                                                                                             & \cellcolor[HTML]{DDEBF7}German  & \multicolumn{1}{c}{\cellcolor[HTML]{DDEBF7}0.9984} & \cellcolor[HTML]{DDEBF7}0.9865 & \cellcolor[HTML]{DDEBF7}10/10 \\ 
\multirow{-4}{*}{\begin{tabular}[c]{@{}c@{}}English\\      French\\      German\end{tabular}}                & Spanish                         & \multicolumn{1}{c}{0.8544}                         & 0.8846                         & 9/10                          \\ \hline
                                                                                                             & \cellcolor[HTML]{DDEBF7}English & \multicolumn{1}{c}{\cellcolor[HTML]{DDEBF7}0.9935} & \cellcolor[HTML]{DDEBF7}0.9877 & \cellcolor[HTML]{DDEBF7}10/10 \\ 
                                                                                                             & \cellcolor[HTML]{DDEBF7}French  & \multicolumn{1}{c}{\cellcolor[HTML]{DDEBF7}0.9988} & \cellcolor[HTML]{DDEBF7}0.9862 & \cellcolor[HTML]{DDEBF7}10/10 \\ 
                                                                                                             & \cellcolor[HTML]{DDEBF7}German  & \multicolumn{1}{c}{\cellcolor[HTML]{DDEBF7}0.9247} & \cellcolor[HTML]{DDEBF7}0.8242 & \cellcolor[HTML]{DDEBF7}10/10 \\ 
\multirow{-4}{*}{\begin{tabular}[c]{@{}c@{}}English\\      French\\      German\\      Spanish\end{tabular}} & \cellcolor[HTML]{DDEBF7}Spanish & \multicolumn{1}{c}{\cellcolor[HTML]{DDEBF7}0.9800} & \cellcolor[HTML]{DDEBF7}0.9856 & \cellcolor[HTML]{DDEBF7}10/10 \\ \hline
\end{tabular}
\label{tb:music-canary-enhance}
\end{table}

\subsubsection{Physical Test}
As described in \autoref{sec:tm}, a more severe attack method involves transmitting adversarial music over the air, as illustrated in \autoref{fig:tm}. To implement this, we integrated simulated air-channel transmission distortions into the adversarial perturbation optimization process. Details of these distortions are provided in Appendix \autoref{sec:ota-simulate}.

We evaluated adversarial music attacks on two models: Seamless Large and Canary. Consumer-grade speakers were used for playback, while a consumer-grade microphone and a smartphone captured the audio to simulate typical over-the-air conditions. The specifications of the devices are detailed in \autoref{fig:device} \footnote{Experiments were conducted in a room measuring 4.37m × 2.35m × 2.95m, with the microphone and speaker placed 50 cm apart.}.

For each attack, six adversarial music samples were generated and tested multiple times to ensure stability, resulting in 60 test samples per target language. The results, summarized in \autoref{tb:ota-seamless}, indicate that adversarial music achieves an attack success rate of approximately 50\% across various models and devices in over-the-air attack scenarios. These findings suggest that adversarial music could be exploited to inject malicious semantics into real-time speech translation conferences or conversations, posing significant security risks.

\begin{table}[htbp]
\centering
\caption{Attack ability of adversarial Music in over-the-air english-targeted translation scenarios.}
\vspace{-.5em}
\fontsize{8}{10}\selectfont
\setlength{\tabcolsep}{10pt} 
\renewcommand{\arraystretch}{1.1}
\begin{tabular}{c|c|c|c}
\hline
\multicolumn{1}{l|}{Target model} & Device                        & Target Language & ASR   \\ \hline
\multirow{12}{*}{Seamless large}   & \multirow{6}{*}{Microphone}   & English         & 31/60 \\  
                                   &                               & Mandarin        & 34/60 \\ 
                                   &                               & German          & 38/60 \\ 
                                   &                               & French          & 33/60 \\ 
                                   &                               & Italian         & 29/60 \\ 
                                   &                               & Spanish         & 27/60 \\ \cline{2-4} 
                                   & \multirow{6}{*}{Cell   Phone} & English         & 28/60 \\ 
                                   &                               & Mandarin        & 35/60 \\ 
                                   &                               & German          & 38/60 \\  
                                   &                               & French          & 34/60 \\ 
                                   &                               & Italian         & 33/60 \\ 
                                   &                               & Spanish         & 25/60 \\ \hline
\multirow{8}{*}{Canary}            & \multirow{4}{*}{Microphone}   & English         & 27/60 \\ 
                                   &                               & French          & 47/60 \\ 
                                   &                               & German          & 30/60 \\ 
                                   &                               & Spanish         & 36/60 \\ \cline{2-4} 
                                   & \multirow{4}{*}{Cell   Phone} & English         & 29/60 \\ 
                                   &                               & French          & 38/60 \\ 
                                   &                               & German          & 34/60 \\ 
                                   &                               & Spanish         & 41/60 \\ \hline
\end{tabular}
\label{tb:ota-seamless}
\vspace{-1.5em}
\end{table}

\subsection{User Study}

In addition to the attack effectiveness and objective quality evaluations, we also conducted subjective experimental assessments on both adversarial perturbations and adversarial music. In this test, 20 participants were invited to rate the quality of speech overlaid with adversarial perturbations and the generated adversarial music. To serve as a baseline, random white noise matching the energy intensity of each adversarial perturbation was generated. Similarly, white noise with equivalent energy intensity was created for each piece of adversarial music.
The detailed scoring criteria are provided in \autoref{tb:mos}. The scoring statistics for the perturbations and music are presented in \autoref{fig:mos} and \autoref{fig:mos-music}, respectively.

As shown in \autoref{fig:mos}, as the perturbation strength increases, the scores tend to decrease. However, adversarial perturbations consistently demonstrate better perceptual quality compared to random perturbations of the same strength, particularly at higher perturbation levels. With the default \(\epsilon = 0.1\), \autoref{tb:mos} indicates that most adversarial perturbations do not significantly affect the perception of speech content.

For the generated adversarial music, as shown in \autoref{fig:mos-music}, adversarial music demonstrates better perceptual quality compared to random perturbations of the same strength. Furthermore, the generated music receives high scores, demonstrating the imperceptibility of the adversarial music.

\section{Defense Attempt}

To evaluate potential countermeasures against the identified security vulnerability, we conducted a series of defense experiments targeting the proposed adversarial perturbations and music-based attacks.
Specifically, various audio signal processing techniques were applied to introduce distortions, aiming to disrupt the adversarial effectiveness of proposed attacks. These techniques included filtering (6 kHz low-pass, denoted as LPF), compression (64 kbps, MP3), noise addition (SNR 64 dB, Noise), quantization (8-bit, Quant), and resampling (12 kHz, Resample). The results of these experiments are presented in \autoref{tb:defense-processing}.

The experimental results indicate that adversarial perturbations and adversarial music exhibit a certain degree of robustness to audio processing. However, certain techniques, particularly quantization and resampling, can significantly impact the attack effectiveness. This finding suggests that, in the absence of cost concerns, resisting adversarial audio attacks is feasible. However, based on the results from perturbation removal experiments, while these processing techniques mitigate the intensity of adversarial attacks, they do not fully restore the semantic integrity of the original speech. Moreover, these methods may interfere with the semantic information of the original audio, thereby reducing its usability.

\section{Conclusion}

In this paper, we explored the vulnerability of ST systems to adversarial attacks and proposed two targeted strategies: perturbation-based attack and an innovative adversarial music optimization approach. We introduced several methods to enhance adversarial attacks on ST models, including Multi-language Enhancement and Target Cycle Optimization. Extensive experiments were conducted using various source and target language pairs, demonstrating the susceptibility of current ST systems to adversarial attacks. We hope our research raises awareness of the security challenges in ST systems and contributes to efforts to improve their robustness.

\bibliographystyle{IEEEtran}
\bibliography{ref, ref-link}

% Generated by IEEEtran.bst, version: 1.14 (2015/08/26)
\begin{thebibliography}{10}
\providecommand{\url}[1]{#1}
\csname url@samestyle\endcsname
\providecommand{\newblock}{\relax}
\providecommand{\bibinfo}[2]{#2}
\providecommand{\BIBentrySTDinterwordspacing}{\spaceskip=0pt\relax}
\providecommand{\BIBentryALTinterwordstretchfactor}{4}
\providecommand{\BIBentryALTinterwordspacing}{\spaceskip=\fontdimen2\font plus
\BIBentryALTinterwordstretchfactor\fontdimen3\font minus
  \fontdimen4\font\relax}
\providecommand{\BIBforeignlanguage}[2]{{%
\expandafter\ifx\csname l@#1\endcsname\relax
\typeout{** WARNING: IEEEtran.bst: No hyphenation pattern has been}%
\typeout{** loaded for the language `#1'. Using the pattern for}%
\typeout{** the default language instead.}%
\else
\language=\csname l@#1\endcsname
\fi
#2}}
\providecommand{\BIBdecl}{\relax}
\BIBdecl

\bibitem{dhawan2022speech}
S.~Dhawan, ``Speech to speech translation: Challenges and future,''
  \emph{International Journal of Computer Applications Technology and
  Research}, vol.~11, no.~03, pp. 36--55, 2022.

\bibitem{wang2022survey}
Y.~Wang, Z.~Su, N.~Zhang, R.~Xing, D.~Liu, T.~H. Luan, and X.~Shen, ``A survey
  on metaverse: Fundamentals, security, and privacy,'' \emph{IEEE
  Communications Surveys \& Tutorials}, vol.~25, no.~1, pp. 319--352, 2022.

\bibitem{seamless-blog}
M.~AI, ``Seamless communication,''
  \url{https://ai.meta.com/blog/seamless-communication/}, 2023, accessed:
  2024-05-21.

\bibitem{lavie1997janus}
A.~Lavie, A.~Waibel, L.~Levin, M.~Finke, D.~Gates, M.~Gavalda, T.~Zeppenfeld,
  and P.~Zhan, ``Janus-iii: Speech-to-speech translation in multiple
  languages,'' in \emph{1997 IEEE International Conference on Acoustics,
  Speech, and Signal Processing}, vol.~1.\hskip 1em plus 0.5em minus
  0.4em\relax IEEE, 1997, pp. 99--102.

\bibitem{wahlster2013verbmobil}
W.~Wahlster, \emph{Verbmobil: foundations of speech-to-speech
  translation}.\hskip 1em plus 0.5em minus 0.4em\relax Springer Science \&
  Business Media, 2013.

\bibitem{nakamura2006atr}
S.~Nakamura, K.~Markov, H.~Nakaiwa, G.-i. Kikui, H.~Kawai, T.~Jitsuhiro, J.-S.
  Zhang, H.~Yamamoto, E.~Sumita, and S.~Yamamoto, ``The atr multilingual
  speech-to-speech translation system,'' \emph{IEEE Transactions on Audio,
  Speech, and Language Processing}, vol.~14, no.~2, pp. 365--376, 2006.

\bibitem{inaguma2020espnet}
H.~Inaguma, S.~Kiyono, K.~Duh, S.~Karita, N.~E.~Y. Soplin, T.~Hayashi, and
  S.~Watanabe, ``Espnet-st: All-in-one speech translation toolkit,''
  \emph{arXiv preprint arXiv:2004.10234}, 2020.

\bibitem{barrault2023seamless}
L.~Barrault, Y.-A. Chung, M.~C. Meglioli, D.~Dale, N.~Dong, M.~Duppenthaler,
  P.-A. Duquenne, B.~Ellis, H.~Elsahar, J.~Haaheim \emph{et~al.}, ``Seamless:
  Multilingual expressive and streaming speech translation,'' \emph{arXiv
  preprint arXiv:2312.05187}, 2023.

\bibitem{wang2024end}
H.~Wang, Z.~Xue, Y.~Lei, and D.~Xiong, ``End-to-end speech translation with
  mutual knowledge distillation,'' in \emph{ICASSP 2024-2024 IEEE International
  Conference on Acoustics, Speech and Signal Processing (ICASSP)}.\hskip 1em
  plus 0.5em minus 0.4em\relax IEEE, 2024, pp. 11\,306--11\,310.

\bibitem{canary}
NVIDIA, ``Canary,'' \url{https://huggingface.co/nvidia/canary-1b}.

\bibitem{ney1999speech}
H.~Ney, ``Speech translation: Coupling of recognition and translation,'' in
  \emph{1999 IEEE International Conference on Acoustics, Speech, and Signal
  Processing. Proceedings. ICASSP99 (Cat. No. 99CH36258)}, vol.~1.\hskip 1em
  plus 0.5em minus 0.4em\relax IEEE, 1999, pp. 517--520.

\bibitem{matusov2005integration}
E.~Matusov, S.~Kanthak, and H.~Ney, ``On the integration of speech recognition
  and statistical machine translation.'' in \emph{Interspeech}, 2005, pp.
  3177--3180.

\bibitem{berard2016listen}
A.~B{\'e}rard, O.~Pietquin, C.~Servan, and L.~Besacier, ``Listen and translate:
  A proof of concept for end-to-end speech-to-text translation,'' \emph{arXiv
  preprint arXiv:1612.01744}, 2016.

\bibitem{bansal2018low}
S.~Bansal, H.~Kamper, K.~Livescu, A.~Lopez, and S.~Goldwater, ``Low-resource
  speech-to-text translation,'' \emph{arXiv preprint arXiv:1803.09164}, 2018.

\bibitem{iranzo2020europarl}
J.~Iranzo-S{\'a}nchez, J.~A. Silvestre-Cerda, J.~Jorge, N.~Rosell{\'o},
  A.~Gim{\'e}nez, A.~Sanchis, J.~Civera, and A.~Juan, ``Europarl-st: A
  multilingual corpus for speech translation of parliamentary debates,'' in
  \emph{ICASSP 2020-2020 IEEE International Conference on Acoustics, Speech and
  Signal Processing (ICASSP)}.\hskip 1em plus 0.5em minus 0.4em\relax IEEE,
  2020, pp. 8229--8233.

\bibitem{barrault2023seamlessm4t}
L.~Barrault, Y.-A. Chung, M.~C. Meglioli, D.~Dale, N.~Dong, P.-A. Duquenne,
  H.~Elsahar, H.~Gong, K.~Heffernan, J.~Hoffman \emph{et~al.},
  ``Seamlessm4t-massively multilingual \& multimodal machine translation,''
  \emph{arXiv preprint arXiv:2308.11596}, 2023.

\bibitem{huang2021defending}
C.-y. Huang, Y.~Y. Lin, H.-y. Lee, and L.-s. Lee, ``Defending your voice:
  Adversarial attack on voice conversion,'' in \emph{2021 IEEE Spoken Language
  Technology Workshop (SLT)}.\hskip 1em plus 0.5em minus 0.4em\relax IEEE,
  2021, pp. 552--559.

\bibitem{yu2023antifake}
Z.~Yu, S.~Zhai, and N.~Zhang, ``Antifake: Using adversarial audio to prevent
  unauthorized speech synthesis,'' in \emph{Proceedings of the 2023 ACM SIGSAC
  Conference on Computer and Communications Security}, 2023, pp. 460--474.

\bibitem{liu2023protecting}
Z.~Liu, Y.~Zhang, and C.~Miao, ``Protecting your voice from speech synthesis
  attacks,'' in \emph{Proceedings of the 39th Annual Computer Security
  Applications Conference}, 2023, pp. 394--408.

\bibitem{carlini2016hidden}
N.~Carlini, P.~Mishra, T.~Vaidya, Y.~Zhang, M.~Sherr, C.~Shields, D.~Wagner,
  and W.~Zhou, ``Hidden voice commands,'' in \emph{25th USENIX security
  symposium (USENIX security 16)}, 2016, pp. 513--530.

\bibitem{yuan2018commandersong}
X.~Yuan, Y.~Chen, Y.~Zhao, Y.~Long, X.~Liu, K.~Chen, S.~Zhang, H.~Huang,
  X.~Wang, and C.~A. Gunter, ``$\{$CommanderSong$\}$: A systematic approach for
  practical adversarial voice recognition,'' in \emph{27th USENIX security
  symposium (USENIX security 18)}, 2018, pp. 49--64.

\bibitem{abdullah2021hear}
H.~Abdullah, M.~S. Rahman, W.~Garcia, K.~Warren, A.~S. Yadav, T.~Shrimpton, and
  P.~Traynor, ``Hear" no evil", see" kenansville": Efficient and transferable
  black-box attacks on speech recognition and voice identification systems,''
  in \emph{2021 IEEE Symposium on Security and Privacy (SP)}.\hskip 1em plus
  0.5em minus 0.4em\relax IEEE, 2021, pp. 712--729.

\bibitem{chen2020metamorph}
T.~Chen, L.~Shangguan, Z.~Li, and K.~Jamieson, ``Metamorph: Injecting inaudible
  commands into over-the-air voice controlled systems,'' in \emph{Network and
  Distributed Systems Security (NDSS) Symposium}, 2020.

\bibitem{schonherr2018adversarial}
L.~Sch{\"o}nherr, K.~Kohls, S.~Zeiler, T.~Holz, and D.~Kolossa, ``Adversarial
  attacks against automatic speech recognition systems via psychoacoustic
  hiding,'' \emph{arXiv preprint arXiv:1808.05665}, 2018.

\bibitem{yu2023smack}
Z.~Yu, Y.~Chang, N.~Zhang, and C.~Xiao, ``$\{$SMACK$\}$: Semantically
  meaningful adversarial audio attack,'' in \emph{32nd USENIX Security
  Symposium (USENIX Security 23)}, 2023, pp. 3799--3816.

\bibitem{chen2021real}
G.~Chen, S.~Chenb, L.~Fan, X.~Du, Z.~Zhao, F.~Song, and Y.~Liu, ``Who is real
  bob? adversarial attacks on speaker recognition systems,'' in \emph{2021 IEEE
  Symposium on Security and Privacy (SP)}.\hskip 1em plus 0.5em minus
  0.4em\relax IEEE, 2021, pp. 694--711.

\bibitem{li2023enrollment}
X.~Li, J.~Ze, C.~Yan, Y.~Cheng, X.~Ji, and W.~Xu, ``Enrollment-stage backdoor
  attacks on speaker recognition systems via adversarial ultrasound,''
  \emph{IEEE Internet of Things Journal}, 2023.

\bibitem{chen2023qfa2sr}
G.~Chen, Y.~Zhang, Z.~Zhao, and F.~Song, ``$\{$QFA2SR$\}$:$\{$Query-Free$\}$
  adversarial transfer attacks to speaker recognition systems,'' in \emph{32nd
  USENIX Security Symposium (USENIX Security 23)}, 2023, pp. 2437--2454.

\bibitem{yakura2018robust}
H.~Yakura and J.~Sakuma, ``Robust audio adversarial example for a physical
  attack,'' \emph{arXiv preprint arXiv:1810.11793}, 2018.

\bibitem{carlini2018audio}
N.~Carlini and D.~Wagner, ``Audio adversarial examples: Targeted attacks on
  speech-to-text,'' in \emph{2018 IEEE security and privacy workshops
  (SPW)}.\hskip 1em plus 0.5em minus 0.4em\relax IEEE, 2018, pp. 1--7.

\bibitem{vaswani2017attention}
A.~Vaswani, N.~Shazeer, N.~Parmar, J.~Uszkoreit, L.~Jones, A.~N. Gomez,
  {\L}.~Kaiser, and I.~Polosukhin, ``Attention is all you need,''
  \emph{Advances in neural information processing systems}, vol.~30, 2017.

\bibitem{wu2023decoder}
J.~Wu, Y.~Gaur, Z.~Chen, L.~Zhou, Y.~Zhu, T.~Wang, J.~Li, S.~Liu, B.~Ren,
  L.~Liu \emph{et~al.}, ``On decoder-only architecture for speech-to-text and
  large language model integration,'' in \emph{2023 IEEE Automatic Speech
  Recognition and Understanding Workshop (ASRU)}.\hskip 1em plus 0.5em minus
  0.4em\relax IEEE, 2023, pp. 1--8.

\bibitem{bapna2022mslam}
A.~Bapna, C.~Cherry, Y.~Zhang, Y.~Jia, M.~Johnson, Y.~Cheng, S.~Khanuja,
  J.~Riesa, and A.~Conneau, ``mslam: Massively multilingual joint pre-training
  for speech and text,'' \emph{arXiv preprint arXiv:2202.01374}, 2022.

\bibitem{xiong2023fundamentals}
X.~Xiong, ``Fundamentals of speech recognition,'' 2023.

\bibitem{cheng2023alif}
P.~Cheng, Y.~Wang, P.~Huang, Z.~Ba, X.~Lin, F.~Lin, L.~Lu, and K.~Ren, ``Alif:
  Low-cost adversarial audio attacks on black-box speech platforms using
  linguistic features,'' in \emph{2024 IEEE Symposium on Security and Privacy
  (SP)}.\hskip 1em plus 0.5em minus 0.4em\relax IEEE Computer Society, 2023,
  pp. 56--56.

\bibitem{kreuk2018fooling}
F.~Kreuk, Y.~Adi, M.~Cisse, and J.~Keshet, ``Fooling end-to-end speaker
  verification with adversarial examples,'' in \emph{2018 IEEE international
  conference on acoustics, speech and signal processing (ICASSP)}.\hskip 1em
  plus 0.5em minus 0.4em\relax IEEE, 2018, pp. 1962--1966.

\bibitem{li2020practical}
Z.~Li, C.~Shi, Y.~Xie, J.~Liu, B.~Yuan, and Y.~Chen, ``Practical adversarial
  attacks against speaker recognition systems,'' in \emph{Proceedings of the
  21st international workshop on mobile computing systems and applications},
  2020, pp. 9--14.

\bibitem{xie2020real}
Y.~Xie, C.~Shi, Z.~Li, J.~Liu, Y.~Chen, and B.~Yuan, ``Real-time, universal,
  and robust adversarial attacks against speaker recognition systems,'' in
  \emph{ICASSP 2020-2020 IEEE international conference on acoustics, speech and
  signal processing (ICASSP)}.\hskip 1em plus 0.5em minus 0.4em\relax IEEE,
  2020, pp. 1738--1742.

\bibitem{zuo2024advtts}
C.-X. Zuo, Z.-J. Jia, and W.-J. Li, ``Advtts: Adversarial text-to-speech
  synthesis attack on speaker identification systems,'' in \emph{ICASSP
  2024-2024 IEEE International Conference on Acoustics, Speech and Signal
  Processing (ICASSP)}.\hskip 1em plus 0.5em minus 0.4em\relax IEEE, 2024, pp.
  4840--4844.

\bibitem{carlini2017towards}
N.~Carlini and D.~Wagner, ``Towards evaluating the robustness of neural
  networks,'' in \emph{2017 ieee symposium on security and privacy (sp)}.\hskip
  1em plus 0.5em minus 0.4em\relax Ieee, 2017, pp. 39--57.

\bibitem{zhang2018mitigating}
B.~H. Zhang, B.~Lemoine, and M.~Mitchell, ``Mitigating unwanted biases with
  adversarial learning,'' in \emph{Proceedings of the 2018 AAAI/ACM Conference
  on AI, Ethics, and Society}, 2018, pp. 335--340.

\bibitem{bender2018data}
E.~M. Bender and B.~Friedman, ``Data statements for natural language
  processing: Toward mitigating system bias and enabling better science,''
  \emph{Transactions of the Association for Computational Linguistics}, vol.~6,
  pp. 587--604, 2018.

\bibitem{beinborn2020semantic}
L.~Beinborn and R.~Choenni, ``Semantic drift in multilingual representations,''
  \emph{Computational Linguistics}, vol.~46, no.~3, pp. 571--603, 2020.

\bibitem{ghosal2023tango}
D.~Ghosal, N.~Majumder, A.~Mehrish, and S.~Poria, ``Text-to-audio generation
  using instruction tuned llm and latent diffusion model,'' \emph{arXiv
  preprint arXiv:2304.13731}, 2023.

\bibitem{liu2023audioldm}
H.~Liu, Z.~Chen, Y.~Yuan, X.~Mei, X.~Liu, D.~Mandic, W.~Wang, and M.~D.
  Plumbley, ``Audioldm: Text-to-audio generation with latent diffusion
  models,'' \emph{arXiv preprint arXiv:2301.12503}, 2023.

\bibitem{melechovsky2023mustango}
J.~Melechovsky, Z.~Guo, D.~Ghosal, N.~Majumder, D.~Herremans, and S.~Poria,
  ``Mustango: Toward controllable text-to-music generation,'' \emph{arXiv
  preprint arXiv:2311.08355}, 2023.

\bibitem{rombach2022high}
R.~Rombach, A.~Blattmann, D.~Lorenz, P.~Esser, and B.~Ommer, ``High-resolution
  image synthesis with latent diffusion models,'' in \emph{Proceedings of the
  IEEE/CVF conference on computer vision and pattern recognition}, 2022, pp.
  10\,684--10\,695.

\bibitem{panayotov2015librispeech}
V.~Panayotov, G.~Chen, D.~Povey, and S.~Khudanpur, ``Librispeech: an asr corpus
  based on public domain audio books,'' in \emph{2015 IEEE international
  conference on acoustics, speech and signal processing (ICASSP)}.\hskip 1em
  plus 0.5em minus 0.4em\relax IEEE, 2015, pp. 5206--5210.

\bibitem{jeub2009binaural}
M.~Jeub, M.~Schafer, and P.~Vary, ``A binaural room impulse response database
  for the evaluation of dereverberation algorithms,'' in \emph{2009 16th
  International Conference on Digital Signal Processing}.\hskip 1em plus 0.5em
  minus 0.4em\relax IEEE, 2009, pp. 1--5.

\bibitem{veaux2017cstr}
C.~Veaux, J.~Yamagishi, K.~MacDonald \emph{et~al.}, ``Cstr vctk corpus: English
  multi-speaker corpus for cstr voice cloning toolkit,'' \emph{University of
  Edinburgh. The Centre for Speech Technology Research (CSTR)}, 2017.

\bibitem{shi2020aishell}
Y.~Shi, H.~Bu, X.~Xu, S.~Zhang, and M.~Li, ``Aishell-3: A multi-speaker
  mandarin tts corpus and the baselines,'' \emph{arXiv preprint
  arXiv:2010.11567}, 2020.

\bibitem{ardila2019common}
R.~Ardila, M.~Branson, K.~Davis, M.~Henretty, M.~Kohler, J.~Meyer, R.~Morais,
  L.~Saunders, F.~M. Tyers, and G.~Weber, ``Common voice: A
  massively-multilingual speech corpus,'' \emph{arXiv preprint
  arXiv:1912.06670}, 2019.

\bibitem{wang-etal-2021-voxpopuli}
\BIBentryALTinterwordspacing
C.~Wang, M.~Riviere, A.~Lee, A.~Wu, C.~Talnikar, D.~Haziza, M.~Williamson,
  J.~Pino, and E.~Dupoux, ``{V}ox{P}opuli: A large-scale multilingual speech
  corpus for representation learning, semi-supervised learning and
  interpretation,'' in \emph{Proceedings of the 59th Annual Meeting of the
  Association for Computational Linguistics and the 11th International Joint
  Conference on Natural Language Processing (Volume 1: Long Papers)}.\hskip 1em
  plus 0.5em minus 0.4em\relax Online: Association for Computational
  Linguistics, Aug. 2021, pp. 993--1003. [Online]. Available:
  \url{https://aclanthology.org/2021.acl-long.80}
\BIBentrySTDinterwordspacing

\bibitem{rix2001perceptual}
A.~W. Rix, J.~G. Beerends, M.~P. Hollier, and A.~P. Hekstra, ``Perceptual
  evaluation of speech quality (pesq)-a new method for speech quality
  assessment of telephone networks and codecs,'' in \emph{2001 IEEE
  international conference on acoustics, speech, and signal processing.
  Proceedings (Cat. No. 01CH37221)}, vol.~2.\hskip 1em plus 0.5em minus
  0.4em\relax IEEE, 2001, pp. 749--752.

\bibitem{casanova2021sc}
E.~Casanova, C.~Shulby, E.~G{\"o}lge, N.~M. M{\"u}ller, F.~S. de~Oliveira,
  A.~C. Junior, A.~d.~S. Soares, S.~M. Aluisio, and M.~A. Ponti, ``Sc-glowtts:
  an efficient zero-shot multi-speaker text-to-speech model,'' \emph{arXiv
  preprint arXiv:2104.05557}, 2021.

\bibitem{timbrewatermarking-ndss2024}
C.~Liu, J.~Zhang, T.~Zhang, X.~Yang, W.~Zhang, and N.~Yu, ``Detecting voice
  cloning attacks via timbre watermarking,'' in \emph{Network and Distributed
  System Security Symposium}, 2024.

\bibitem{jemine2019real}
C.~Jemine, ``Real-time-voice-cloning,'' \emph{University of Li{\'e}ge,
  Li{\'e}ge, Belgium}, 2019.

\bibitem{reimers2019sentence}
N.~Reimers, ``Sentence-bert: Sentence embeddings using siamese bert-networks,''
  \emph{arXiv preprint arXiv:1908.10084}, 2019.

\bibitem{rei2020comet}
R.~Rei, C.~Stewart, A.~C. Farinha, and A.~Lavie, ``Comet: A neural framework
  for mt evaluation,'' \emph{arXiv preprint arXiv:2009.09025}, 2020.

\bibitem{liu2019roberta}
Y.~Liu, ``Roberta: A robustly optimized bert pretraining approach,''
  \emph{arXiv preprint arXiv:1907.11692}, 2019.

\bibitem{wang2018glue}
A.~Wang, ``Glue: A multi-task benchmark and analysis platform for natural
  language understanding,'' \emph{arXiv preprint arXiv:1804.07461}, 2018.

\end{thebibliography}

\newpage
\section{Appendix}

\subsection{GPT-4 Interaction Example for Semantic Similarity}
\begin{figure}[htbp]
\centering
\includegraphics[width=0.9\columnwidth]{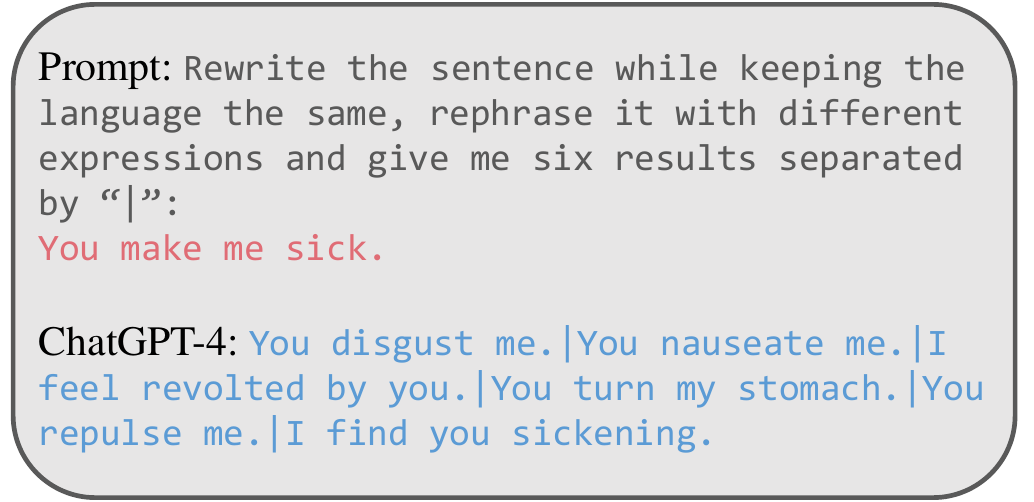}
\caption{Prompts and example outputs for generating semantically equivalent expressions using ChatGPT-4.}
\label{fig:prompt}
\end{figure}

To define appropriate semantic similarity thresholds for measuring Attack Success Rate (ASR), we employed sentence embedding similarity scores, which typically yield very low values for semantically unrelated sentences (examples in Table \ref{tb:threshold}). For each target semantic, we used ChatGPT-4 to generate six variations with the same meaning. These variations, such as ``shame on you" and ``you should be ashamed of yourself," were semantically consistent but structurally diverse. We can calculate similarity thresholds based on the similarity between the original text and the variations with the same semantic generated by GPT-4, using the ESIM and NSCORE values. The specific prompts and examples used for generating these variations are shown in \autoref{fig:prompt}.

\subsection{Updated Targets after Target Cycle Optimization}
\label{sec:perception-perturbation}
As described in \autoref{sec:cycle} and \autoref{fig:method-cycle-enhance}, we can perform Target Cycle Optimization (TCO) on the attack targets to generate semantically similar targets that are easier to attack. \autoref{tb:target-update} shows the changes in the text sequences corresponding to the same target semantics before and after applying Cycle Optimization, the semantic whose corresponding sentence changes during the updating are used for enhancement testing.

\begin{table}[htbp]
\centering
\caption{Updated Target after Cycle Optimization.}
\vspace{-.5em}
\resizebox{\columnwidth}{!}{
\label{tb:target-update}
\begin{tabular}{c|c|c}
\hline
Attack With & Original Target & Optimized Target \\ \hline
\makecell[c]{English \\ Mandarin \\ German \\ French} & 
\makecell[l]{``You make me sick." \\ \\ ``Shame on you." \\ \\ ``Are you insane?" \\ \\ ``Who do you think \\ you're talking to?" \\ \\ ``Don't waste my \\ time anymore."} & 
\makecell[l]{``You made me sick." \\ \\ ``You should be ashamed." \\ \\ ``Are you crazy?" \\ \\ ``Who do you think \\ you're talking to?" \\ \\ ``Don't waste my \\ time anymore."} \\ \hline
\end{tabular}
}
\end{table}

\subsection{More Music Style Prompt Exploring}
\label{sec:prompts}

Specifically, we selected three types of prompts: Techno, Classical, and Orchestral, to generate adversarial music. The test results are presented in \autoref{tb:prompt-select}. The results indicate that all music styles are effective in performing adversarial music attacks on speech translation systems. The experiments in previous sections used Techno as the default setting.

\begin{table}[!htbp]
\centering
\caption{Analysis of different music generation prompts on adversarial music, highlighting variations in attack performance and robustness across prompt categories. Blue-highlighted areas indicate tests conducted on \textbf{Seen} languages.}
\vspace{-.5em}
\fontsize{8}{10}\selectfont
\setlength{\tabcolsep}{9.5pt} 
\renewcommand{\arraystretch}{1.1}
\begin{tabular}{c|c|cc|c}
\hline
                             &                                 & \multicolumn{2}{c|}{Similarity With Target}                                          &                               \\ \cline{3-4}
\multirow{-2}{*}{Style}      & \multirow{-2}{*}{Target}        & \multicolumn{1}{c}{ESIM}                           & NSCORE                         & \multirow{-2}{*}{ASR}         \\ \hline
                             & \cellcolor[HTML]{DDEBF7}English & \multicolumn{1}{c}{\cellcolor[HTML]{DDEBF7}0.7879} & \cellcolor[HTML]{DDEBF7}0.7507 & \cellcolor[HTML]{DDEBF7}9/10  \\ 
                             & Mandarin                        & \multicolumn{1}{c}{0.5152}                         & 0.4257                         & 6/10                          \\ 
                             & German                          & \multicolumn{1}{c}{0.5706}                         & 0.4236                         & 6/10                          \\ 
                             & French                          & \multicolumn{1}{c}{0.4643}                         & 0.5759                         & 7/10                          \\ 
                             & Italian                         & \multicolumn{1}{c}{0.4877}                         & 0.6616                         & 7/10                          \\ 
\multirow{-6}{*}{Techno}     & Spanish                         & \multicolumn{1}{c}{0.4408}                         & 0.4661                         & 4/10                          \\ \hline
                             & \cellcolor[HTML]{DDEBF7}English & \multicolumn{1}{c}{\cellcolor[HTML]{DDEBF7}0.9788} & \cellcolor[HTML]{DDEBF7}0.9849 & \cellcolor[HTML]{DDEBF7}10/10 \\ 
                             & Mandarin                        & \multicolumn{1}{c}{0.4460}                         & 0.3418                         & 5/10                          \\ 
                             & German                          & \multicolumn{1}{c}{0.5288}                         & 0.3940                         & 7/10                          \\ 
                             & French                          & \multicolumn{1}{c}{0.5271}                         & 0.5235                         & 6/10                          \\ 
                             & Italian                         & \multicolumn{1}{c}{0.5531}                         & 0.6150                         & 8/10                          \\ 
\multirow{-6}{*}{Classical}  & Spanish                         & \multicolumn{1}{c}{0.5820}                         & 0.5776                         & 7/10                          \\ \hline
                             & \cellcolor[HTML]{DDEBF7}English & \multicolumn{1}{c}{\cellcolor[HTML]{DDEBF7}0.8353} & \cellcolor[HTML]{DDEBF7}0.7409 & \cellcolor[HTML]{DDEBF7}8/10  \\ 
                             & Mandarin                        & \multicolumn{1}{c}{0.4421}                         & 0.1909                         & 5/10                          \\ 
                             & German                          & \multicolumn{1}{c}{0.5890}                         & 0.4969                         & 7/10                          \\ 
                             & French                          & \multicolumn{1}{c}{0.5267}                         & 0.4562                         & 8/10                          \\ 
                             & Italian                         & \multicolumn{1}{c}{0.3812}                         & 0.2625                         & 4/10                          \\ 
\multirow{-6}{*}{Orchestral} & Spanish                         & \multicolumn{1}{c}{0.4964}                         & 0.1883                         & 5/10                          \\ \hline
\end{tabular}
\label{tb:prompt-select}
\end{table}

\begin{figure}[htbp]
\centering
\includegraphics[width=\columnwidth]{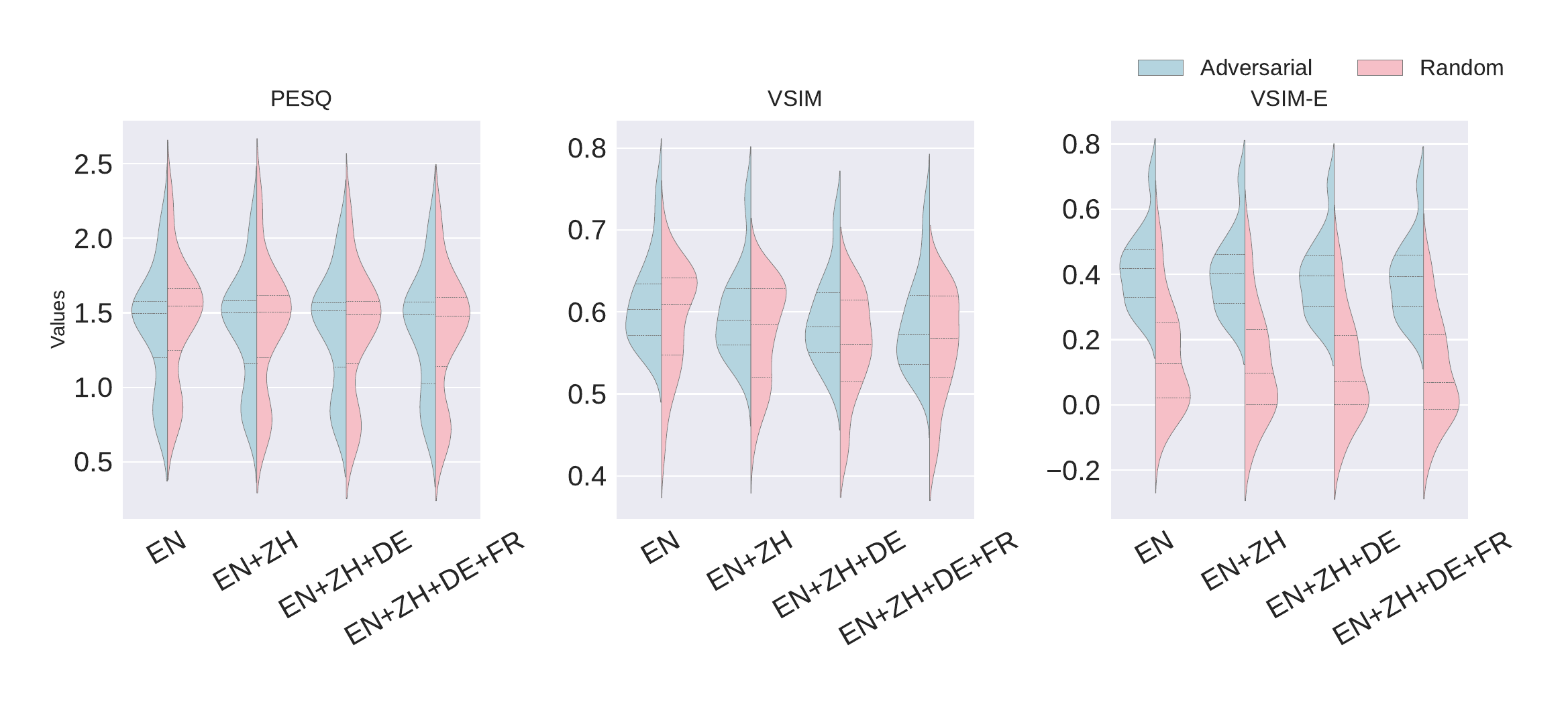}
% \vspace{-1em}
\footnotesize Note: EN=English, ZH=Mandarin, DE=German, FR=French
\caption{Violin plots comparing the perception influence of Adversarial and Random perturbations across different attack languages with $\epsilon = 0.1$. The internal black lines represent quartiles.}
\label{fig:distribution-perturbation-eps0.1}
\end{figure}

\subsection{More Discussion on perception of perturbation}
\label{sec:perception-perturbation}
We present the perceptual impact comparison between Adversarial and Random perturbations across different attack languages in \autoref{fig:distribution-perturbation-eps0.1} with standard violin plots.
Specifically, we introduce random noise with the same energy intensity as the adversarial perturbations in the original speech as a baseline. The effects of adding these perturbations or noise on the quality of the original speech are demonstrated using PESQ, VSIM, and VSIM-E metrics. The distributions in the figure indicate that adversarial perturbations result in better perceptual quality than random noise with the same energy intensity, especially in terms of the Seamless speech features (VSIM-E), where the quality degradation from adversarial perturbations is significantly lower. This is because our perturbations are specifically designed to avoid both high and low-frequency bands, as explained in \autoref{sec:attack-with-perturbation}. This design strategy effectively minimizes the impact on the core content of the speech (PESQ) while preserving speech style (VSIM, VSIM-E).

\subsection{More Tests on Different Perturbation Strength and Models}
\label{sec:ap-more-eps}
\cref{tb:perturbation-seamless-enhance-eps0.5,tb:perturbation-seamless-enhance-eps0.01} shows the results of Perturbation-based adversarial attacks on Seamless Large under conditions of $\epsilon=0.5$ and $0.01$, while \autoref{tb:perturbation-canary-enhance-eps0.1} shows the attack performance on Canary under different perturbation intensities. The results of Enhancement based on More Seen Languages are consistent with those in \autoref{tb:perturbation-seamless-enhance-eps0.1}, further indicating that a larger number of Seen languages enhances the generalization of adversarial perturbations across languages.

\begin{table}[htbp]
\centering
\fontsize{8}{10}\selectfont
\setlength{\tabcolsep}{9.5pt} 
\renewcommand{\arraystretch}{1.1}
\caption{Attack ability of adversarial perturbation($\epsilon=0.5$). Blue-highlighted areas indicate tests conducted on \textbf{Seen} languages.}
\vspace{-.5em}
\label{tb:perturbation-seamless-enhance-eps0.5}
\begin{tabular}{c|c|cc|c}
\hline
                                                                                                              &                                  & \multicolumn{2}{c|}{Similarity With Target}                                          &                               \\ \cline{3-4}
\multirow{-2}{*}{Attack   with}                                                                               & \multirow{-2}{*}{Target}         & \multicolumn{1}{c}{ESIM}                           & NSCORE                         & \multirow{-2}{*}{ASR}         \\ \hline
                                                                                                              & \cellcolor[HTML]{DDEBF7}English  & \multicolumn{1}{c}{\cellcolor[HTML]{DDEBF7}0.9612} & \multicolumn{1}{c|}{\cellcolor[HTML]{DDEBF7}0.9198} & \cellcolor[HTML]{DDEBF7}59/60 \\ 
                                                                                                              & Mandarin                         & \multicolumn{1}{c}{0.4329}                         & \multicolumn{1}{c|}{0.2300}                         & 20/60                         \\ 
                                                                                                              & German                           & \multicolumn{1}{c}{0.3496}                         & \multicolumn{1}{c|}{0.1749}                         & 27/60                         \\ 
                                                                                                              & French                           & \multicolumn{1}{c}{0.3772}                         & \multicolumn{1}{c|}{0.2000}                         & 32/60                         \\ 
                                                                                                              & Italian                          & \multicolumn{1}{c}{0.3897}                         & \multicolumn{1}{c|}{0.1880}                         & 17/60                         \\ 
\multirow{-6}{*}{English}                                                                                     & Spanish                          & \multicolumn{1}{c}{0.3803}                         & \multicolumn{1}{c|}{0.1497}                         & 22/60                         \\ \hline
                                                                                                              & \cellcolor[HTML]{DDEBF7}English  & \multicolumn{1}{c}{\cellcolor[HTML]{DDEBF7}0.9408} & \multicolumn{1}{c|}{\cellcolor[HTML]{DDEBF7}0.9293} & \cellcolor[HTML]{DDEBF7}58/60 \\ 
                                                                                                              & \cellcolor[HTML]{DDEBF7}Mandarin & \multicolumn{1}{c}{\cellcolor[HTML]{DDEBF7}0.9962} & \multicolumn{1}{c|}{\cellcolor[HTML]{DDEBF7}0.9737} & \cellcolor[HTML]{DDEBF7}60/60 \\ 
                                                                                                              & German                           & \multicolumn{1}{c}{0.5848}                         & \multicolumn{1}{c|}{0.4724}                         & 46/60                         \\ 
                                                                                                              & French                           & \multicolumn{1}{c}{0.6883}                         & \multicolumn{1}{c|}{0.5089}                         & 51/60                         \\ 
                                                                                                              & Italian                          & \multicolumn{1}{c}{0.5386}                         & \multicolumn{1}{c|}{0.4769}                         & 43/60                         \\ 
\multirow{-6}{*}{\begin{tabular}[c]{@{}c@{}}English\\      Mandarin\end{tabular}}                             & Spanish                          & \multicolumn{1}{c}{0.6566}                         & \multicolumn{1}{c|}{0.4779}                         & 50/60                         \\ \hline
                                                                                                              & \cellcolor[HTML]{DDEBF7}English  & \multicolumn{1}{c}{\cellcolor[HTML]{DDEBF7}0.9885} & \multicolumn{1}{c|}{\cellcolor[HTML]{DDEBF7}0.9649} & \cellcolor[HTML]{DDEBF7}60/60 \\ 
                                                                                                              & \cellcolor[HTML]{DDEBF7}Mandarin & \multicolumn{1}{c}{\cellcolor[HTML]{DDEBF7}0.9906} & \multicolumn{1}{c|}{\cellcolor[HTML]{DDEBF7}0.9758} & \cellcolor[HTML]{DDEBF7}59/60 \\ 
                                                                                                              & \cellcolor[HTML]{DDEBF7}German   & \multicolumn{1}{c}{\cellcolor[HTML]{DDEBF7}0.9537} & \multicolumn{1}{c|}{\cellcolor[HTML]{DDEBF7}0.9423} & \cellcolor[HTML]{DDEBF7}59/60 \\ 
                                                                                                              & French                           & \multicolumn{1}{c}{0.7526}                         & \multicolumn{1}{c|}{0.6487}                         & 56/60                         \\ 
                                                                                                              & Italian                          & \multicolumn{1}{c}{0.6612}                         & \multicolumn{1}{c|}{0.6764}                         & 53/60                         \\ 
\multirow{-6}{*}{\begin{tabular}[c]{@{}c@{}}English\\      Mandarin \\      German\end{tabular}}              & Spanish                          & \multicolumn{1}{c}{0.7087}                         & \multicolumn{1}{c|}{0.5904}                         & 54/60                         \\ \hline
                                                                                                              & \cellcolor[HTML]{DDEBF7}English  & \multicolumn{1}{c}{\cellcolor[HTML]{DDEBF7}0.9135} & \multicolumn{1}{c|}{\cellcolor[HTML]{DDEBF7}0.9374} & \cellcolor[HTML]{DDEBF7}59/60 \\ 
                                                                                                              & \cellcolor[HTML]{DDEBF7}Mandarin & \multicolumn{1}{c}{\cellcolor[HTML]{DDEBF7}0.9435} & \multicolumn{1}{c|}{\cellcolor[HTML]{DDEBF7}0.9093} & \cellcolor[HTML]{DDEBF7}58/60 \\ 
                                                                                                              & \cellcolor[HTML]{DDEBF7}German   & \multicolumn{1}{c}{\cellcolor[HTML]{DDEBF7}0.9179} & \multicolumn{1}{c|}{\cellcolor[HTML]{DDEBF7}0.9073} & \cellcolor[HTML]{DDEBF7}60/60 \\ 
                                                                                                              & \cellcolor[HTML]{DDEBF7}French   & \multicolumn{1}{c}{\cellcolor[HTML]{DDEBF7}0.9837} & \multicolumn{1}{c|}{\cellcolor[HTML]{DDEBF7}0.9676} & \cellcolor[HTML]{DDEBF7}59/60 \\ 
                                                                                                              & Italian                          & \multicolumn{1}{c}{0.6616}                         & \multicolumn{1}{c|}{0.7699}                         & 56/60                         \\ 
\multirow{-6}{*}{\begin{tabular}[c]{@{}c@{}}English\\      Mandarin\\      German\\      French\end{tabular}} & Spanish                          & \multicolumn{1}{c}{0.7281}                         & \multicolumn{1}{c|}{0.7050}                         & 55/60                         \\ \hline
\end{tabular}

\end{table}

\begin{table}[htbp]
\centering
\fontsize{8}{10}\selectfont
\setlength{\tabcolsep}{9.5pt} 
\renewcommand{\arraystretch}{1.1}
\caption{Attack ability of adversarial perturbation($\epsilon=0.01$). Blue-highlighted areas indicate tests conducted on \textbf{Seen} languages.}
\vspace{-.5em}

\label{tb:perturbation-seamless-enhance-eps0.01}
\begin{tabular}{c|c|cc|c}
\hline
                                                                                                              &                                  & \multicolumn{2}{c|}{Similarity With Target}                                          &                               \\ \cline{3-4}
\multirow{-2}{*}{Attack   with}                                                                               & \multirow{-2}{*}{Target}         & \multicolumn{1}{c}{ESIM}                           & NSCORE                         & \multirow{-2}{*}{ASR}         \\ \hline
                                                                                                              & \cellcolor[HTML]{DDEBF7}English  & \multicolumn{1}{c}{\cellcolor[HTML]{DDEBF7}0.5503} & \multicolumn{1}{c|}{\cellcolor[HTML]{DDEBF7}0.4207} & \cellcolor[HTML]{DDEBF7}32/60 \\ 
                                                                                                              & Mandarin                         & \multicolumn{1}{c}{0.2371}                         & \multicolumn{1}{c|}{0.0978}                         & 10/60                         \\ 
                                                                                                              & German                           & \multicolumn{1}{c}{0.2186}                         & \multicolumn{1}{c|}{0.0529}                         & 12/60                         \\ 
                                                                                                              & French                           & \multicolumn{1}{c}{0.2473}                         & \multicolumn{1}{c|}{0.0993}                         & 17/60                         \\ 
                                                                                                              & Italian                          & \multicolumn{1}{c}{0.2252}                         & \multicolumn{1}{c|}{0.1177}                         & 15/60                         \\ 
\multirow{-6}{*}{English}                                                                                     & Spanish                          & \multicolumn{1}{c}{0.2403}                         & \multicolumn{1}{c|}{0.0969}                         & 20/60                         \\ \hline
                                                                                                              & \cellcolor[HTML]{DDEBF7}English  & \multicolumn{1}{c}{\cellcolor[HTML]{DDEBF7}0.6510} & \multicolumn{1}{c|}{\cellcolor[HTML]{DDEBF7}0.5460} & \cellcolor[HTML]{DDEBF7}38/60 \\ 
                                                                                                              & \cellcolor[HTML]{DDEBF7}Mandarin & \multicolumn{1}{c}{\cellcolor[HTML]{DDEBF7}0.7140} & \multicolumn{1}{c|}{\cellcolor[HTML]{DDEBF7}0.6113} & \cellcolor[HTML]{DDEBF7}42/60 \\ 
                                                                                                              & German                           & \multicolumn{1}{c}{0.3739}                         & \multicolumn{1}{c|}{0.2037}                         & 27/60                         \\ 
                                                                                                              & French                           & \multicolumn{1}{c}{0.4415}                         & \multicolumn{1}{c|}{0.2127}                         & 30/60                         \\ 
                                                                                                              & Italian                          & \multicolumn{1}{c}{0.3200}                         & \multicolumn{1}{c|}{0.1909}                         & 17/60                         \\ 
\multirow{-6}{*}{\begin{tabular}[c]{@{}c@{}}English\\      Mandarin\end{tabular}}                             & Spanish                          & \multicolumn{1}{c}{0.3527}                         & \multicolumn{1}{c|}{0.1618}                         & 24/60                         \\ \hline
                                                                                                              & \cellcolor[HTML]{DDEBF7}English  & \multicolumn{1}{c}{\cellcolor[HTML]{DDEBF7}0.6399} & \multicolumn{1}{c|}{\cellcolor[HTML]{DDEBF7}0.4652} & \cellcolor[HTML]{DDEBF7}38/60 \\ 
                                                                                                              & \cellcolor[HTML]{DDEBF7}Mandarin & \multicolumn{1}{c}{\cellcolor[HTML]{DDEBF7}0.6041} & \multicolumn{1}{c|}{\cellcolor[HTML]{DDEBF7}0.4386} & \cellcolor[HTML]{DDEBF7}36/60 \\ 
                                                                                                              & \cellcolor[HTML]{DDEBF7}German   & \multicolumn{1}{c}{\cellcolor[HTML]{DDEBF7}0.5569} & \multicolumn{1}{c|}{\cellcolor[HTML]{DDEBF7}0.3972} & \cellcolor[HTML]{DDEBF7}35/60 \\ 
                                                                                                              & French                           & \multicolumn{1}{c}{0.4555}                         & \multicolumn{1}{c|}{0.3002}                         & 30/60                         \\ 
                                                                                                              & Italian                          & \multicolumn{1}{c}{0.3636}                         & \multicolumn{1}{c|}{0.2536}                         & 22/60                         \\ 
\multirow{-6}{*}{\begin{tabular}[c]{@{}c@{}}English\\      Mandarin \\      German\end{tabular}}              & Spanish                          & \multicolumn{1}{c}{0.4397}                         & \multicolumn{1}{c|}{0.2442}                         & 33/60                         \\ \hline
                                                                                                              & \cellcolor[HTML]{DDEBF7}English  & \multicolumn{1}{c}{\cellcolor[HTML]{DDEBF7}0.7435} & \multicolumn{1}{c|}{\cellcolor[HTML]{DDEBF7}0.6826} & \cellcolor[HTML]{DDEBF7}47/60 \\ 
                                                                                                              & \cellcolor[HTML]{DDEBF7}Mandarin & \multicolumn{1}{c}{\cellcolor[HTML]{DDEBF7}0.7212} & \multicolumn{1}{c|}{\cellcolor[HTML]{DDEBF7}0.5687} & \cellcolor[HTML]{DDEBF7}42/60 \\ 
                                                                                                              & \cellcolor[HTML]{DDEBF7}German   & \multicolumn{1}{c}{\cellcolor[HTML]{DDEBF7}0.5928} & \multicolumn{1}{c|}{\cellcolor[HTML]{DDEBF7}0.4469} & \cellcolor[HTML]{DDEBF7}38/60 \\ 
                                                                                                              & \cellcolor[HTML]{DDEBF7}French   & \multicolumn{1}{c}{\cellcolor[HTML]{DDEBF7}0.7131} & \multicolumn{1}{c|}{\cellcolor[HTML]{DDEBF7}0.6209} & \cellcolor[HTML]{DDEBF7}47/60 \\ 
                                                                                                              & Italian                          & \multicolumn{1}{c}{0.4997}                         & \multicolumn{1}{c|}{0.3942}                         & 37/60                         \\ 
\multirow{-6}{*}{\begin{tabular}[c]{@{}c@{}}English\\      Mandarin\\      German\\      French\end{tabular}} & Spanish                          & \multicolumn{1}{c}{0.5204}                         & \multicolumn{1}{c|}{0.3496}                         & 42/60                         \\ \hline
\end{tabular}

\end{table}

\begin{table*}[t]
\centering
\caption{Rating rules of subjective evaluation}
\vspace{-.5em}

\label{tb:mos}
\begin{tabular}{>{\centering\arraybackslash}p{1cm}|p{5.5cm}|p{5.5cm}}
\hline
\textbf{Rating} & \textbf{Speech} & \textbf{Music} \\
\hline
1 & Very Poor: \newline Audio content is incomprehensible due to severe distortion or issues. & Very Poor: \newline Extremely antagonizing, completely intolerable, want to turn it off immediately. \\
\hline
2 & Poor: \newline Audio has noticeable defects, making it difficult to understand the content. & Poor: \newline Strongly impactful, really unpleasant to listen to. \\
\hline
3 & Fair: \newline Audio meets minimum standards, content is understandable. & Fair: \newline Moderately stimulating, starting to cause discomfort. \\
\hline
4 & Good: \newline Audio is clear, with only minor defects if any. & Good: \newline Slight discomfort, a bit annoying, but still tolerable. \\
\hline
5 & Excellent: \newline Audio quality is very high, sound is clear and content is fully comprehensible. & Excellent: \newline No noticeable impact felt. \\
\hline
\end{tabular}
\vspace{-1em}
\end{table*}

\begin{table}[htbp]
\vspace{4.8em}
\centering
\fontsize{7}{8.5}\selectfont
\setlength{\tabcolsep}{9pt} 

\caption{Attack ability on Canary~\cite{canary} of adversarial perturbation. Blue-highlighted areas indicate tests conducted on \textbf{Seen} languages.}
\vspace{-.5em}
\label{tb:perturbation-canary-enhance-eps0.1}
\begin{tabular}{c|c|c|cc|c}
\hline
                             &                                                                                                              &                                 & \multicolumn{2}{c|}{Similarity With Target}                                          &                               \\ \cline{4-5}
\multirow{-2}{*}{$\epsilon$} & \multirow{-2}{*}{Attack with}                                                                                & \multirow{-2}{*}{Target}        & \multicolumn{1}{c}{ESIM}                           & NSCORE                         & \multirow{-2}{*}{ASR}         \\ \hline
                             &                                                                                                              & \cellcolor[HTML]{DDEBF7}English & \multicolumn{1}{c}{\cellcolor[HTML]{DDEBF7}0.4797} & \cellcolor[HTML]{DDEBF7}0.2294 & \cellcolor[HTML]{DDEBF7}4/10  \\  
                             &                                                                                                              & French                          & \multicolumn{1}{c}{0.2652}                         & 0.0459                         & 2/10                          \\  
                             &                                                                                                              & German                          & \multicolumn{1}{c}{0.2127}                         & 0.0483                         & 2/10                          \\  
                             & \multirow{-4}{*}{English}                                                                                    & Spanish                         & \multicolumn{1}{c}{0.2704}                         & 0.1376                         & 1/10                          \\ \cline{2-6} 
                             &                                                                                                              & \cellcolor[HTML]{DDEBF7}English & \multicolumn{1}{c}{\cellcolor[HTML]{DDEBF7}0.8119} & \cellcolor[HTML]{DDEBF7}0.7101 & \cellcolor[HTML]{DDEBF7}8/10  \\  
                             &                                                                                                              & \cellcolor[HTML]{DDEBF7}French  & \multicolumn{1}{c}{\cellcolor[HTML]{DDEBF7}0.7237} & \cellcolor[HTML]{DDEBF7}0.5347 & \cellcolor[HTML]{DDEBF7}7/10  \\  
                             &                                                                                                              & German                          & \multicolumn{1}{c}{0.4688}                         & 0.3797                         & 6/10                          \\  
                             & \multirow{-4}{*}{\begin{tabular}[c]{@{}c@{}}English\\      French\end{tabular}}                              & Spanish                         & \multicolumn{1}{c}{0.4708}                         & 0.3127                         & 5/10                          \\ \cline{2-6} 
                             &                                                                                                              & \cellcolor[HTML]{DDEBF7}English & \multicolumn{1}{c}{\cellcolor[HTML]{DDEBF7}0.9698} & \cellcolor[HTML]{DDEBF7}0.8947 & \cellcolor[HTML]{DDEBF7}10/10 \\  
                             &                                                                                                              & \cellcolor[HTML]{DDEBF7}French  & \multicolumn{1}{c}{\cellcolor[HTML]{DDEBF7}0.9129} & \cellcolor[HTML]{DDEBF7}0.7865 & \cellcolor[HTML]{DDEBF7}10/10 \\  
                             &                                                                                                              & \cellcolor[HTML]{DDEBF7}German  & \multicolumn{1}{c}{\cellcolor[HTML]{DDEBF7}0.9318} & \cellcolor[HTML]{DDEBF7}0.8851 & \cellcolor[HTML]{DDEBF7}10/10 \\  
                             & \multirow{-4}{*}{\begin{tabular}[c]{@{}c@{}}English\\      French\\      German\end{tabular}}                & Spanish                         & \multicolumn{1}{c}{0.6151}                         & 0.5134                         & 6/10                          \\ \cline{2-6} 
                             &                                                                                                              & \cellcolor[HTML]{DDEBF7}English & \multicolumn{1}{c}{\cellcolor[HTML]{DDEBF7}1.0000} & \cellcolor[HTML]{DDEBF7}0.9846 & \cellcolor[HTML]{DDEBF7}10/10 \\  
                             &                                                                                                              & \cellcolor[HTML]{DDEBF7}French  & \multicolumn{1}{c}{\cellcolor[HTML]{DDEBF7}0.9919} & \cellcolor[HTML]{DDEBF7}0.9821 & \cellcolor[HTML]{DDEBF7}10/10 \\  
                             &                                                                                                              & \cellcolor[HTML]{DDEBF7}German  & \multicolumn{1}{c}{\cellcolor[HTML]{DDEBF7}0.9331} & \cellcolor[HTML]{DDEBF7}0.8861 & \cellcolor[HTML]{DDEBF7}10/10 \\  
\multirow{-16}{*}{0.5}       & \multirow{-4}{*}{\begin{tabular}[c]{@{}c@{}}English\\      French\\      German\\      Spanish\end{tabular}} & \cellcolor[HTML]{DDEBF7}Spanish & \multicolumn{1}{c}{\cellcolor[HTML]{DDEBF7}0.9409} & \cellcolor[HTML]{DDEBF7}0.9074 & \cellcolor[HTML]{DDEBF7}10/10 \\ \hline
                             &                                                                                                              & \cellcolor[HTML]{DDEBF7}English & \multicolumn{1}{c}{\cellcolor[HTML]{DDEBF7}0.5712} & \cellcolor[HTML]{DDEBF7}0.3437 & \cellcolor[HTML]{DDEBF7}4/10  \\  
                             &                                                                                                              & French                          & \multicolumn{1}{c}{0.2770}                         & 0.1654                         & 4/10                          \\  
                             &                                                                                                              & German                          & \multicolumn{1}{c}{0.2953}                         & 0.2561                         & 6/10                          \\  
                             & \multirow{-4}{*}{English}                                                                                    & Spanish                         & \multicolumn{1}{c}{0.2844}                         & 0.0991                         & 3/10                          \\ \cline{2-6} 
                             &                                                                                                              & \cellcolor[HTML]{DDEBF7}English & \multicolumn{1}{c}{\cellcolor[HTML]{DDEBF7}0.7306} & \cellcolor[HTML]{DDEBF7}0.6853 & \cellcolor[HTML]{DDEBF7}8/10  \\  
                             &                                                                                                              & \cellcolor[HTML]{DDEBF7}French  & \multicolumn{1}{c}{\cellcolor[HTML]{DDEBF7}0.7085} & \cellcolor[HTML]{DDEBF7}0.4940 & \cellcolor[HTML]{DDEBF7}7/10  \\  
                             &                                                                                                              & German                          & \multicolumn{1}{c}{0.4919}                         & 0.2257                         & 4/10                          \\  
                             & \multirow{-4}{*}{\begin{tabular}[c]{@{}c@{}}English\\      French\end{tabular}}                              & Spanish                         & \multicolumn{1}{c}{0.5169}                         & 0.2816                         & 4/10                          \\ \cline{2-6} 
                             &                                                                                                              & \cellcolor[HTML]{DDEBF7}English & \multicolumn{1}{c}{\cellcolor[HTML]{DDEBF7}0.9863} & \cellcolor[HTML]{DDEBF7}0.9850 & \cellcolor[HTML]{DDEBF7}10/10 \\  
                             &                                                                                                              & \cellcolor[HTML]{DDEBF7}French  & \multicolumn{1}{c}{\cellcolor[HTML]{DDEBF7}0.7552} & \cellcolor[HTML]{DDEBF7}0.6675 & \cellcolor[HTML]{DDEBF7}8/10  \\  
                             &                                                                                                              & \cellcolor[HTML]{DDEBF7}German  & \multicolumn{1}{c}{\cellcolor[HTML]{DDEBF7}0.9024} & \cellcolor[HTML]{DDEBF7}0.8009 & \cellcolor[HTML]{DDEBF7}10/10 \\  
                             & \multirow{-4}{*}{\begin{tabular}[c]{@{}c@{}}English\\      French\\      German\end{tabular}}                & Spanish                         & \multicolumn{1}{c}{0.6242}                         & 0.3079                         & 5/10                          \\ \cline{2-6} 
                             &                                                                                                              & \cellcolor[HTML]{DDEBF7}English & \multicolumn{1}{c}{\cellcolor[HTML]{DDEBF7}1.0000} & \cellcolor[HTML]{DDEBF7}0.9846 & \cellcolor[HTML]{DDEBF7}10/10 \\  
                             &                                                                                                              & \cellcolor[HTML]{DDEBF7}French  & \multicolumn{1}{c}{\cellcolor[HTML]{DDEBF7}0.8968} & \cellcolor[HTML]{DDEBF7}0.7989 & \cellcolor[HTML]{DDEBF7}9/10  \\  
                             &                                                                                                              & \cellcolor[HTML]{DDEBF7}German  & \multicolumn{1}{c}{\cellcolor[HTML]{DDEBF7}0.9267} & \cellcolor[HTML]{DDEBF7}0.8861 & \cellcolor[HTML]{DDEBF7}10/10 \\  
\multirow{-16}{*}{0.1}       & \multirow{-4}{*}{\begin{tabular}[c]{@{}c@{}}English\\      French\\      German\\      Spanish\end{tabular}} & \cellcolor[HTML]{DDEBF7}Spanish & \multicolumn{1}{c}{\cellcolor[HTML]{DDEBF7}0.9232} & \cellcolor[HTML]{DDEBF7}0.9540 & \cellcolor[HTML]{DDEBF7}10/10 \\ \hline
                             &                                                                                                              & \cellcolor[HTML]{DDEBF7}English & \multicolumn{1}{c}{\cellcolor[HTML]{DDEBF7}0.3020} & \cellcolor[HTML]{DDEBF7}0.2100 & \cellcolor[HTML]{DDEBF7}3/10  \\  
                             &                                                                                                              & French                          & \multicolumn{1}{c}{0.2899}                         & 0.0037                         & 1/10                          \\  
                             &                                                                                                              & German                          & \multicolumn{1}{c}{0.1405}                         & 0.0137                         & 1/10                          \\  
                             & \multirow{-4}{*}{English}                                                                                    & Spanish                         & \multicolumn{1}{c}{0.1940}                         & 0.0244                         & 1/10                          \\ \cline{2-6} 
                             &                                                                                                              & \cellcolor[HTML]{DDEBF7}English & \multicolumn{1}{c}{\cellcolor[HTML]{DDEBF7}0.7560} & \cellcolor[HTML]{DDEBF7}0.5972 & \cellcolor[HTML]{DDEBF7}8/10  \\  
                             &                                                                                                              & \cellcolor[HTML]{DDEBF7}French  & \multicolumn{1}{c}{\cellcolor[HTML]{DDEBF7}0.6483} & \cellcolor[HTML]{DDEBF7}0.2896 & \cellcolor[HTML]{DDEBF7}6/10  \\  
                             &                                                                                                              & German                          & \multicolumn{1}{c}{0.4350}                         & 0.1764                         & 4/10                          \\  
                             & \multirow{-4}{*}{\begin{tabular}[c]{@{}c@{}}English\\      French\end{tabular}}                              & Spanish                         & \multicolumn{1}{c}{0.5015}                         & 0.0645                         & 4/10                          \\ \cline{2-6} 
                             &                                                                                                              & \cellcolor[HTML]{DDEBF7}English & \multicolumn{1}{c}{\cellcolor[HTML]{DDEBF7}0.9228} & \cellcolor[HTML]{DDEBF7}0.8862 & \cellcolor[HTML]{DDEBF7}9/10  \\  
                             &                                                                                                              & \cellcolor[HTML]{DDEBF7}French  & \multicolumn{1}{c}{\cellcolor[HTML]{DDEBF7}0.7723} & \cellcolor[HTML]{DDEBF7}0.5921 & \cellcolor[HTML]{DDEBF7}8/10  \\  
                             &                                                                                                              & \cellcolor[HTML]{DDEBF7}German  & \multicolumn{1}{c}{\cellcolor[HTML]{DDEBF7}0.8610} & \cellcolor[HTML]{DDEBF7}0.7900 & \cellcolor[HTML]{DDEBF7}9/10  \\  
                             & \multirow{-4}{*}{\begin{tabular}[c]{@{}c@{}}English\\      French\\      German\end{tabular}}                & Spanish                         & \multicolumn{1}{c}{0.6412}                         & 0.5914                         & 6/10                          \\ \cline{2-6} 
                             &                                                                                                              & \cellcolor[HTML]{DDEBF7}English & \multicolumn{1}{c}{\cellcolor[HTML]{DDEBF7}0.8160} & \cellcolor[HTML]{DDEBF7}0.7941 & \cellcolor[HTML]{DDEBF7}8/10  \\  
                             &                                                                                                              & \cellcolor[HTML]{DDEBF7}French  & \multicolumn{1}{c}{\cellcolor[HTML]{DDEBF7}0.8868} & \cellcolor[HTML]{DDEBF7}0.8594 & \cellcolor[HTML]{DDEBF7}8/10  \\  
                             &                                                                                                              & \cellcolor[HTML]{DDEBF7}German  & \multicolumn{1}{c}{\cellcolor[HTML]{DDEBF7}0.8992} & \cellcolor[HTML]{DDEBF7}0.8293 & \cellcolor[HTML]{DDEBF7}9/10  \\  
\multirow{-16}{*}{0.01}      & \multirow{-4}{*}{\begin{tabular}[c]{@{}c@{}}English\\      French\\      German\\      Spanish\end{tabular}} & \cellcolor[HTML]{DDEBF7}Spanish & \multicolumn{1}{c}{\cellcolor[HTML]{DDEBF7}0.7928} & \cellcolor[HTML]{DDEBF7}0.6876 & \cellcolor[HTML]{DDEBF7}7/10  \\ \hline
\end{tabular}

\end{table}

\subsection{MOS Test Details}
\label{sec:mos}
In addition to the objective quality assessments, we also conducted subjective experiments on both adversarial perturbations and adversarial music.
\autoref{tb:mos} provides a detailed scoring criteria for assessing adversarial perturbations in speech and the quality of the generated adversarial music. The MOS scores represent the perceptual quality of both speech and music, with specific ratings for the level of distortion caused by adversarial perturbations and the overall audio quality. 

For the evaluation, 20 participants were invited to rate the quality of speech overlaid with adversarial perturbations and the generated adversarial music.
To establish a baseline, random white noise matching the energy intensity of each adversarial perturbation was generated. Similarly, white noise with the same energy intensity was created for each piece of adversarial music.

\begin{figure}[t]
\centering
\includegraphics[width=\columnwidth]{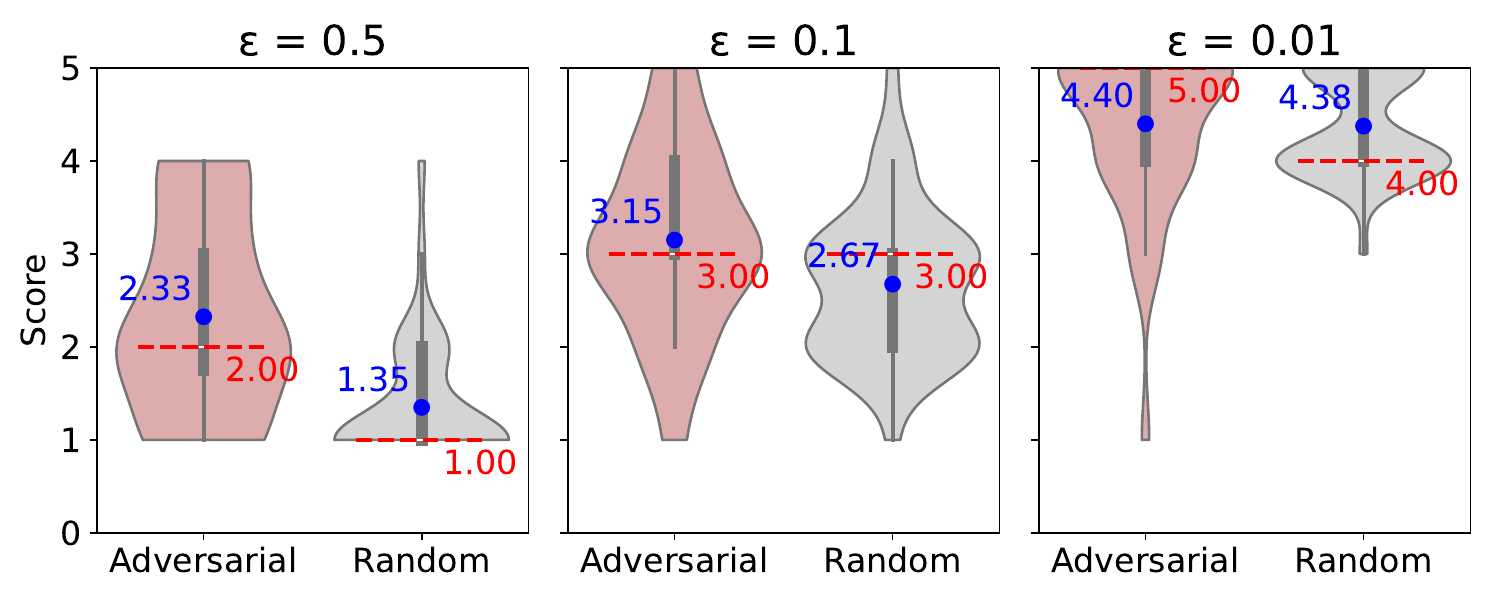}
\caption{MOS test score of perturbations. The red line represents the median, and the blue dot represents the mean.}
\label{fig:mos}
\end{figure}

\begin{figure}[!htb]
\centering
\includegraphics[width=0.8\columnwidth]{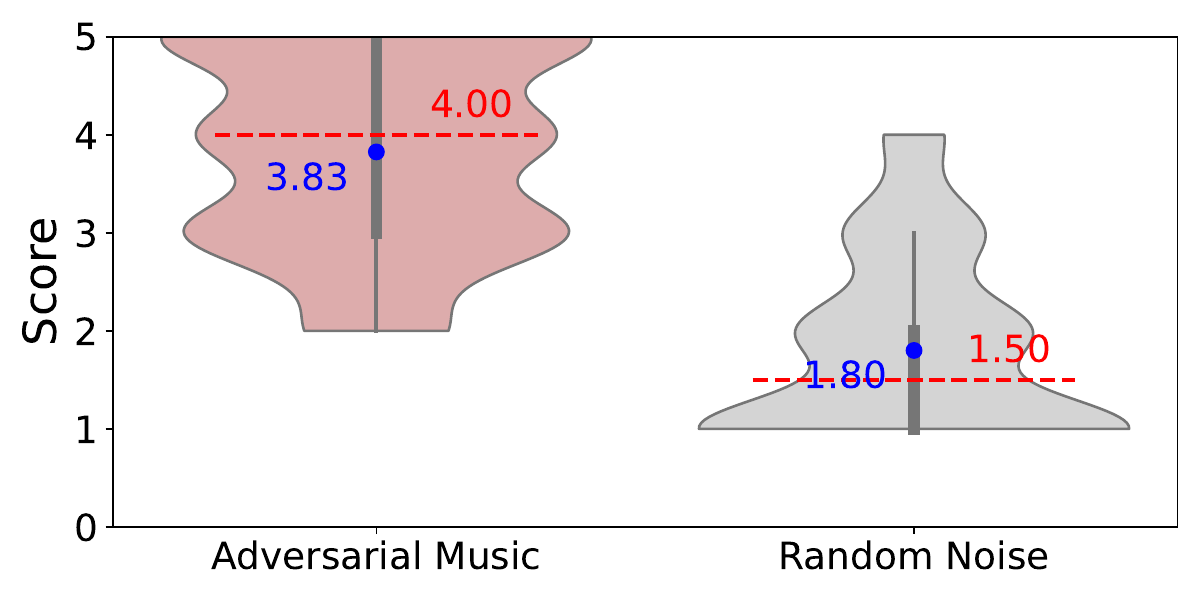}
\caption{MOS test score of adversarial musics, the red line represents the median, and the blue dot represents the mean.}
\label{fig:mos-music}
\vspace{-.5em}
\end{figure}

As shown in \autoref{fig:mos}, increasing the perturbation strength generally leads to lower scores. However, adversarial perturbations consistently exhibit better perceptual quality than random perturbations of the same strength, especially at higher perturbation levels. With the default \(\epsilon = 0.1\), \autoref{tb:mos} demonstrates that most adversarial perturbations do not significantly affect the perception of speech content. Regarding the generated adversarial music, as illustrated in \autoref{fig:mos-music}, it shows superior perceptual quality compared to random perturbations of the same strength. Additionally, the generated music receives high ratings, indicating its imperceptibility.

\subsection{Details of Over-the-air Simulation}
\label{sec:ota-simulate}
To ensure that the adversarial music exhibits over-the-air robustness, enabling attacks in real-world environments, we introduce simulated air transmission distortions and environmental noise before passing the generated adversarial music to the target model for inference and gradient acquisition. Specifically, in each optimization step, we sample a segment of human speech from the Librispeech dataset \cite{panayotov2015librispeech} and overlay it onto the adversarial music to simulate a noisy speech environment. 
Additionally, we use the Aachen Impulse Response Database \cite{jeub2009binaural} to simulate environmental reverberation. During each optimization step, an impulse response is randomly sampled from the dataset with a certain probability and convolved with the input generated adversarial music.
Moreover, we add small random white noise to the reverberated audio.

\subsection{Devices Details}
\label{sec:device}
\begin{figure}[!htbp]
\centering
\includegraphics[width=0.8\columnwidth]{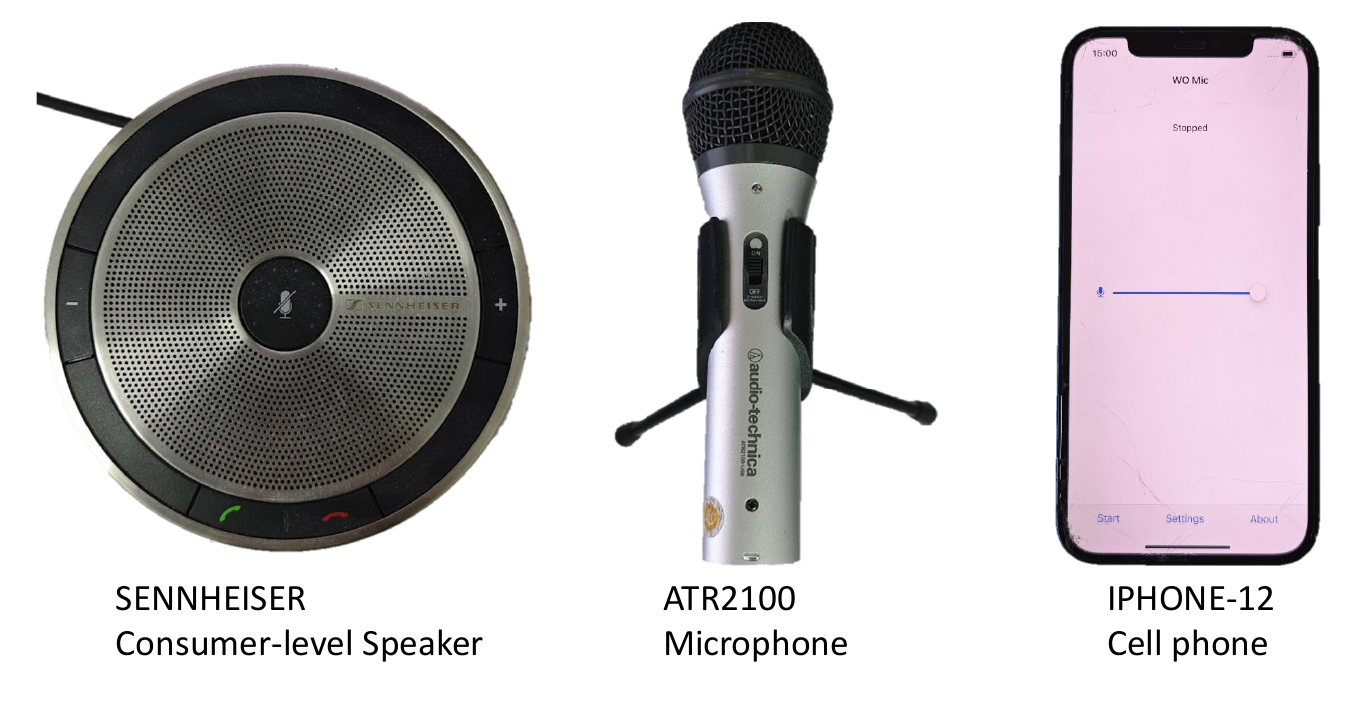}
\vspace{-1em}
\caption{Audio playing and recording devices used in physical tests.}
\label{fig:device}
\end{figure}

\autoref{fig:device} lists the audio playback and recording devices used in our physical world Over-the-air attacks. Specifically, we use the consumer-level speaker SENNHEISER SP10 for audio playback, and the consumer-grade microphone ATR2100 along with an iPhone 12 as the recording devices.

\end{document}